\documentclass[10pt,a4paper]{article}
\pdfoutput=1

\usepackage{jheppub}
\usepackage{booktabs}
\usepackage{multirow}
\usepackage{epsfig}
\usepackage{lscape}
\usepackage{rotating}
\usepackage{epsf}
\usepackage{pstricks}
\usepackage{slashed}
\usepackage{bm}
\usepackage{amsmath}
\usepackage{amssymb}
\usepackage{axodraw4j}
\usepackage{mathtools}
\usepackage{booktabs}
\usepackage{array}
\usepackage{caption}
\usepackage{subcaption}
\usepackage[utf8]{inputenc}
\usepackage{float}

\newcolumntype{L}[1]{>{\raggedright\let\newline\\\arraybackslash\hspace{0pt}}m{#1}}
\newcolumntype{C}[1]{>{\centering\let\newline\\\arraybackslash\hspace{0pt}}m{#1}}
\newcolumntype{R}[1]{>{\raggedleft\let\newline\\\arraybackslash\hspace{0pt}}m{#1}}

\allowdisplaybreaks

\newcommand{\beq}{\begin{align}}
\newcommand{\eeq}{\end{align}}

\newcommand{\MET}{\textbf{\textit{E}$_{\rm T}^{\rm miss}$}}
\graphicspath{{./figs/}}

\title{Dark Matter characterization at the LHC in the Effective Field Theory approach}

\author[a,b,1]{Alexander Belyaev\note{e-mail: a.belyaev@soton.ac.uk}}
\author[a,b,2]{Luca Panizzi\note{e-mail: l.panizzi@soton.ac.uk}}
\author[c,3]{Alexander Pukhov\note{e-mail: pukhov@lapp.in2p3.fr}}
\author[a,b,4]{Marc Thomas\note{e-mail: m.c.thomas@soton.ac.uk}}

\affiliation[a]{School of Physics and Astronomy, University of Southampton, Southampton SO17 1BJ, UK}
\affiliation[b]{Particle Physics Department, Rutherford Appleton Laboratory, Didcot, Oxon OX11 0QX, UK}
\affiliation[c]{Skobeltsyn Inst. of Nuclear Physics, Moscow State Univ., Moscow 119992, Russia}

\abstract{
We have studied the complete set of dimension 5 and dimension 6 effective operators involving the interaction of scalar, fermion and vector Dark Matter (DM) with SM quarks and gluons,
to explore the possibility to distinguish these operators and characterise the spin of DM at the LHC.
We have found that three factors --  the effective dimension of the operator, 
the structure of the SM part of the operator and the parton densities of the SM particles connected to the operator
-- uniquely define the shape of the (unobservable) invariant mass distribution of the DM pair and, consequently, the shape of the (observable) \MET{} 
distribution related to it.
Using  $\chi^2$ analysis, we found that at the LHC, with a luminosity of 300 fb$^{-1}$, certain classes of EFT operators can be distinguished from each other.
Hence, since DM spin is partly correlated with the factors defining the shape of \MET{}, the LHC can potentially shed a light also on DM spin. 
We have also observed a drastic difference in the efficiencies (up to two orders of magnitude) for large \MET{} cuts scenarios with different DM spin, thus indicating that the DM discovery potential strongly depends on it.
The study we perform here can be applied more generally than within the EFT paradigm, where the DM mediator is not produced on-the-mass-shell, such as the case of t-channel mediator or mediator with mass below $2M_{DM}$, where the invariant mass of the DM pair is not fixed.
}

\newcommand{\tcr}[1]{\textcolor{red}{#1}}
\definecolor{darkgreen}{RGB}{50,200,50}

\newcommand{\VDM}{$V_{DM}$ }

\preprint{}
\makeindex

\begin{document}
\maketitle
\newpage


\section{Introduction}\label{sec:intro}

The determination of the nature of Dark Matter (DM) is one of the most fundamental problems of  particle physics and cosmology.  If DM is light enough and interacts with Standard Model (SM) particles directly or via some mediators with a strength beyond the gravitational one, it can be directly produced at the Large Hadron Collider (LHC) or future  particle accelerators. The typical signature from DM produced in particles collisions is 
missing transverse energy, \MET, due to the fact that they escape undetected from the experimental apparatus.

Despite strong  experimental efforts by the ATLAS and CMS collaborations at the LHC, no excess of \MET{} over the SM background has been detected so far (as an example of the analyses most relevant to this paper see \cite{Aad:2015zva,Khachatryan:2014rra} with 8 TeV data and \cite{Aaboud:2016tnv,CMS:2016flr} with 13 TeV data).

The stability of the DM is usually associated with a discrete symmetry, most commonly a $\mathbb{Z}_2$ parity under which the SM particles are even, while the DM is the lightest odd particle. This mechanism is invoked in most theoretically motivated scenarios which predict a DM candidate, such as SUSY with R-parity \cite{Goldberg:1983nd,Ellis:1983ew}, Universal Extra Dimensions \cite{Antoniadis:1990ew, Appelquist:2000nn, Servant:2002aq, Csaki:2003sh}, Little Higgs \cite{ArkaniHamed:2002qx, Cheng:2003ju, Cheng:2004yc, Low:2004xc, Hubisz:2004ft, Cheng:2005as, Hubisz:2005tx} or Technicolor~\cite{Nussinov:1985xr,Barr:1990ca,Gudnason:2006ug}.

At the moment we do not know any information about the properties of DM
(except the fact that it does exists), such as its spin, mass, symmetry responsible for its stability,
interactions it is involved in (except gravitational), how many components it is made of, 
and which particles mediate the interactions between DM and the SM.
One such property, namely the \textit{spin} of the DM, could play a special role 
in discriminating between DM models via collider observables such as \MET{}
and kinematical properties of the  SM particles produced in association with DM particles.
For example, if a signal is found to be associated with a bosonic DM, the class of models predicting a fermionic DM (such as minimal SUSY models where the fermionic DM is a neutralino) would be excluded, while if DM is determined  to be a fermionic one, models of Universal Extra-Dimensions which generally predict bosonic DM would be 
ruled out.

At the LHC, mono-jet signatures which are events with a high-pT hadronic jet and a large \MET{} are generally considered as a ``discovery channel'' for DM. It is the purpose of this paper to analyse the kinematical properties of mono-jet signatures 
for models with DM  of different spin and study the LHC potential to differentiate them.

To effectively perform a phenomenological analysis on the characterisation of DM properties, two main model-independent approaches are generally used:
\begin{enumerate}
\item \textbf{\textit{The Effective Field Theory (EFT) approach}}, where the interactions between DM and the SM particles are described by higher dimensional (non-renormalizable) operators (see e.g. \cite{Fox:2011pm,Rajaraman:2011wf,Goodman:2010ku,Bai:2010hh,Beltran:2010ww,Goodman:2010yf}). These operators arise after integrating out heavy mediators and are therefore suppressed by a large UV scale. The advantage of this approach is that the only free parameters are the coefficients of the operators and the mass of the DM. However, this approach is consistent and accurately describes particle interactions if the energy scale of the interactions is small in comparison with the mediator mass, and can lead to an over- or underestimate of the cross-section depending on the precise relation between the mediator mass and energy transfer if this is not the case (see e.g. \cite{Buchmueller:2013dya}). Whilst this condition is  always satisfied for direct detection searches (where the energy transfer is $\mathcal{O}({\rm KeV})$), at the LHC the energy transfer is much larger necessitating $M_{med} \gtrsim \mathcal{O} (\text{few TeV})$ for the EFT description to agree with the underlying UV model. Furthermore the range of validity of the EFT approach is further constrained by requiring that the simplest UV completion is perturbative \cite{Goodman:2010ku,Fox:2011fx,Shoemaker:2011vi,Fox:2012ru,Haisch:2012kf,Busoni:2013lha,Busoni:2014haa,Busoni2014a,Abercrombie:2015wmb}, and that scattering processes are unitary \cite{Shoemaker:2011vi,Endo:2014mja}.
  
\item \textbf{\textit{The simplified-models approach}} 
see e.g. \cite{Buchmueller:2013dya,Busoni:2014sya,Busoni:2014haa,Buchmueller:2014yoa,Buckley:2014fba,Abdallah:2015ter,Abdallah:2014hon,Abercrombie:2015wmb}), which goes one step beyond EFT, by adding a single mediator and a single DM particle to the SM, and usually requiring the Lagrangian operators to be renormalizable. It makes one step towards a more UV-complete model, overcoming the EFT requirement of a heavy mediator, at the expense of introducing more parameters. Depending on the spin of the mediator as well as whether it's even or odd under the $\mathbb{Z}_2$ parity which stabilises the DM, the mediator can propagate either in the s-channel or t-channel or can even be pair produced. There is also the possibility that a SM particle
(e.g. Higgs or Z boson) plays a role of the mediator. It should be noted that simplified models are also not necessarily valid at all energies, suffering (for certain models) from a lack of gauge invariance and perturbative unitarity \cite{Englert:2016joy,Kahlhoefer:2015bea}.
\end{enumerate}

In this analysis we will focus on the EFT approach. Whilst this approach has the limitations discussed above, the advantages of having fewer parameters make EFT the most suitable choice for a first exploration of the effects of spin and their correlations with  kinematic observables. 
We have studied the complete set of dimension 5 and dimension 6 effective operators involving the interactions between scalar, fermion and vector DM with SM quarks and gluons, implemented the respective models and made them publicly available.
We perform our study at the parton and detector simulation levels and show that the pattern of \MET{} distributions initially observed at the parton level does not change at the detector level.

We have found that the invariant mass of the DM pair, $M_{\rm inv}(DM,DM)$ (defined by the EFT operator and the DM spin) and the structure of the SM bilinear entering the EFT operator uniquely define \MET{} shape. Thus we show that the \MET{} distribution depends on the spin of the DM and can characterise it at least for some EFT operators, as we quantitatively prove using a $\chi^2$ analysis, hence making a new step towards the characterisation of the DM including its spin.
The study we report here could be generically applicable for scenarios which are different from the EFT approach, \textit{e.g.} where the mediator is not produced on-the-mass-shell, such as the case of t-channel mediators, or the mediator has a mass below $2M_{DM}$, such that $M_{\rm inv}(DM,DM)$ is not fixed.
We have found drastic differences in the efficiencies (up to two orders of magnitude) for large \MET{} cuts for the cases of different DM spin, thus stressing that the DM discovery potential
strongly depends on it. This makes another step forward beyond the findings obtained at the LHC DM forum~\cite{Abercrombie:2015wmb}.

The structure of the paper is as follows: in Sect.~\ref{sec:contact_int} we setup our framework and notations and review the effective operators usually considered in literature and introduce new operators (not independent from the minimal set, but useful for a reinterpretation of our results in terms of the underlying UV completion); in Sect.~\ref{sec:signalgeneration} we describe the tool we use and the parameters we set to generate and analyse the signals coming from the different operators; in Sect.~\ref{sec:kinematicanalysis} we focus on the peculiar kinematic properties associated with different DM spins and operators; in Sect.~\ref{sec:comparison} we compare our results against LHC data at 8 TeV, 13 TeV and projections at higher luminosities
and demonstrate the  LHC potential to  distinguish certain classes of EFT operators between each other.

\section{DM Effective Field Theory Operators}\label{sec:contact_int}

Higher dimensional operators involving DM have been extensively discussed in the literature, see e.g. \cite{Fox:2011pm,Rajaraman:2011wf,Goodman:2010ku,Bai:2010hh,Beltran:2010ww,Goodman:2010yf}. 
In Table~\ref{tab:EFToperators} we have summarised a minimal set of independent dimension-5 and dimension-6 operators for complex scalar, Dirac fermion and complex vector DM coupling to quarks and gluons, adopting the widely used notations of \cite{Goodman:2010ku,Kumar:2015wya}.
For the case of vector DM, in addition to the DM-DM-quark-quark interactions studied in \cite{Kumar:2015wya},
we have added the V11 and V12 operators involving interactions between DM and gluons: these operators are also relevant for the phenomenology of vectorial DM at the LHC.

\begin{table}[h]
  \centering
  \begin{minipage}[t]{.4\textwidth}
    \begin{align*}
      \begin{array}{l@{\quad}l}
        \toprule
        \multicolumn{2}{c}{\text{Complex Scalar DM}}\\
        \midrule
        \frac{\tilde m}{\Lambda^2} \phi^{\dagger}\phi \bar{q}  q		    				& [C1] \\[2pt]
        \frac{\tilde m}{\Lambda^2} \phi^{\dagger}\phi \bar{q} i \gamma^5 q    					& [C2] \\[2pt]
        \frac{1}{\Lambda^2} \phi^{\dagger} i \overleftrightarrow{\partial_\mu} \phi \bar{q}\gamma^\mu q		& [C3] \\[2pt]
        \frac{1}{\Lambda^2} \phi^{\dagger} i \overleftrightarrow{\partial_\mu} \phi \bar{q}\gamma^\mu\gamma^5 q	& [C4] \\[2pt]
        \midrule
        \frac{1}{\Lambda^2} \phi^{\dagger}\phi G^{\mu\nu} G_{\mu\nu}						& [C5] \\[2pt]
        \frac{1}{\Lambda^2} \phi^{\dagger}\phi \tilde{G}^{\mu\nu} G_{\mu\nu}					& [C6]\\
        \bottomrule
      \end{array}
    \end{align*}
    \begin{align*}
      \begin{array}{l@{\quad}l}
        \toprule
        \multicolumn{2}{c}{\text{Dirac Fermion DM}}\\
        \midrule
        \frac{\tilde m}{\Lambda^3} \bar{\chi}\chi \bar{q} q 						& [\textrm{D1}] \\[2pt]
        \frac{\tilde m}{\Lambda^3} \bar{\chi} i \gamma^5\chi \bar{q} q 			       	& [\textrm{D2}] \\[2pt]
        \frac{\tilde m}{\Lambda^3} \bar{\chi}\chi \bar{q} i \gamma^5q 					& [\textrm{D3}] \\[2pt]
        \frac{\tilde m}{\Lambda^3} \bar{\chi}\gamma^5\chi \bar{q} \gamma^5q 				& [\textrm{D4}] \\[2pt]
        \frac{1}{\Lambda^2} \bar{\chi}\gamma^\mu\chi \bar{q}\gamma_\mu q			& [\textrm{D5}] \\[2pt]
        \frac{1}{\Lambda^2} \bar{\chi}\gamma^\mu\gamma^5\chi \bar{q}\gamma_\mu q 		& [\textrm{D6}] \\[2pt]
        \frac{1}{\Lambda^2} \bar{\chi}\gamma^\mu\chi \bar{q}\gamma_\mu \gamma^5 q		& [\textrm{D7}] \\[2pt]
        \frac{1}{\Lambda^2} \bar{\chi}\gamma^\mu \gamma^5\chi \bar{q}\gamma_\mu \gamma^5 q	& [\textrm{D8}] \\[2pt]
        \frac{1}{\Lambda^2} \bar{\chi}\sigma^{\mu\nu}\chi \bar{q}\sigma_{\mu\nu} q		& [\textrm{D9}] \\[2pt]
        \frac{1}{\Lambda^2} \bar{\chi}\sigma^{\mu\nu}i\gamma^5\chi \bar{q}\sigma_{\mu\nu} q	& [\textrm{D10}] \\
        \bottomrule
      \end{array}
    \end{align*}
  \end{minipage}\hfill
  \begin{minipage}[t]{0.6\textwidth}
    \begin{align*}
      \begin{array}{l@{\quad}l}
        \toprule
        \multicolumn{2}{c}{\text{Complex Vector DM}}\\
        \midrule
        \frac{\tilde m}{\Lambda^2} V^\dagger_\mu V^\mu \bar{q}  q													& [\textrm{V1}] \\[2pt]
        \frac{\tilde m}{\Lambda^2} V^\dagger_\mu V^\mu \bar{q} i \gamma^5 q												& [\textrm{V2}] \\[2pt]
        \frac{1}{2\Lambda^2} (V^\dagger_\nu\partial_\mu V^\nu - V^\nu \partial_\mu V^\dagger_\nu ) \bar{q} \gamma^\mu q 						& [\textrm{V3}] \\[2pt]
        \frac{1}{2\Lambda^2} (V^\dagger_\nu\partial_\mu V^\nu - V^\nu \partial_\mu V^\dagger_\nu ) \bar{q} i \gamma^\mu \gamma^5 q					& [\textrm{V4}] \\[2pt]
        \frac{\tilde m}{\Lambda^2} V^\dagger_\mu V_\nu \bar{q} i \sigma^{\mu\nu} q											& [\textrm{V5}] \\[2pt]
        \frac{\tilde m}{\Lambda^2} V^\dagger_\mu V_\nu \bar{q} \sigma^{\mu\nu} \gamma^5 q										& [\textrm{V6}] \\[2pt]
        \frac{1}{2\Lambda^2} (V^\dagger_\nu\partial^\nu V_\mu + V^\nu \partial^\nu V^\dagger_\mu ) \bar{q} \gamma^\mu q							& [\textrm{V7P}] \\[2pt]
        \frac{1}{2\Lambda^2} (V^\dagger_\nu\partial^\nu V_\mu - V^\nu \partial^\nu V^\dagger_\mu ) \bar{q} i \gamma^\mu q						& [\textrm{V7M}] \\[2pt]
        \frac{1}{2\Lambda^2} (V^\dagger_\nu\partial^\nu V_\mu + V^\nu \partial^\nu V^\dagger_\mu ) \bar{q} \gamma^\mu \gamma^5 q					& [\textrm{V8P}] \\[2pt]
        \frac{1}{2\Lambda^2} (V^\dagger_\nu\partial^\nu V_\mu - V^\nu \partial^\nu V^\dagger_\mu ) \bar{q} i \gamma^\mu \gamma^5 q					& [\textrm{V8M}] \\[2pt]
        \frac{1}{2\Lambda^2} \epsilon^{\mu\nu\rho\sigma} (V^\dagger_\nu\partial_\rho V_\sigma + V_\nu \partial_\rho V^\dagger_\sigma ) \bar{q} \gamma_\mu q		& [\textrm{V9P}] \\[2pt]
        \frac{1}{2\Lambda^2} \epsilon^{\mu\nu\rho\sigma} (V^\dagger_\nu\partial^\nu V_\mu - V^\nu \partial^\nu V^\dagger_\mu ) \bar{q} i \gamma_\mu q			& [\textrm{V9M}] \\[2pt]
        \frac{1}{2\Lambda^2} \epsilon^{\mu\nu\rho\sigma} (V^\dagger_\nu\partial_\rho V_\sigma + V_\nu \partial_\rho V^\dagger_\sigma ) \bar{q} \gamma_\mu \gamma^5 q	& [\textrm{V10P}] \\[2pt]
        \frac{1}{2\Lambda^2} \epsilon^{\mu\nu\rho\sigma} (V^\dagger_\nu\partial^\nu V_\mu - V^\nu \partial^\nu V^\dagger_\mu ) \bar{q} i \gamma_\mu \gamma^5 q		& [\textrm{V10M}] \\[2pt]
        \midrule
        \frac{1}{\Lambda^2} V^\dagger_\mu V^\mu G^{\rho\sigma} G_{\rho\sigma}			                                                    			& [V11] \\[2pt]
        \frac{1}{\Lambda^2} V^\dagger_\mu V^\mu \tilde{G}^{\rho\sigma} G_{\rho\sigma}			                                                		& [V12]\\[2pt]
        \bottomrule
      \end{array}
    \end{align*}
  \end{minipage}
  \caption{
  List of a minimal basis of EFT operators (dimension $\le$ 6) involving only complex scalar DM ($\phi$), Dirac fermion DM ($\chi$) or complex vector DM ($V^\mu$) interacting with SM quarks ($q$) or gluons (through the field strength tensor $G^{\mu\nu}$ and its dual $\tilde G^{\mu\nu}$.).}
  \label{tab:EFToperators}
\end{table}

A subset of operators in Table~\ref{tab:EFToperators} can also be used to describe interactions of real DM states. The only difference with respect to operators for complex DM is a factor two in the cross section for real DM  production for those operators which do not vanish\footnote{This factor of two comes from the $2^2=4$ factor from Feynman rules with identical particles, times the $1/2$ symmetrization factor which occurs at the level of the cross section evaluation.} . More specifically, the operators C1-C2 can be applied to real scalar DM, D1-D4 to Majorana fermion DM, and V1-V2 to real vector DM. However, the kinematic properties of the final states corresponding to complex DM are unaltered in comparison with the real DM case. Therefore, without loss of generality, in the following we will not discuss the real DM scenario.
 
Some operators, which involve scalar and pseudo-scalar SM quark operators, such as C1-C2, D1-D4, V1-V2 and tensor SM quark operators for V5-V6, are effectively originated from higher-dimensional operators with a dimensionful coupling. For all these operator we have made explicit the dimensionful coupling in Table~\ref{tab:EFToperators}. The origin of this coupling may be different, depending on the underlying physics. For example, it may originate from the vacuum expectation value of a scalar field or from a trilinear scalar coupling. In order to maintain a model-independent approach, we do not restrict ourselves to specific theoretical scenarios which may explain the physical origins of the different coefficients. Instead, we will just consider two scenarios which are simply related to the energy scales of the problem: a) the parameter does not depend on the UV scale $\Lambda$ (it can be proportional to the SM quark mass (as in Ref.~\cite{Goodman:2010ku})\footnote{In case the coefficient is proportional to the quark mass, in the following we will add the suffix ``Q'' to the operator, e.g. C1$\to$C1Q.} or to the mass of the DM) or b) it is proportional to the UV scale $\Lambda$, thus making the coefficient of the operator proportional to $1/\Lambda$ (for C1-C2, V1-V2 and V5-V6) or $1/\Lambda^2$ (for D1-D4). This also allows us to go beyond previous phenomenological studies of EFT operators: for scalar and fermionic DM operators, only scenario a) has been considered in Ref.~\cite{Goodman:2010yf}; for vectorial DM operators, only the b) case was explored in Ref.~\cite{Kumar:2015wya}.

We note that there are a number of other dimension-6 operators, which can be related to the operators of this minimal set either by equations of motions (EOM) \cite{Buchmuller:1985jz,Arzt:1993gz} or by Fierz identities, and they are therefore not independent. However, some of these alternative operators are worth studying in addition to those presented in Table~\ref{tab:EFToperators} because they have direct connections to the simplified models and allows one to make a straightforward respective interpretation of the experimental limits. 
In particular, we would like to introduce and study 4 additional operators (D1T, D2T, D3T, D4T) which are presented in Table~\ref{tab:EFT_Extra_Operators}. We stress that these are \textit{not} independent of those in Table~\ref{tab:EFToperators}, however it is instructive to explore them as they are the high mediator mass limit of simplified models with a fermion DM and a scalar t-channel mediator.
\begin{table}
  \begin{minipage}{1.0\textwidth}
    \begin{align*}
      \begin{array}{l@{\qquad}l}
        \toprule
        \multicolumn{2}{c}{\text{Dirac Fermion DM}}\\
        \midrule
        \frac{1}{\Lambda^2} \bar{\chi}q \bar{q}\chi				            & [\textrm{D1T}] \\
        \frac{i}{2\Lambda^2} (\bar{\chi}\gamma^5 q \bar{q}\chi +\bar{\chi}q \bar{q}\gamma^5 \chi) & [\textrm{D2T}] \\
        \frac{1}{2\Lambda^2} (\bar{\chi}\gamma^5 q \bar{q}\chi -\bar{\chi}q \bar{q}\gamma^5 \chi) & [\textrm{D3T}] \\
	\frac{1}{\Lambda^2} \bar{\chi} \gamma^5 q \bar{q} \gamma^5\chi			       & [\textrm{D4T}] \\
        \bottomrule
      \end{array}
    \end{align*}
  \end{minipage}
  \caption{Additional EFT operators, non-linearly-independent from those in Table~\ref{tab:EFToperators}. }
  \label{tab:EFT_Extra_Operators}
\end{table}
These D1T - D4T operators
can be expressed in terms of linear combinations of the minimal basis (D1 to D10) operators using the Fierz identities as follows:
\begin{eqnarray}
\begin{array}{l@{\qquad}l}
\textrm{[D1T]} & \bar{\chi}q \bar{q}\chi = \frac{1}{4}\left(\bar{\chi}\chi \bar{q} q  + \bar{\chi}\gamma^5\chi \bar{q} \gamma^5q + \bar{\chi}\gamma^\mu\chi \bar{q}\gamma_\mu q - \bar{\chi}\gamma^\mu \gamma^5\chi \bar{q}\gamma_\mu \gamma^5 q + \frac{1}{2}\bar{\chi}\sigma^{\mu\nu}\chi \bar{q}\sigma_{\mu\nu} q \right) \\[2pt] 
\textrm{[D2T]} & \frac{i}{2} (\bar{\chi}\gamma^5 q \bar{q}\chi +\bar{\chi}q \bar{q}\gamma^5 \chi) = \frac{1}{4}\left(\bar{\chi}i\gamma^5\chi \bar{q} q  + \bar{\chi}\chi \bar{q} i\gamma^5 q + \frac{1}{2}\bar{\chi}\sigma^{\mu\nu} i \gamma^5\chi \bar{q}\sigma_{\mu\nu} q \right) \\[2pt]
\textrm{[D3T]} & \frac{1}{2} (\bar{\chi}\gamma^5 q \bar{q}\chi -\bar{\chi}q \bar{q}\gamma^5 \chi) = \frac{1}{4}\left( \bar{\chi} \gamma^\mu \chi \bar{q} \gamma_\mu \gamma^5 q - \bar{\chi}\gamma^\mu\gamma^5\chi \bar{q} \gamma_\mu q  \right) \\[2pt]
\textrm{[D4T]} & \bar{\chi} \gamma^5 q \bar{q} \gamma^5\chi = \frac{1}{4} \left(\bar{\chi}\chi \bar{q} q  + \bar{\chi}\gamma^5\chi \bar{q} \gamma^5q - \bar{\chi}\gamma^\mu\chi \bar{q}\gamma_\mu q + \bar{\chi}\gamma^\mu \gamma^5\chi \bar{q}\gamma_\mu \gamma^5 q + \frac{1}{2}\bar{\chi}\sigma^{\mu\nu}\chi \bar{q}\sigma_{\mu\nu} q \right).
\end{array}
\label{eqn:D1T}
\end{eqnarray}
More details of the derivation of these Fierz identities are given in Appendix~\ref{app:Fierz}.

\noindent For completeness, the  examples of EOM-redundant dimension-6 operators which we will not consider are: 
\begin{itemize}
  \item $\phi^{\dagger}\phi (\bar{q}  i \overleftrightarrow{\slashed{D}} q)$, which can be related by the EOM $i \slashed{D} q = m q $ to $C1$.
  \item $\partial_\mu (\phi^{\dagger}\phi) \bar{q}\gamma^\mu q$, which can be seen to vanish by integrating by parts, using the relation $\partial_\mu (\bar{q}\gamma^\mu q) = (D_\mu \bar{q}) \gamma^\mu q + \bar{q} \gamma^\mu (D_\mu q)$, followed by application of the EOM $i \slashed{D} q = m q $.
\end{itemize}

\noindent It is important to notice that  EFT operators for  vector DM (\VDM) should be treated specially.
The subtlety is related to the fact that for \VDM the EFT energy asymptotics is different from the naively expected one, as we discuss below.
The cross section for the generic $qq(gg)\to DM DM$ ($2\to2$) scattering with a given power of the energy asymptotics ${\Delta_\sigma}$ can written as:
\begin{equation}
\sigma_{2\to 2}\propto \frac{1}{\Lambda^2}\times \left(\frac{E}{\Lambda}\right)^{\Delta_\sigma} .
\end{equation}
On the other hand,  ${\Delta_\sigma}$ is  related to the {\it effective} energy dimension, $D$, of the EFT operator as follows
\begin{equation}
\Delta_\sigma = 2(D-5)  \implies  D=\Delta_\sigma/2+5.
\end{equation}
We call $D$ as an {\it effective} energy dimension since formally the dimension  of vector DM (\VDM) operators is 
$d=5$ (V1,V2,V5,V6) or $d=6$ (V3,V4,V7-V12); however for each (allowed) \VDM longitudinal polarization  there is an additional 
$E/M_{DM}$ factor which leads to the  energy scaling of \VDM EFT operator different from the naive one, which we denoted by $d$.
In particular, (V1,V2,V5,V6) operators with $d=5$ behave as effective dimension $D=7$ operators, while (V3,V4,V7M,V8M,V11,V12) operators with $d=6$ 
behave as effective dimension $D=8$ operators, so the amplitude for the $qq(gg)\to DM DM$ process for both groups is enhanced by a $(E/M_{DM})^2$ factor.
For (V7P,V8P,V9,V10) operators only one longitudinal \VDM is allowed for $qq\to DM DM$ scattering, therefore its amplitude is enhanced with a $E/M_{DM}$ factor and the operators behave as effective dimension $D=7$ operators. 
This behaviour was noted in Ref.~\cite{Kumar:2015wya}.
In our paper we would like to suggest a new parameterisation of
\VDM operators. Our point is that since for scalar and fermionic DM operators the collider energy $E$ and the collider limit $\Lambda$ 
are of the same order, as we will see below, for vector DM it is natural to use an additional $M_{DM}/\Lambda$ factor for each power 
of $E/M_{DM}$ enhancement such that collider limits in this new parameterisation 
are not artificially enhanced  and will be of the same order as limits for other operators.
Therefore, for a given enhancement $\left(\frac{E}{M_{DM}}\right)^{D-d}$ 
the respective new factor for each \VDM EFT operator
will take the form:
\begin{equation}
\frac{1}{\Lambda_D^{d-4}}  \left(\frac{M_{DM}}{\Lambda_D}\right)^{D-d}  = \frac{M_{DM}^{D-d}}{\Lambda_D^{D-4}}
\label{eq:newparam}
\end{equation}
In Table \ref{tab:VDM-newpar} we summarise  the values of $D$ and $\Delta_\sigma$
together with the new parameterisation for each \VDM operator characterised by $\Lambda_D$
in comparison to   $\Lambda_d$ from the old parameterisation. From now we  omit $D$ from $\Lambda_D$
subscript and will denote it as $\Lambda$ while will keep  $\Lambda_d$ whenever we compare them together.

\begin{table}[h]
  \centering
    \begin{align*}
      \begin{array}{r@{\quad}|l@{\quad}l@{\quad}|l@{\quad}l@{\quad}|c | c}
        \toprule
        \text{\VDM Operator}		&\Lambda_{d}		 & d	& \quad\Lambda_{D} &  D	      & \Delta_\sigma (\sigma_{2\to 2} \propto E^{\Delta_\sigma})  & \text{Amplitude Enhancement}	\\
        \midrule
         \textrm{V1,V2,V5,V6}		 &\frac{1}{\Lambda} 	 &5  	& \quad\frac{M_{DM}^2}{\Lambda^3} & 7	        & 4&	(E/M_{DM})^2     	\\[2pt]
         \textrm{V3,V4,V7M,V8M,V11,V12}  &\frac{1}{\Lambda^2}  	 &6  	& \quad\frac{M_{DM}^2}{\Lambda^4} & 8	        & 6&	(E/M_{DM})^2     	\\[2pt]
         \textrm{V7P,V8P,V9,V10}	 &\frac{1}{\Lambda^2}	 &6  	& \quad\frac{M_{DM}}{\Lambda^3}	  & 7  		& 4&	E/M_{DM}   		\\[2pt]
        \bottomrule
      \end{array}
    \end{align*}
  \caption{
  The values of $d$,$D$,$\Delta_\sigma$ and the amplitude enhancement factors for  $qq(gg)\to DM DM$  process together with the new parameterisation for each \VDM operator characterised by $\Lambda_D$ in comparison to   $\Lambda_d$ from the old parameterisation.
Below we  omit $D$ from $\Lambda_D$
subscript and will denote it as $\Lambda \equiv\Lambda_D $.
}
  \label{tab:VDM-newpar}
\end{table}
The respective connection between $\Lambda_{d}$ and $\Lambda_{D}\equiv \Lambda$ is given by the following equation
\begin{equation}
\Lambda_D =  \left(\Lambda_d^{d-4} M_{DM}^{D-d}\right)^\frac{1}{D-4}.
\label{eq:LD}
\end{equation}

\section{Setup for the Signal Simulation}
\label{sec:signalgeneration}

\begin{figure}[t!]
  \centering
    \includegraphics[width=0.99\textwidth]{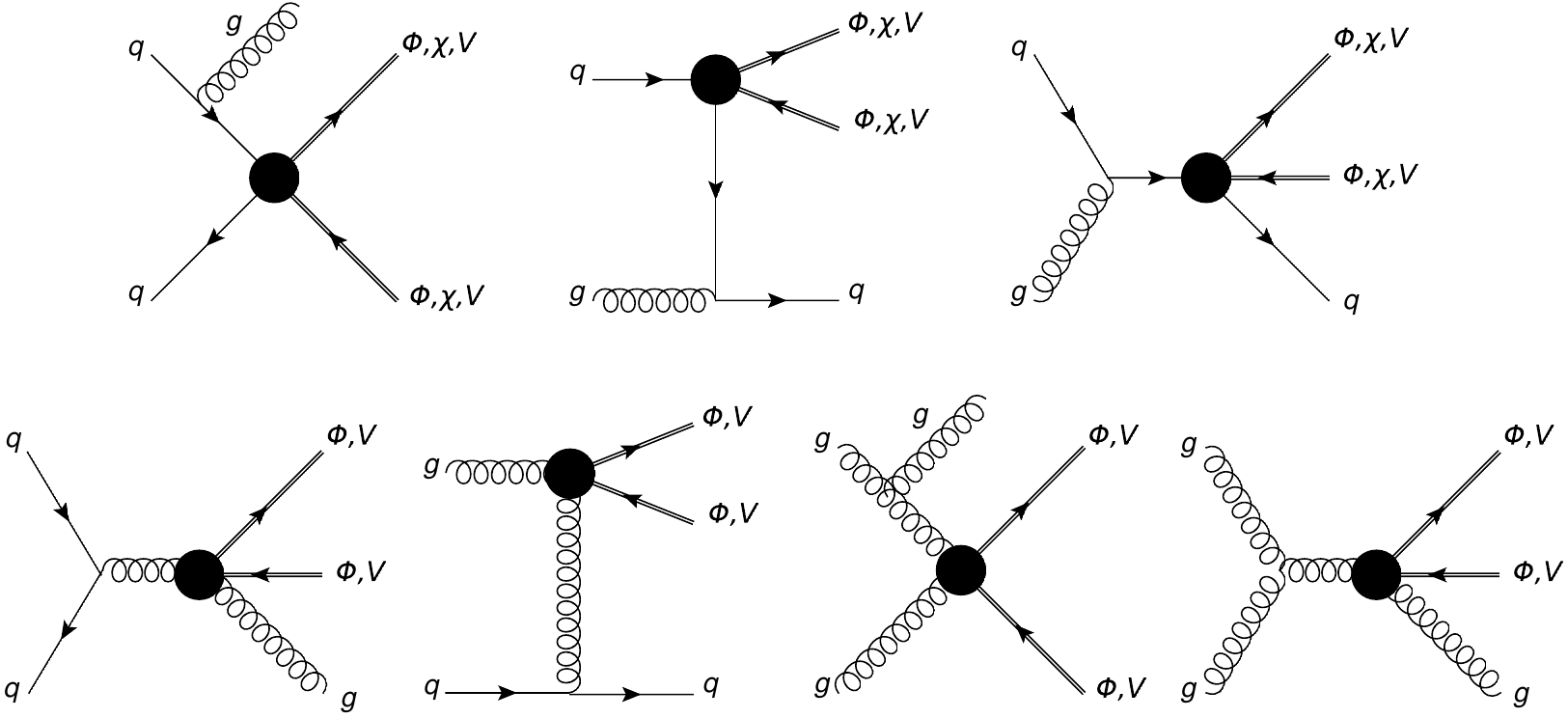}
  \caption{\label{fig:FeynDiagMonojet} Feynman diagrams for monojet processes for the operators listed in Table~\ref{tab:EFToperators}. All 7 diagrams are possible for scalar and vector DM, whilst only the top 3 diagrams occur for fermion DM as we do not consider the $GG\chi\chi$ (dimension-7) vertex.}\label{fig:FeynDiag}
\end{figure}

The aim of our study is to explore the possibility of distinguishing the different EFT operators from Tables~\ref{tab:EFToperators} and \ref{tab:EFT_Extra_Operators} via kinematic distributions for the monojet + \MET{} signature, where a DM pair recoils against a high-pT jet. In order to study  the effects for different DM masses, the analysis is performed for the representative benchmarks $M_{DM}=\{10, 100, 1000\}$ GeV.
The Feynman diagrams for this process are shown in Fig.~\ref{fig:FeynDiag}.

The analysis of the kinematic distributions is performed at both parton and detector level:
\begin{itemize}
\item the \textit{parton level} analysis is used to explore the difference in kinematic distributions  which occurs because different operators can have both different energy dependence and relations between incoming and outgoing momenta, and also have different weights of the various initial state subprocesses. If this difference is significant for operators with different  DM spin then this can be potentially used to chracterize of the DM spin.
\item the \textit{detector level} analysis is used to explore if the kinematic differences at the parton level are preserved after taking into account
hadronisation and detector effects, and thus understand if it is possible to effectively distinguish different operators at the experimental level.
\end{itemize}

Due to the different weights of the gluon-gluon, gluon-quark and quark-antiquark initial states at different energies, the possible identification of different EFT operators may depend significantly on the collider energy, and we therefore performed our analyses at both 8 and 13 TeV. However, for the sake of simplicity and clarity, results for the kinmatics distributions will only be discussed for the 13 TeV case. 

Our simulations and analysis have been performed using the {\sc MadGraph 5}~\cite{Alwall:2011uj,Alwall:2014hca} and {\sc CalcHEP}~\cite{Belyaev:2012qa} frameworks, and results have been cross-checked for every operator to ensure consistency and reproducibility. The plots and tables have been obtained with the settings described below.
The model files have been independently implemented into {\sc CalcHEP} using the {\sc LanHEP}~\cite{Semenov:2010qt} package and into {\sc MadGraph 5} using the {\sc Feynrules}~\cite{Alloul:2013bka} package and have been thoroughly cross-checked against each other. These models are public and available at {\sc HEPMDB}\footnote{The CalCHEP models for EFT DIM6  operators with scalar, fermion and vector DM 
are respectively available under hepmdb:0715.0185,  hepmdb:0715.0186 and hepmdb:1016.0214 IDs at HEPMDB (https://hepmdb.soton.ac.uk).
The respective MadGraph model is available under hepmdb:1016.0216, and is a single model containing EFT operators for scalar, fermion and vector DM.}~\cite{hepmdb,hepmdbmodel}.

In our analysis we have used the {\sc cteq6l1}~\cite{Pumplin:2002vw} PDF set. For both QCD renormalisation and PDF factorization scales we used $Q=(\sqrt{M_{\rm{inv}}(DM,DM)^2 +p_T(DM,DM)^2} + p_T^j)/2$. This choice is motivated by NLO DM studies performed in~\cite{Backovic:2015soa},
where it was found reasonably small differences in \MET{}  shapes between LO and NLO.

The bottom quark has been always included in both the definition of the proton and of the jets. The hadronization and parton showering were performed through {\sc Pythia} v6.4~\cite{Sjostrand:2006za}, with subsequent
fast detector simulation performed using  {\sc Delphes\,3}~\cite{deFavereau:2013fsa} and {\sc FastJet} v.3.1.3~\cite{Cacciari:2005hq,Cacciari:2011ma} with a cone radius $\Delta R = 0.4$ for the jet reconstruction. The detector level analysis was performed using {\sc CheckMATE} v1.2.2~\cite{Drees:2013wra}.

\section{Kinematic analysis}
\label{sec:kinematicanalysis}

\subsection{Spin-related features at parton level}

\begin{figure}[htb]
\centering
\includegraphics[width=0.95\textwidth]{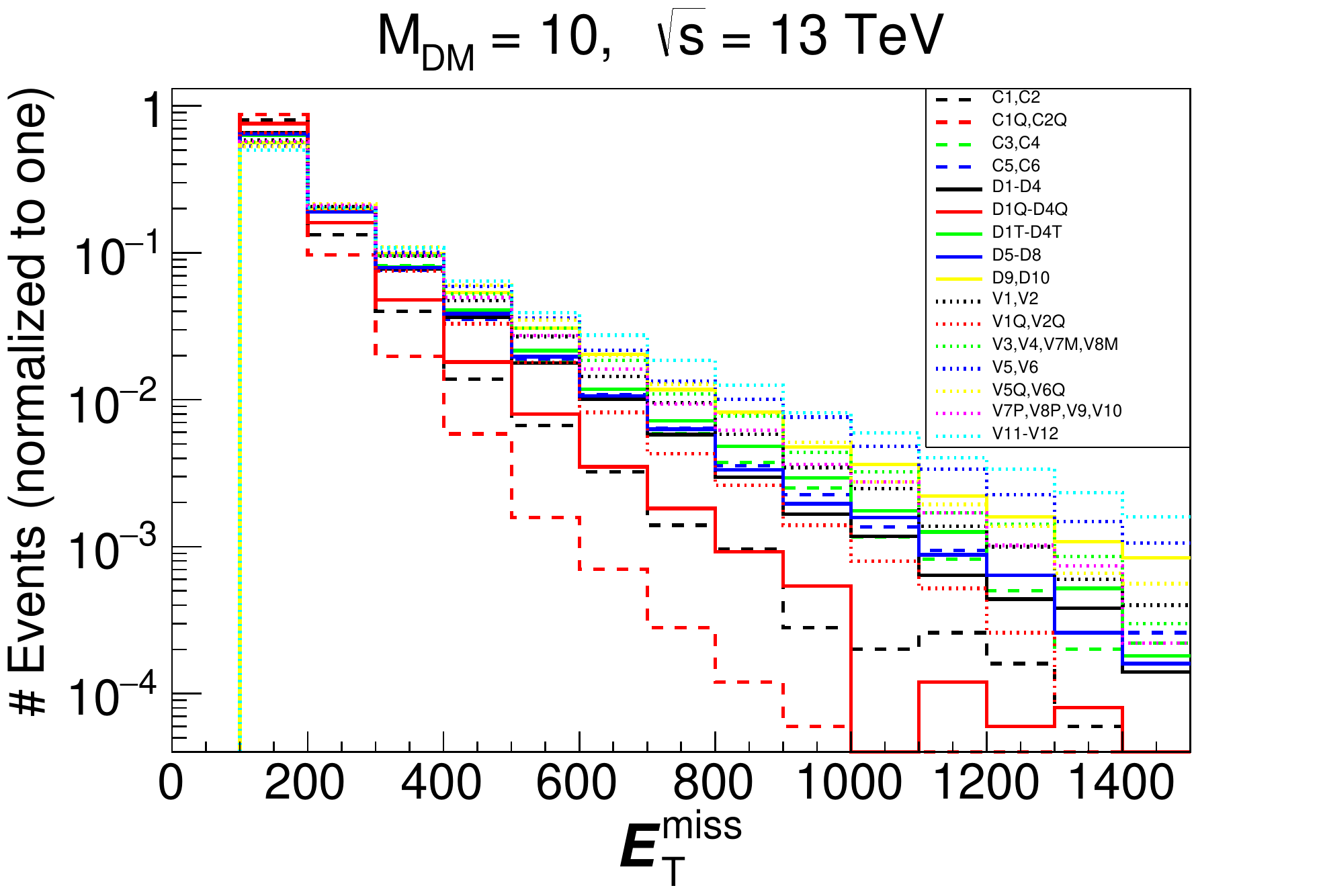}\\
\includegraphics[width=0.95\textwidth]{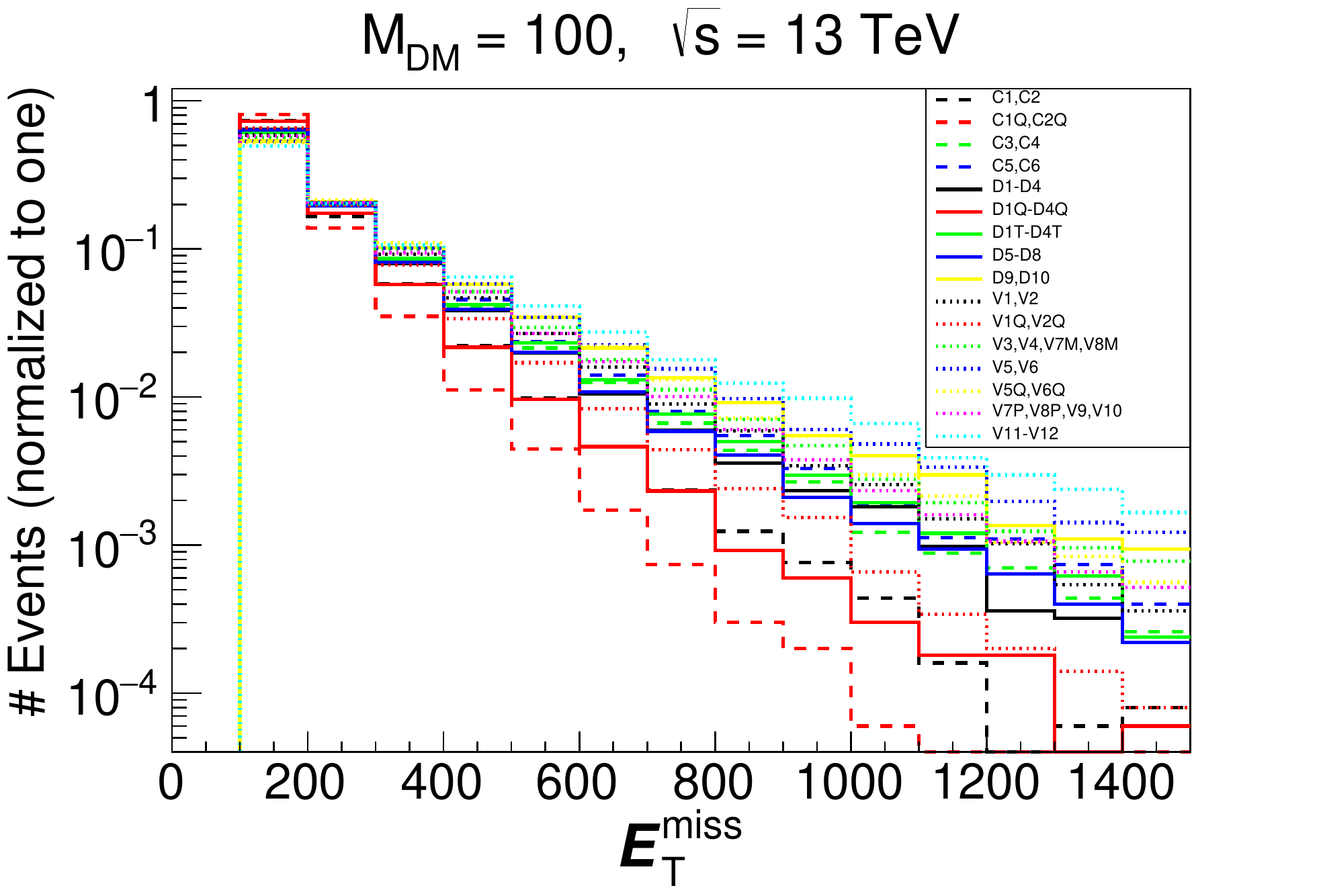}
\caption{\label{fig:AllOperators-10-100-parton}  \MET{} parton level distributions for a representative subset of the EFT operators listed in Tables~\ref{tab:EFToperators} and \ref{tab:EFT_Extra_Operators} for 13 TeV LHC energy. The panels differ by the mass of the DM candidate: $M_{\rm{DM}} = 10$ GeV (top), 100 GeV (bottom). A $p_{T,jet} \geq 100$~GeV cut has been applied in both plots.}
\end{figure}

In Fig.~\ref{fig:AllOperators-10-100-parton} we present the parton-level \MET{} distributions for DM with masses of 10 GeV and 100 GeV, for LHC@13TeV and for a representative subset of the EFT operators listed in Tables~\ref{tab:EFToperators} and \ref{tab:EFT_Extra_Operators}. The distributions are normalised to unity, in order to compare only shapes at this stage of the analysis. 

The subset of operators in Fig.~\ref{fig:AllOperators-10-100-parton} has been derived through the following logic. First of all, operators can be grouped when they just differ by a $\gamma^5$ in the SM bilinears, as C1-C2, D1-D3, V1-V2 and so on. Operators grouped in this way lead to virtually identical distributions, since their \MET distributions differ only by factors $(m_q^2/E_q^2)$
which are negligibly small in the high $\MET$ region and therefore in the high $p_T^q$ region of our interest, where $q$ denotes the SM quarks in the initial state proton.
For scenarios involving fermionic DM, we have further grouped operators which differ by the presence of a $\gamma^5$ in the DM bilinears after having numerically checked that they give also virtually identical $\MET$ distribution \textit{shapes}. Still for fermionic DM, the operators D1T-D4T contain bilinears which couple the DM state with SM quarks; we have numerically checked that the shapes are analogous, and therefore we have grouped all of them in the plots. Finally, for all DM candidates, we have grouped operators involving interactions with gluons as we have checked that the shapes of their distributions are again very similar. Therefore, as a result of this grouping, in Fig.~\ref{fig:AllOperators-10-100-parton} and following, we will present results for the following subset of operators : (C1,C2), (C1Q,C2Q), (C3,C4), (C5,C6), (D1-D4), (D1Q-D4Q), (D1T-D4T), (D5-D8), (D9,D10), (V1,V2), (V1Q,V2Q), (V3,V4, V7M,V8M), (V5-V6), (V5Q-V6Q),  (V7P,V8P,V9,V10)  and (V11,V12). 

One can immediatelly  observe a large difference between \MET{} distributions, ranging from the most steeply falling ones
for (C1,C2) or (C1Q,C2Q) and operators to the most flat ones for  (V11,V12) operators. For the bins with largest \MET{} values the differences  between operators  can be even more than one order of magnitude. Furthermore, we can identify the following main features, according to decreasing steepness of the shapes:
\begin{itemize}
\item[I)] {Operators for which the coefficient is proportional to $m_q$ (those labelled with a ``Q'' suffix) -- (C1Q,C2Q), (V1Q,V2Q), (V5Q,V6Q) -- fall always significantly more steeply than the same operators when the coefficient is proportional to a constant mass scale. The reason for this behaviour is that, being such operators proportional to $m_q$, the main contribution to the cross-section comes from the sea $s,c$ and $b$-quarks, the PDF of which fall more rapidly with the increase of  $x$,
the  fraction of proton momenta carried by quarks (and related \MET{}), than that of valence $u$- and $d$-quark, which give the main contribution to the \MET{} shapes for the other operators. Even if this behaviour is interestingly different, as we will see in the following these operators have very small cross-sections and therefore their investigation is of limited phenomenological interest.}
\item[II)] Among the rest of the operators, (C1,C2) for scalar DM exhibit distributions with the steepest shapes, and are quite clearly distinguishable from all other operators. As a justification for this behaviour, we notice that this operator has dimension
$D=d=5$ and the respective $\Delta_\sigma=0$,
so it has the least energy dependence and the respective \MET{} falling with a steeper slope compared to other operators. 
\item[III)] The subsequent group of operators, i.e. operators which exhibit a less steep \MET{} distribution with respect to the previous group but similar behaviour among themselves, is represented by the set (C3-C6), (D1-D4), (D1T-D4T) and (D5-D8). All these operators  have the same  dimension $D=d=6$  and the respective $\Delta_\sigma=2$, leading to the similar  \MET{} distributions.
One should note that (C5,C6) operators from this group involving  gluons (and not quarks as other operators from this group)
behave similar for low DM mass, however for the large DM mass its shape becomes distinguishable from the rest of the operators of this group as we discuss below.

\item[IV)] The next group of operators with a flatter \MET{} tail includes only vector DM ones, involving quark-anti-quark (pseudo-)scalar or (axial-)vector currents in their SM part:
 (V1,V2), (V3,V4) (V7M,V8M), (V5,V6), (V7P,V8P,V9,V10). 
The different energy behaviour (and therefore \MET{} shape) for these operators is related to the enhancement $\frac{E}{M_{DM}}$ from each  longitudinal vector DM.
The effective dimension for these operators as stated in Table~\ref{tab:VDM-newpar}, $D=7$ for  (V1,V2), (V5,V6),  (V7P,V8P,V9,V10)
and $D=8$ for  (V3,V4, V7M,V8M) with $\Delta_\sigma=2$ and 4 respectively.
Operators with $D=8$ eventually provide slightly flatter \MET{} distribution then those with $D=7$ but this difference is not significant.
\item[V)] The (D9,D10) fermion DM operators exhibit less steep  \MET{} tail than the previous group of operators and can be distinguished from others
 because of the tensorial structure in the bilinears $\sigma^{\mu\nu}$ which represents interactions of magnetic-type.
\item[VI)] Analogously to the previous group, the vector DM operators with $\sigma^{\mu\nu}$ magnetic-type interactions (V5,V6) exhibit even flatter  \MET{} tail and can be distinguished from the rest of the operators.
\item[VII)] Finally, the last set of operators in this sequence is composed of (V11,V12), which involve the gluon strength-tensor in the SM sector coupled to vector DM.
\end{itemize}

\begin{figure}[htb]
\centering
\includegraphics[width=0.95\textwidth]{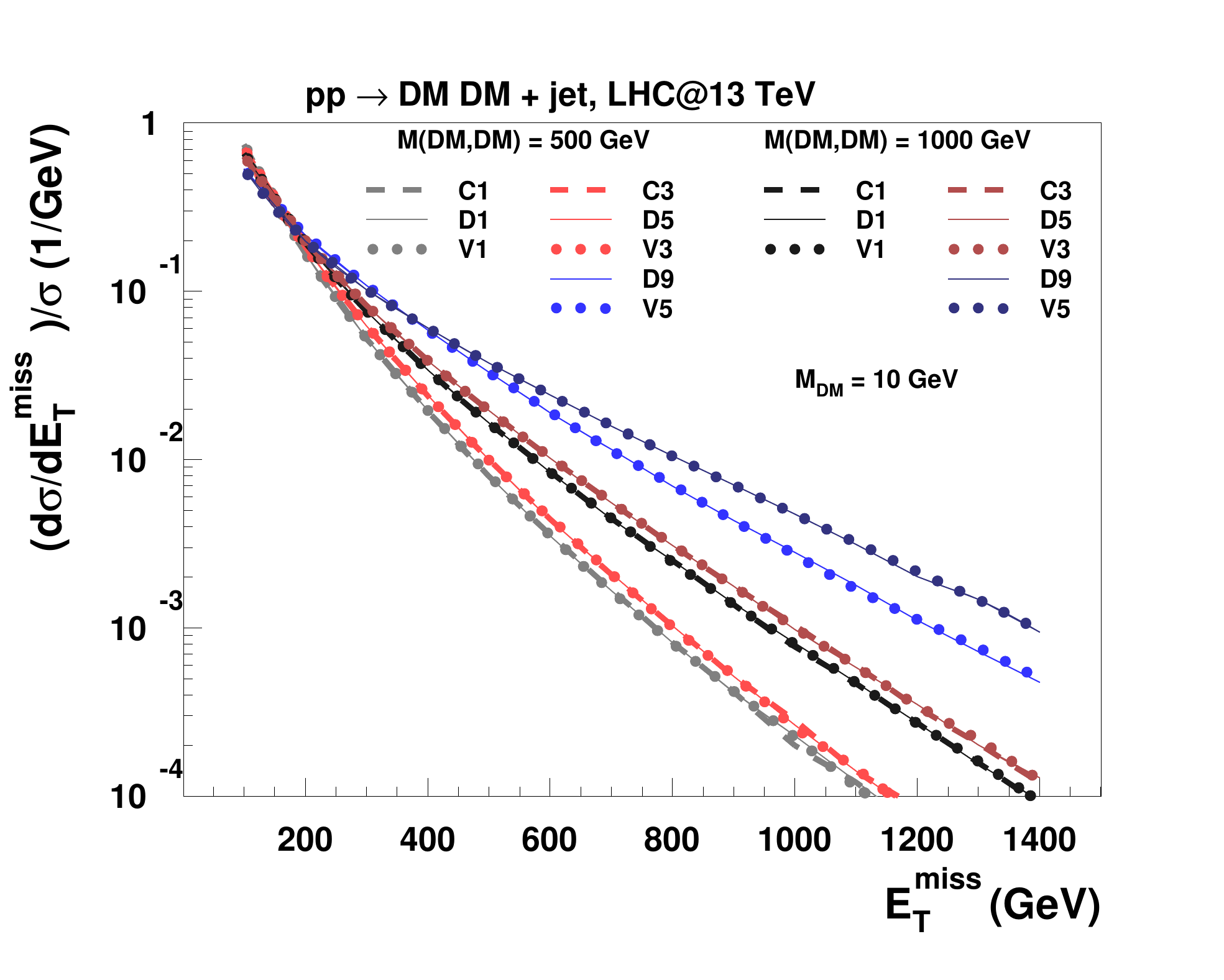}
\caption{\label{fig:DMDM-MET}  $d\sigma/d\MET{}$ parton level distribution normalised to unity for
the fixed invariant mass of the DM pair,$M_{\rm inv}(DM,DM)$=500 and 1000 GeV
for (C1,D1,V1), (C3,D5,V3) and (D9,V5) groups of representative operators 
with the  scalar, vector and tensor structure of SM part respectively.}
\end{figure}

The  \MET{} shapes from different operators are uniquely determined by the combination of three factors: the effective dimension of the operator, $D$, the structure of the SM part of the operator and the parton densities of the SM particles connected to the operator.
Furthermore, it is important to note that the Lorentz structure of the SM part of the EFT operators
and  the  invariant mass distribution of the DM pair, $M_{\rm inv}(DM,DM)$, also uniquely define the shape of the \MET{} distribution, independently of the spin of the DM.\footnote{{In case of  D1T-D4T operators, where the bilinears connect a SM state with the DM candidate, as discussed in Section~\ref{sec:contact_int}, operators can be rewritten through Fierz transformations as a linear combination of operators in the basis of Table~\ref{tab:EFToperators}, where the SM bilinears are always separated from the DM ones.}} Moreover, with the increase of $M_{\rm inv}(DM,DM)$, the \MET{} shape falls less and less steeply (again, for a given SM component of the EFT operator). 
This is a quite remarkable result and is presented in Fig.~\ref{fig:DMDM-MET}.

In this figure we present the $d\sigma/d\MET{}$ parton level distribution normalised to unity for
a fixed invariant mass of the DM pair, $M_{\rm inv}(DM,DM)$=500 and 1000 GeV
for the (C1,D1,V1), (C3,D5,V3) and (D9,V5) groups of representative operators 
which feature a scalar, vector and tensor structure in the SM component of the operator, respectively.
This figure clearly demonstrates that within each group of operators
the shape of the \MET{} distribution is identical for a fixed $M_{\rm inv}(DM,DM)$ value.
At the same time, one can see that with the increase of  $M_{\rm inv}(DM,DM)$
the slope of the \MET{} distribution decreases.
The decrease of the \MET{} slope as $M_{\rm inv}(DM,DM)$ increases can be qualitatively explained by phase space and parton density effects: when $M_{\rm inv}(DM,DM)$ is small, the radiation of a high $P_T$ jet will ``cost'' a {\it large relative shift in $x$}, the proton momentum fraction carried by the  parton, leading to a rapidly falling \MET{} distribution; on the contrary, when $M_{\rm inv}(DM,DM)$ is large, the radiation of a high $P_T$ jet will ``cost" a {\it small relative shift in $x$}, which will lead to a more slowly falling \MET{} distribution in comparison to the first case.
This effect can be used to understand the reason for the different \MET{} shapes presented in Fig.~\ref{fig:AllOperators-10-100-parton}.

\begin{figure}[htb]
\centering
\includegraphics[width=\textwidth]{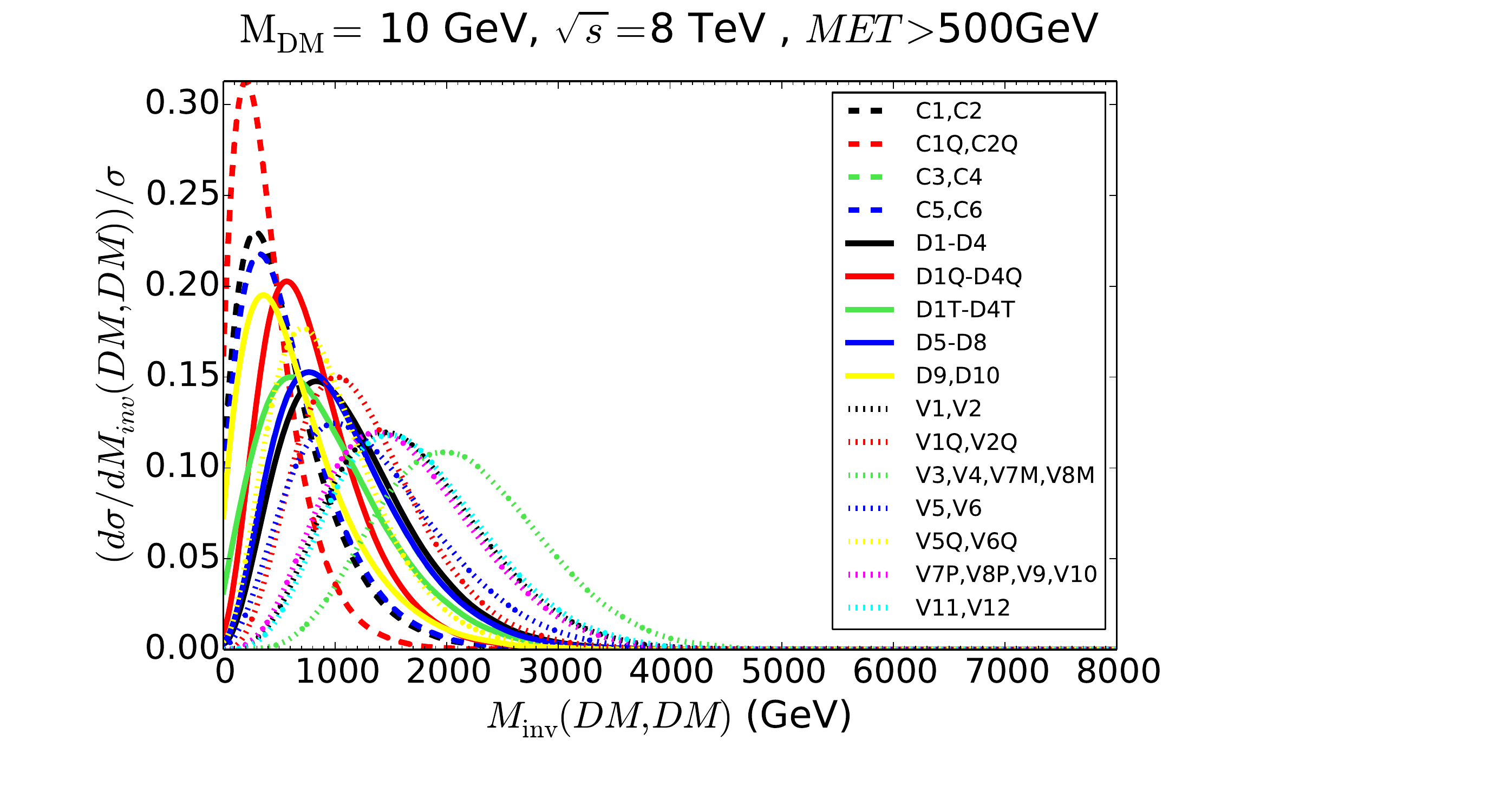}\\
\includegraphics[width=\textwidth]{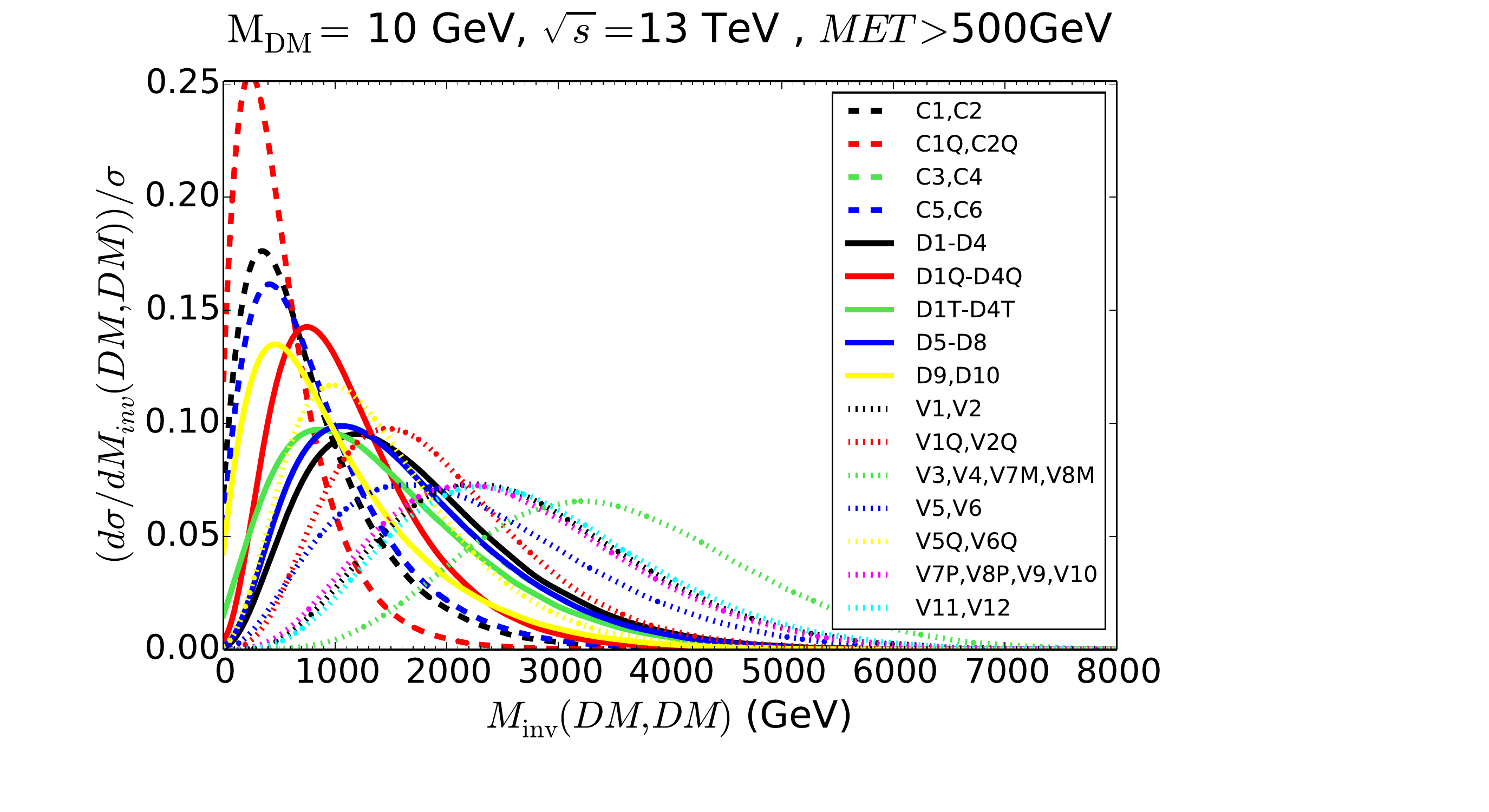}\\
\caption{\label{fig:inv-mass} Invariant mass of DM pair distributions normalised to unity  for  EFT operators listed in Tables~\ref{tab:EFToperators} and \ref{tab:EFT_Extra_Operators} for 8 TeV (top) and 13 TeV (bottom)  LHC energy, $M_{\rm{DM}} = 10$ GeV and  $p_{T,jet} \geq 500$~GeV cut applied.}
\end{figure}

In Fig.~\ref{fig:inv-mass} we present $M_{\rm inv}(DM,DM)$ distributions for the operators under study,again normalised to unity
for 8TeV (top) and 13 TeV(bottom) LHC.
The $M_{\rm inv}(DM,DM)$ variable is not observable, however, these distributions are very informative in understanding
the \MET{} distributions for the EFT operators under study, given the relationship between the two discussed above.
From Fig.~\ref{fig:inv-mass}, one can see that the $M_{\rm inv}(DM,DM)$ distributions are even better separated for different operators in comparison to the \MET{} distributions (Fig.~\ref{fig:AllOperators-10-100-parton}),
although the grouping of similar distributions is slightly different.
Similarly to Fig.~\ref{fig:AllOperators-10-100-parton}, the (C1,C2) operators from group II have the lowest $M_{\rm inv}(DM,DM)$ distribution tails.
However, whilst the high mass tail of $M_{\rm inv}(DM,DM)$ for (C5,C6) is above that of (C1,C2), it is also split from the rest of the group III operators, which all have similar $M_{\rm inv}(DM,DM)$ distributions. The distribution for group V operators (D9,D10) is slightly below this.
One can also see that group  (V3,V4, V7M,V8M) has the highest $M_{\rm inv}(DM,DM)$ mean value 
and the respective shape which very different from that of  (V1,V2), (V5,V6) and (V11,V12) operators.
One can see that
the effective dimension of the operator, $D$, the structure of the SM part of the operator and the SM particles connected to the operator
-- the factors defining the shape of \MET{} -- are even more clearly conected to the $M_{\rm inv}(DM,DM)$ distribution.

We would like to stress the fact      
that $M_{\rm inv}(DM,DM)$ distributions for DM EFT operators for different DM  spins form different groups.
The only exception is the (C3,C4) distributions which is very similar to the one for (D5-D8) operators.
It is not a coincidence, however: the shape of $M_{\rm inv}(DM,DM)$ is primarily defined by the dimension $D$ of the operator,
its structure and the  SM particles  entering the operator. These factors are correlated with the DM spin 
for a given formal dimension $d$ of the EFT operators, chosen to be the minimal one -- 5 and 6.
It is also important to note that in the low $M_{\rm inv}(DM,DM)$ region, the shape of the distributions are qualitatively different depending on DM spin:
for larger DM spins, the $M_{\rm inv}(DM,DM)$ distribution falls more rapidly towards zero 
{as $M_{\rm inv}(DM,DM)$ decreases.}.
Also we found that as the number of $\gamma$ matrices in the quark operator increases, the $M_{\rm inv}(DM,DM)$ distribution {\it falls more rapidly} towards 
zero {as $M_{\rm inv}(DM,DM)$ decreases} for scalar DM and {\it falls less rapidly} for the fermion and vector DM cases.

If $M_{\rm inv}(DM,DM)$ could be fixed or concentrated around specific different values for each operator, then \MET{} would allow us to perfectly distinguish between different models.
Unfortunately, this is not the case, and the resulting \MET{} distribution comes eventually from the integral over $M_{\rm inv}(DM,DM)$, which partly masks
the difference between EFT operators. Nevertheless, the resulting \MET{} distribution presented in Fig.~\ref{fig:AllOperators-10-100-parton}
demonstrates the correlation with Fig.~\ref{fig:inv-mass} (keeping in mind the $M_{\rm inv}(DM,DM)$ ``re-weighting" after the integration mentioned above)
and the corresponding potential to distinguish some EFT operators and related DM spin.

\begin{figure}[htb]
\centering
\includegraphics[width=0.95\textwidth]{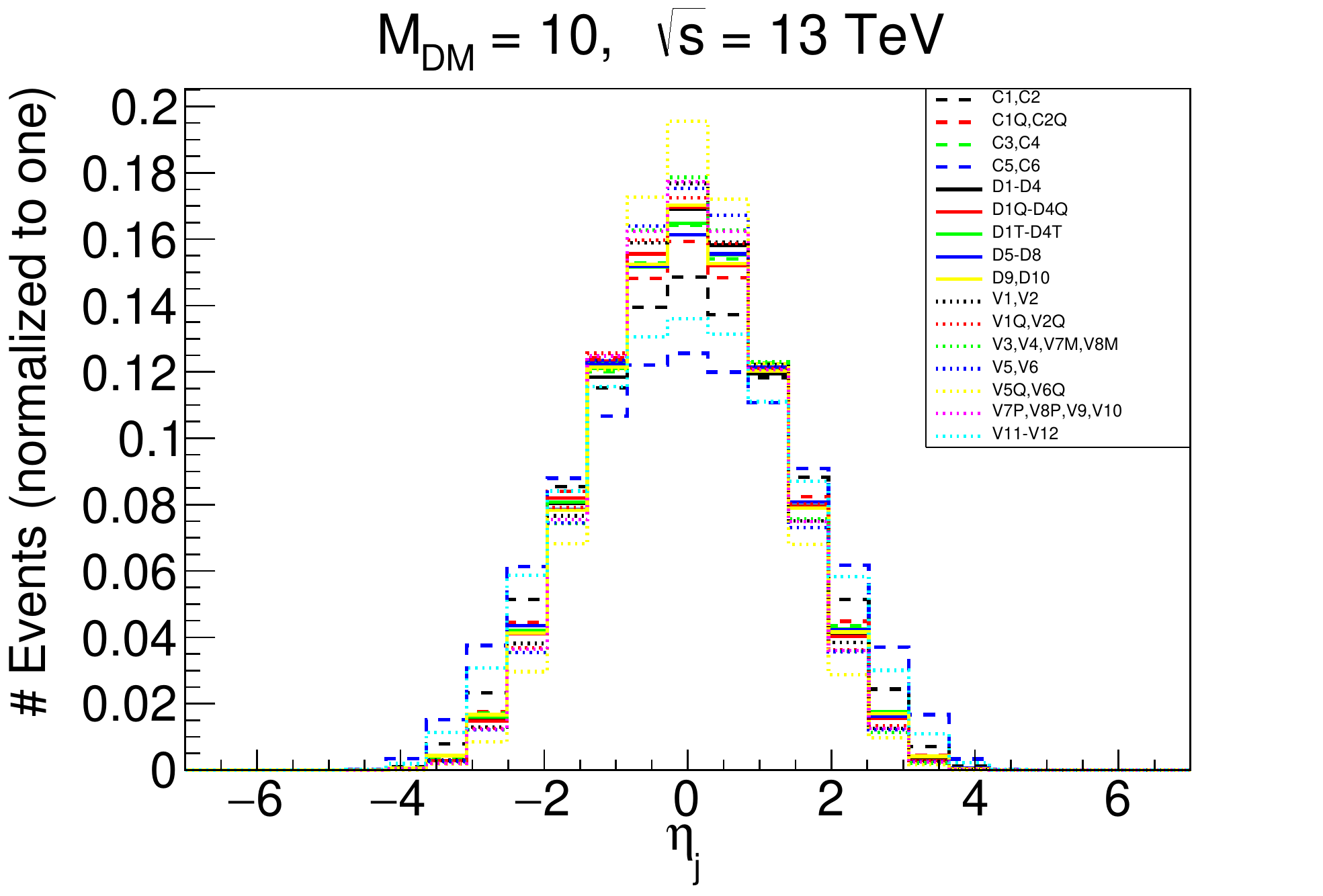}\\
\includegraphics[width=0.95\textwidth]{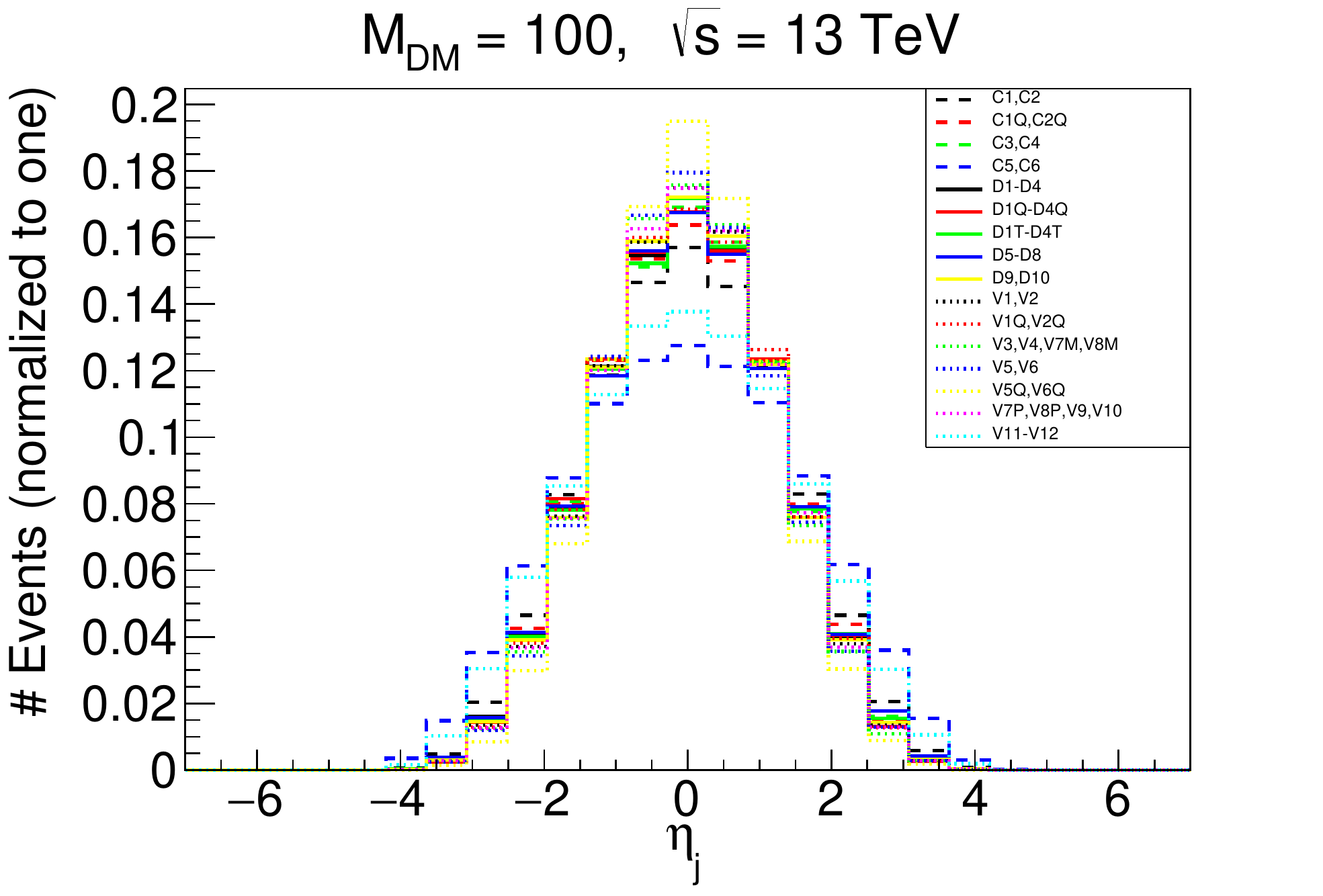}\\
\caption{\label{fig:eta} Pseudorapidity of the mono-jet distributions normalised to unity  for  EFT operators listed in Tables~\ref{tab:EFToperators} and \ref{tab:EFT_Extra_Operators} for 13 TeV LHC energy, $M_{\rm{DM}} = 10$(top) and 100(bottom) GeV and  $p_{T,jet} \geq 100$~GeV cut applied.}
\end{figure}

In Fig.\ref{fig:eta} we also present pseudo-rapidity distributions 
of the mono-jet for the EFT operators under study,  normalised to unity and with an energy of 13 TeV.
One can see that differences between operators are also manifest there.
It is interesting to notice that in this case the distributions for (C1,C2), (C5,C6) and (V11,V12) are less central
than all the other operators. This different grouping of mono-jet pseudo-rapidity distributions in comparison to the \MET{} can be exploited to differentiate between operators with similar \MET{} distributions.

\begin{figure}[htb]
\centering
\includegraphics[width=0.95\textwidth]{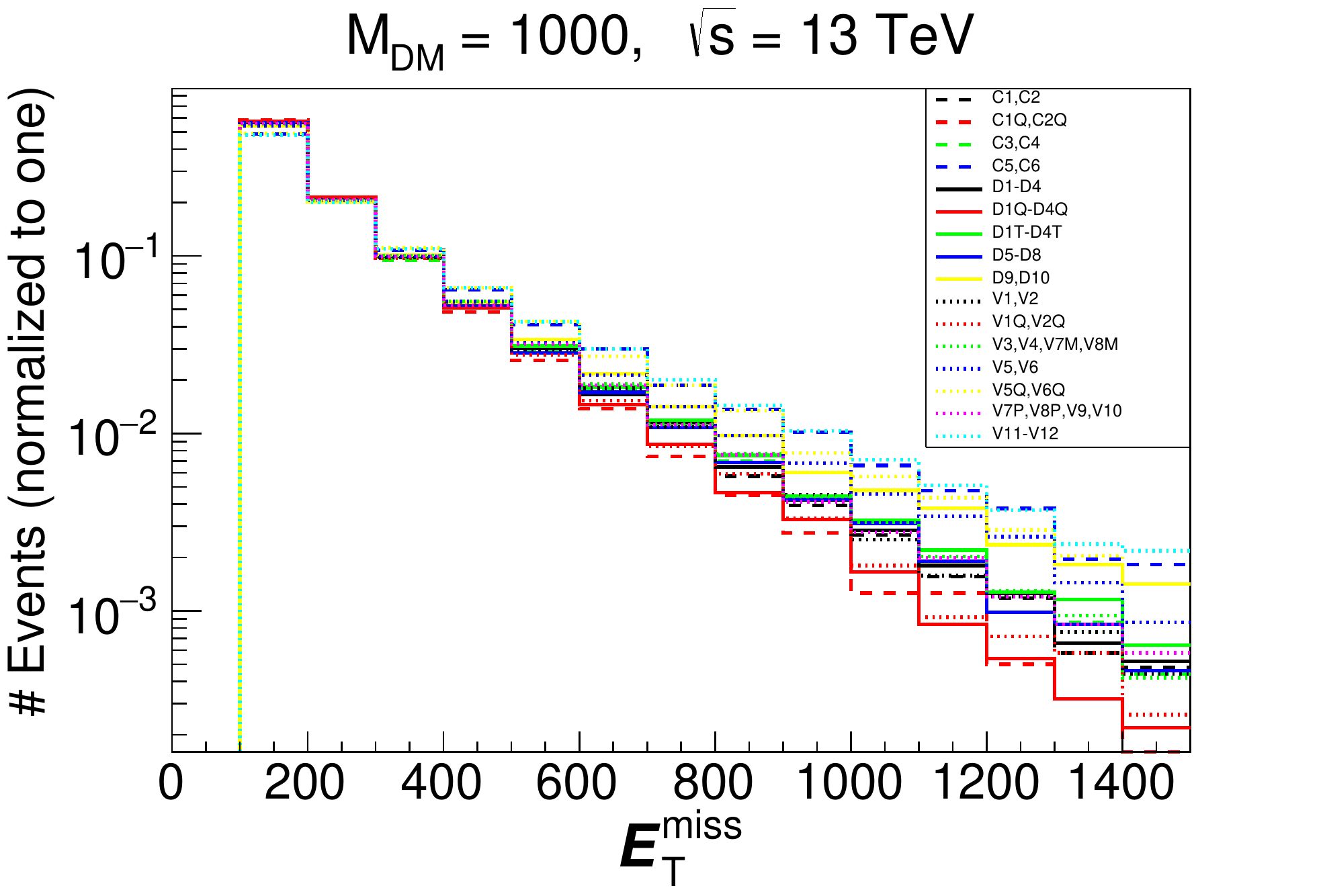}\\
\includegraphics[width=0.95\textwidth]{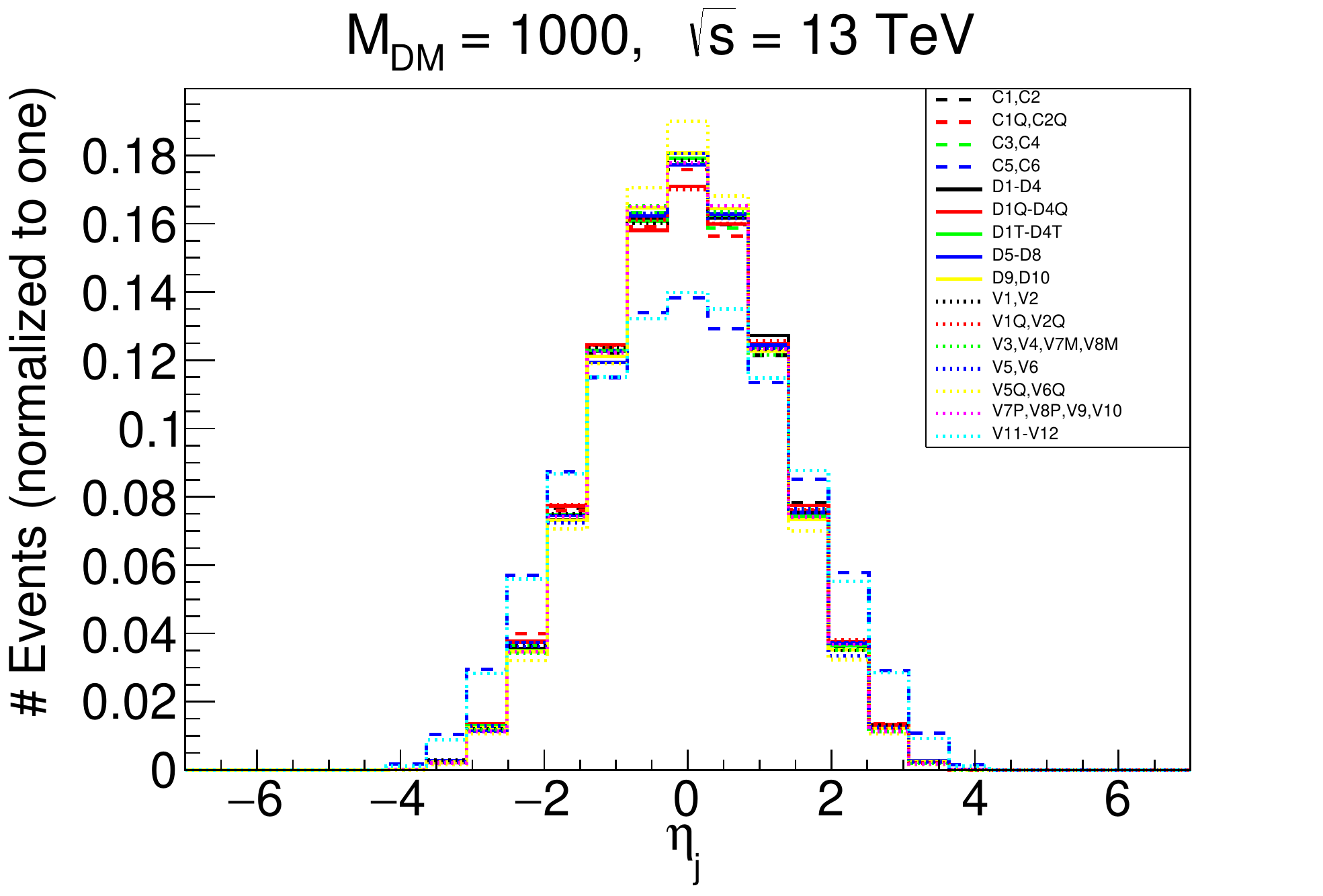}
\caption{\label{fig:MDM-1000} \MET{} (top) and  pseudorapidity of the mono-jet (bottom) distributions normalised to unity  for  EFT operators listed in Tables~\ref{tab:EFToperators} and \ref{tab:EFT_Extra_Operators} for 13 TeV LHC energy, $M_{\rm{DM}} = 1000$  GeV and  $p_{T,jet} \geq 100$~GeV cut applied.}
\end{figure}

While the general picture of distributions for different operators is very similar for the $M_{\rm{DM}}$ range between 10 and 100 GeV
-- a range which is likely to be accessible at the LHC -- we also study here the distributions behaviour for the 
extreme case with  $M_{\rm{DM}}=1000$~GeV. As we will see in the following, however, such large DM masses are unlikely to be testable at the LHC in the EFT regime.
In Fig.~\ref{fig:MDM-1000} we present \MET{} (top) and $\eta_j$ (bottom) distributions for  $M_{\rm{DM}}=1000$~GeV
analogous to those presented above for the lighter DM case.

Fig.~\ref{fig:MDM-1000}
demonstrates that for $M_{\rm{DM}}=1000$~GeV all \MET{} shapes from different 
operators become more similar. The reason for this is the very limited phase space 
when the DM mass is very large and respectively the small ratio of DM momentum over its mass. In this case the difference 
between \MET{} and $\eta_j$ distributions is mainly dictated by the SM operator --- by its structure and type of partons.
This is why for $M_{\rm{DM}}=1000$~GeV only three groups of operators are observed in the \MET{} distributions:
\begin{itemize}
\item the least steep distribution comes from scalar and vector DM operators with gluons -- (C5,C6) and  (V11,V12) operators
\item the  group with the intermediate slope comes from operators whose SM part contains
the quark current  with tensor $\sigma^{\mu\nu}$ interaction
\item the group with the steepest \MET{} slope contains  the rest of the operators,
whose SM part contains (pseudo)scalar or (pseudo)vector quark currents.
\end{itemize}
While it is possible to recognise these three groups (with relatively small differences in distributions) in this set of \MET{} distributions,
only two groups are observed in the $\eta_j$ distributions: the group with operators containing SM gluons -- (C5,C6) and (V11,V12) --- which has a slightly wider $\eta_j$ shape,
and the the group with the rest of operators, for which the SM bilinears contain quarks.

Let us note that in scenarios with large DM masses, like in the previous example, even if data would allow us to measure a signal with large enough statistics, it would be difficult to distinguish between groups of operators because of the similar shapes of the \MET{} and $\eta_j$ distributions. Furthermore, conclusively distinguishing between DM spins in this very heavy DM case would be virtually impossible since the differences between the distributions are driven only by the structure of the SM operators.


\subsection{Beyond the parton level effects.}

After having assessed the possibility of distinguishing between DM EFT operators at the parton level, it is crucial to understand how much the effects of parton showering and hadronization, as well as the smearing effects due to the detector properties affect our conclusions. 

\begin{figure}[htb]
    \includegraphics[width=0.95\textwidth]{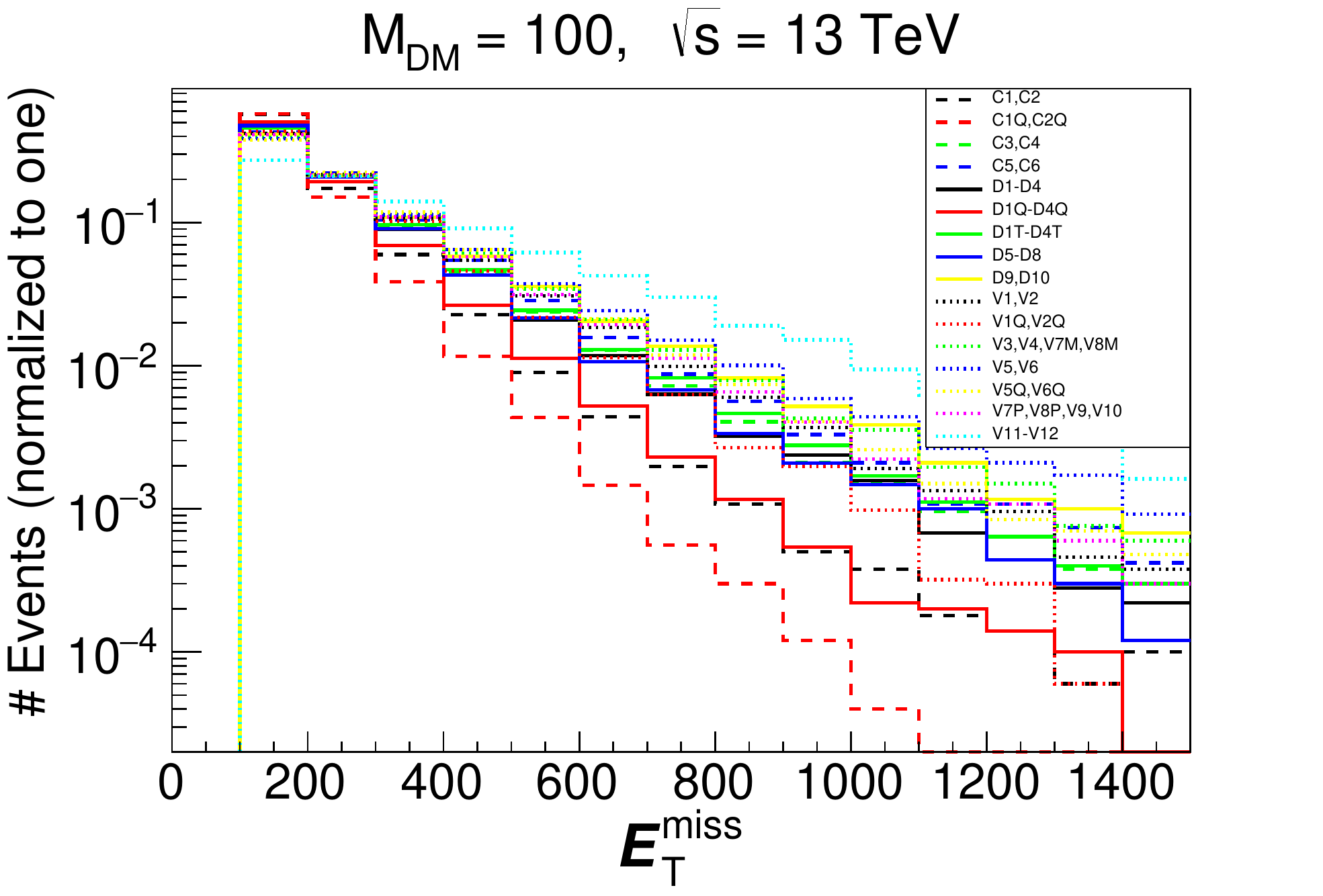}\\
    \includegraphics[width=0.95\textwidth]{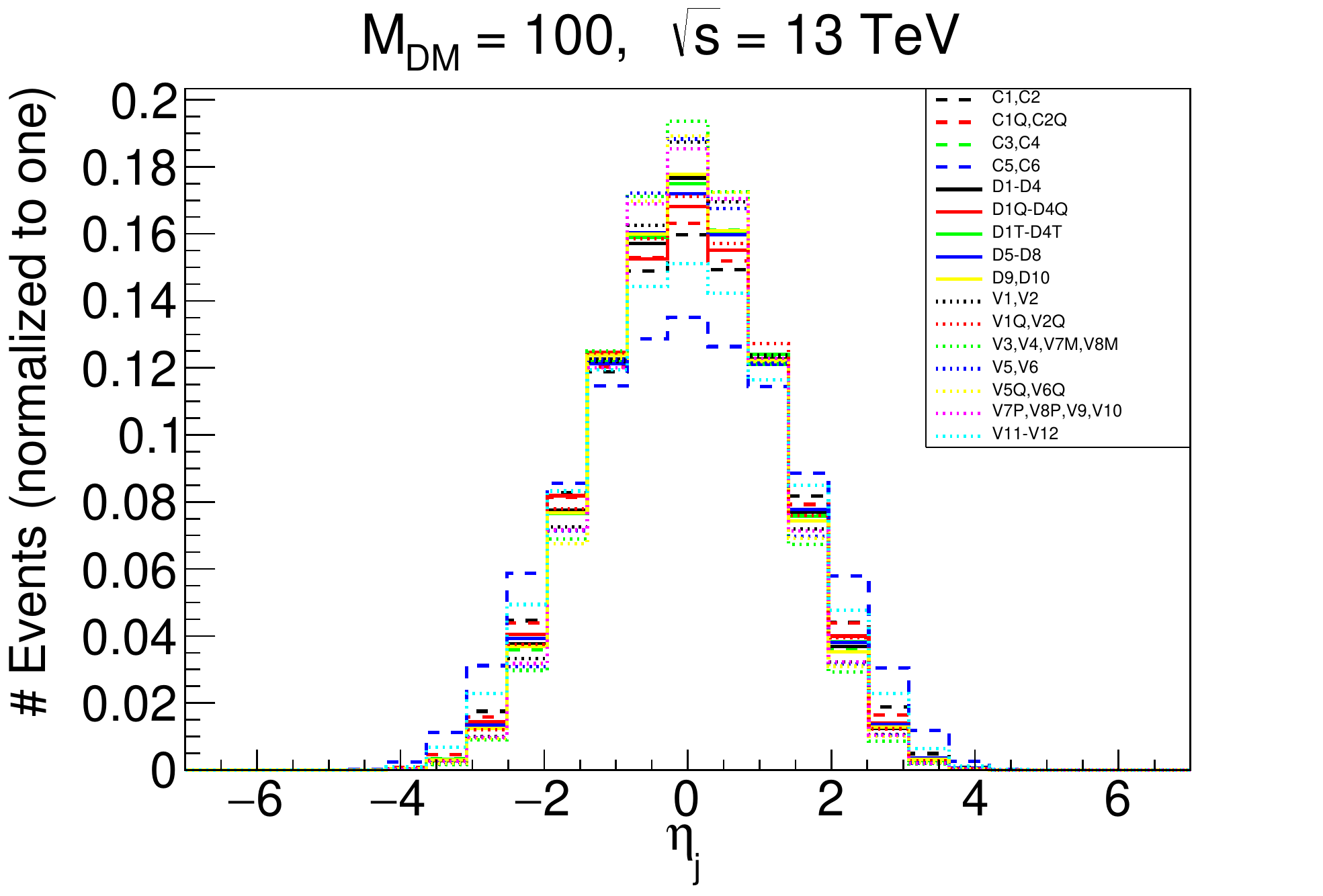}%
  \caption{\label{fig:ComparisonDetector}\MET{} and Jet pseudo-rapidity distributions at the  detector level, for representative EFT operators from the classes in Tables~\ref{tab:EFToperators} and \ref{tab:EFT_Extra_Operators}. In all plots $M_{\rm DM} = 100$ GeV and the collision energy is 13 TeV.}
\end{figure}

In Fig.\ref{fig:ComparisonDetector} we present \MET{} and $\eta_j$ distributions at the detector level for $M_{\rm{DM}}=100$~GeV.
we remind here that the detector effects have been simulated using {\sc Delphes\,3} tuned to model the ATLAS detector and implemented within {\sc CheckMATE} v1.2.2.

One can clearly see that beyond-the-parton-level effects which include parton showering, hadronization, and detector simulation, do not visibly change the shape of any of the distributions under study. The same conclusions apply for all the masses we have tested.

Therefore the \MET{} and leading jet pseudo-rapidity distributions can potentially be used to distinguish some operators, and therefore to characterise the spin of the DM in some cases.
Even if it is not possible to {\it unequivocally} associate certain distributions to a specific DM spin, some operators exhibit peculiar behaviours; therefore, if the DM interacts through such operators, its properties should be clearly distinguishable at the level of the respective shapes.

One should also note that the shape of distributions for a given operator also depends on the mass of the DM, and that this distinction can only be effectively made once the mass of the DM is inferred either by looking at the correlation between the cross-section of the process and the shape of the distribution or, possibly, by complementary observations.
One should also note that for DM masses in the range of 10-100 GeV, which is likely to be the scope of the LHC, the shape for any given operator does not differ significantly and therefore does not depend strongly on the DM mass.

We will present in the following sections how to perform a quantitative analysis for distinguishing between different operators and also between same operators with different DM masses for a given collider luminosity.

\section{LHC sensitivity to the Dark Matter  EFT operators}
\label{sec:comparison}

Thus far we have only explored the \textit{shape} of the distributions for  EFT operators
and demonstrated  differences for some of their classes.
In this section we explore the LHC potential to differentiate these operators.
The main problem here is to study which operators with the strength allowed by present data
can be not only  discovered at high luminosity at the LHC, but also distinguished between each other.
To do this we establish current LHC limits on $\Lambda$  for these operators and verify
if at high luminosity operators with such value of  $\Lambda$  could  provide 
a large enough signal and could be differentiated between each other.

\begin{figure}[htbp]
    \includegraphics[width=\textwidth]{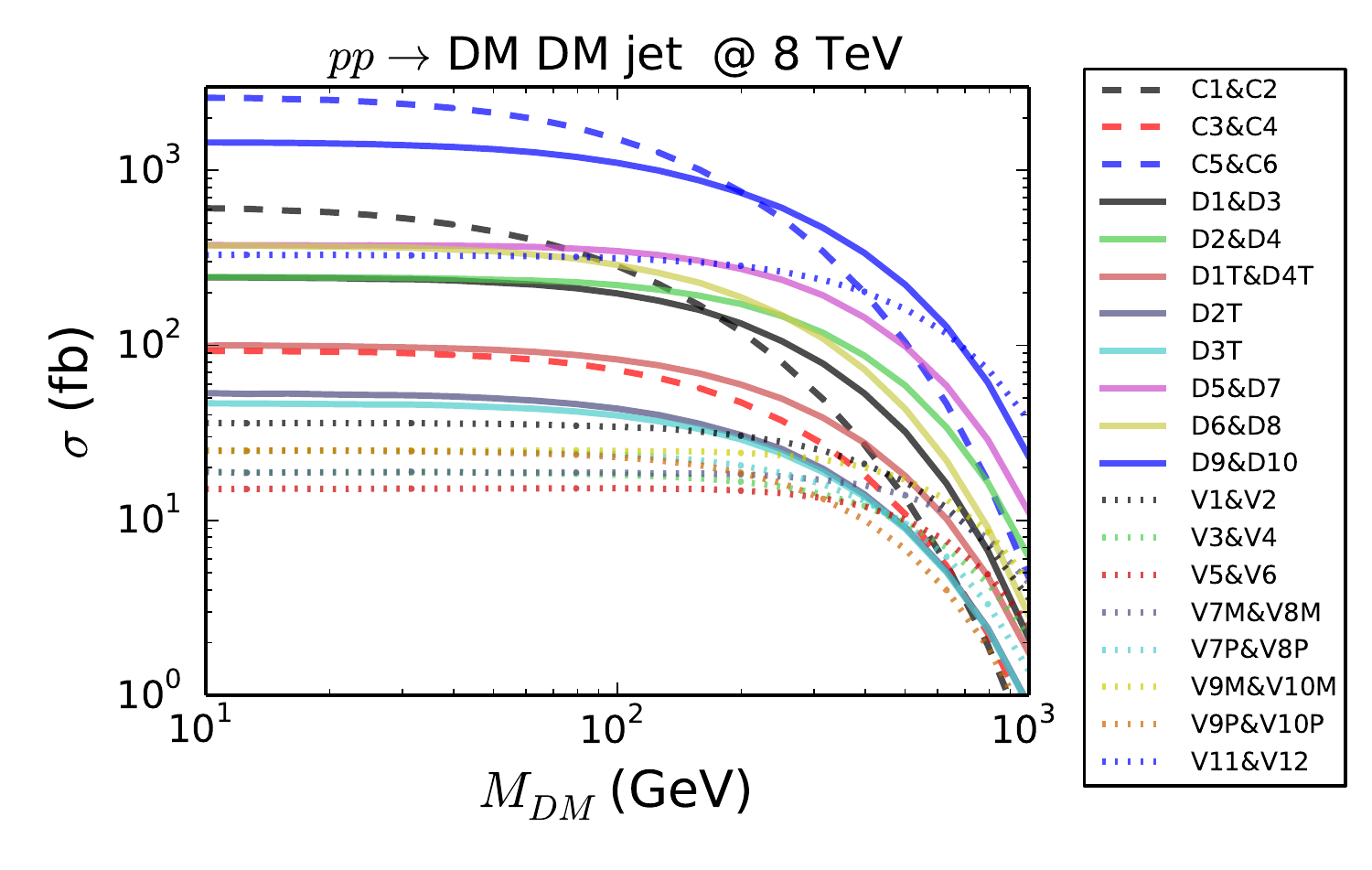}\\
    \includegraphics[width=\textwidth]{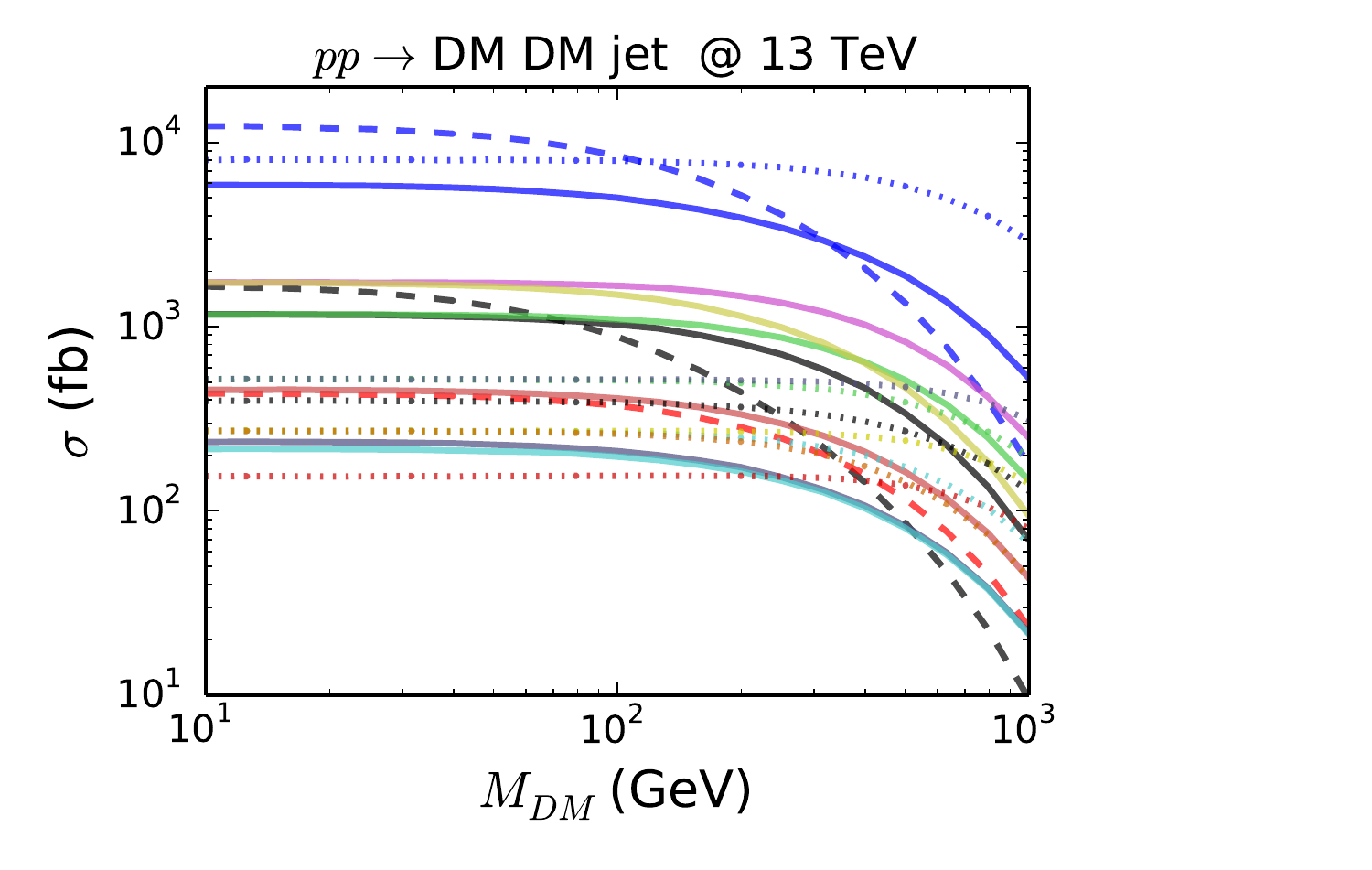}
  \caption{\label{fig:sigmas}
  Mono-jet cross-sections in fb for the EFT operators at 8 TeV and 13 TeV versus DM mass for
  $\Lambda=1~\rm{TeV}$ for all operators, presented in Table~\ref{tab:sigmas} }
\end{figure}

We first find limits on the operators for experimental data at 8 TeV and 13 TeV (with the current luminosity) and then provide the 13 TeV projections at higher luminosities.
The cross-sections for the EFT operators at both 8 TeV and 13 TeV are presented in Table~\ref{tab:sigmas} and Fig.~\ref{fig:sigmas}.
These cross-sections are evaluated for  $\Lambda=1$~TeV.
The coefficients for each scalar and fermionic operator are chosen to be $1/\Lambda^2$ with 
the exceptions of (C1,C2) and (D1,D4) operators for which the coefficient depends on the mass parameter in the numerator. As already anticipated in Section~\ref{sec:contact_int}, for this dimensionful coupling we considered different hypotheses: 1) it corresponds to $\Lambda$, where the relevant scale is the UV cut-off itself; 2)  it is equivalent to $M_{DM}$, which assumes that the relevant mass scale of the coupling corresponds to the DM mass (this scenario is not reported in Table~\ref{tab:sigmas} as its cross-section is a simple re-scaling of the previous scenario); 3) it is equivalent to the SM quark mass, $m_q$ assuming Yukawa couplings-type origin of these $SU(2)_L$ breaking terms. For the operators of vector DM we have considered the coefficients reported in Tab.\ref{tab:VDM-newpar}, and for the operators (V1,V2) and (V5,V6) we have considered also the cases where the numerator corresponds either to $\Lambda_D$ or to $m_q$.

\begin{table}
\small
\setlength{\tabcolsep}{2pt}
\centering
\begin{tabular}{ccc|ccc|ccc}
\toprule
&\multirow{6}{*}{\textbf{Operators}} & \multirow{6}{*}{\textbf{Coefficient}} & \multicolumn{6}{c}{\textbf{Cross Section (fb)}} \\
\cmidrule{4-9}
&&& \multicolumn{6}{c}{LHC Energy} \\
&&& \multicolumn{3}{c|}{8 TeV} & \multicolumn{3}{c}{13 TeV} \\
\cmidrule{4-9}
&&& \multicolumn{6}{c}{DM mass} \\
&                          &                    & 10 GeV               & 100 GeV              & 1000 GeV             & 10 GeV               & 100 GeV              & 1000 GeV             \\
\midrule                                                                                                                                                                                   
\midrule                                                                                                                                                                                   
\multirow{6}{*}{\begin{sideways} \begin{tabular}{c} Complex \\ Scalar DM\end{tabular} \end{sideways}}                                                                                      
&\multirow{2}{*}{C1 \& C2} & $1/\Lambda$        & $6.17\times10^{2}$               & $2.86\times10^{2}$  & $5.09\times10^{-1}$  & $16.9\times10^{2}$   & $8.90\times10^{2}$ & $9.57$  \\
&                          & $m_q/\Lambda^2$    & $2.15\times10^{-4}$  & $6.02\times10^{-5}$  & $4.88\times10^{-9}$ & $8.55\times10^{-4}$  & $3.03\times10^{-4}$  & $2.62\times10^{-7}$ \\
\cmidrule{2-9}                                                                                                                                                        
& C3 \& C4                 & $1/\Lambda^2$      & $9.37\times10$  & $7.28\times10$  & $7.09\times10^{-1}$  & $4.37\times10^{2}$  & $3.75\times10^{2}$  & $2.35\times10$  \\
\cmidrule{2-9}                                                                                                                                                                              
& C5 \& C6                 & $1/\Lambda^2$      & $2.60\times10^{3}$  & $1.52\times10^{3}$  & $4.61	$  & $1.23\times10^{4}$               & $8.42\times10^{3}$  & $1.76\times10^{2}$  \\
\midrule                                                                                                                                                                                    
\midrule                                                                                                                                                                                    
\multirow{13}{*}{\begin{sideways} Dirac Fermion DM \end{sideways}}                                                                                                                          
&\multirow{2}{*}{D1 \& D3} & $1/\Lambda^2$      & $2.45\times10^{2}$  & $1.99\times10^{2}$  & $2.11$  & $1.17\times10^{3}$  & $1.03\times10^{3}$  & $6.98\times10$  \\
&                      & $m_q/\Lambda^3$    & $3.19\times10^{-5}$ & $1.95\times10^{-5}$ & $1.50\times10^{-8}$ & $2.03\times10^{-4}$ & $1.47\times10^{-4}$ & $1.26\times10^{-6}$ \\                                                                                                                        
\cmidrule{2-9}                                                                                                                                                                           
&\multirow{2}{*}{D2 \& D4} & $1/\Lambda^2$      & $2.46\times10^{2}$  & $2.22\times10^{2}$  & $6.18\times10$  & $1.17\times10^{3}$  & $1.10\times10^{3}$  & $1.46\times10^{2}$  \\
&                          & $m_q/\Lambda^3$    & $3.21\times10^{-5}$ & $2.43\times10^{-5}$ & $5.41\times10^{-8}$ & $2.04\times10^{-4}$ & $1.71\times10^{-4}$ & $3.36\times10^{-6}$ \\                                                                                                                           
\cmidrule{2-9}                                                                                                                                                                              
&D1T \& D4T                & $1/\Lambda^2$      & $1.00\times10^{2}$  & $8.35\times10$  & $1.75	$  & $4.58\times10^{2}$  & $4.11\times10^{2}$  & $4.31\times10$  \\
\cmidrule{2-9}                                                                                                                                                                              
&D2T                       & $1/\Lambda^2$      & $5.36\times10$  & $4.36\times10$  & $8.78\times10^{-1}$  & $2.39\times10^{2}$  & $2.12\times10^{2}$  & $2.17\times10$  \\
\cmidrule{2-9}                                                                                                                                                                              
&D3T                       & $1/\Lambda^2$      & $4.69\times10$  & $3.99\times10$  & $8.67\times10^{-1}$  & $2.19\times10^{2}$  & $1.99\times10^{2}$  & $2.14\times10$  \\
\cmidrule{2-9}                                                                                                                                                                              
&D5 \& D7                  & $1/\Lambda^2$      & $3.77\times10^{2}$  & $3.47\times10^{2}$  & $1.11\times10$  & $1.75\times10^{3}$  & $1.68\times10^{3}$  & $2.50\times10^{2}$  \\
\cmidrule{2-9}                                                                                                                                                                             
&D6 \& D8                  & $1/\Lambda^2$      & $3.75\times10^{2}$  & $2.91\times10^{2}$  & $2.83$  & $1.75\times10^{3}$  & $1.50\times10^{3}$  & $9.38\times10$  \\
\cmidrule{2-9}                                                                                                                                                                             
&D9 \& D10                 & $1/\Lambda^2$      & $1.46\times10^{3}$  & $1.11\times10^{3}$  & $2.31\times10$  & $5.96\times10^{3}$  & $5.04\times10^{3}$  & $5.26\times10^{2}$  \\
\midrule                                                                                                                                                                                   
\midrule                                                                                                                                                                                   
\multirow{15}{*}{\begin{sideways} Complex Vector DM \end{sideways}}                                                                                                                        
&\multirow{2}{*}{V1 \& V2} & $M_{DM}^2/\Lambda_D^3$     & $3.60\times10$  & $3.43\times10$  & $3.59$  & $3.95\times10^{2}$  & $3.89\times10^{2}$  & $1.29\times10^{2}$  \\
&                      & $m_q M_{DM}^2/\Lambda_D^4$ & $1.66\times10^{-6}$ & $1.47\times10^{-6}$ & $2.76\times10^{-8}$ & $2.15\times10^{-5}$ & $2.03\times10^{-5}$ & $2.19\times10^{-6}$ \\                                                                                                                              
\cmidrule{2-9}                                                                                                                                                                            
&V3 \& V4                  & $M_{DM}^2/\Lambda_D^4$     & $1.88\times10$  & $1.82\times10$  & $2.17$  & $5.17\times10^{2}$  & $5.11\times10^{2}$  & $1.97\times10^{2}$  \\
\cmidrule{2-9}                                                                                                                                                                            
&\multirow{2}{*}{V5 \& V6} & $M_{DM}^2/\Lambda_D^3$     & $1.51\times10$  & $1.53\times10$  & $2.52$  & $1.54\times10^{2}$  & $1.55\times10^{2}$  & $8.00\times10$  \\
&                          & $M_{DM}^2 m_q/\Lambda_D^4$ & $1.31\times10^{-6}$ & $1.28\times10^{-6}$ & $2.42\times10^{-8}$ & $1.36\times10^{-5}$ & $1.36\times10^{-5}$ & $1.80\times10^{-6}$ \\                                                                                                                              
\cmidrule{2-9}                                                                                                                                                                            
&V7M \& V8M                & $M_{DM}^2/\Lambda_D^4$     & $1.88\times10$  & $1.87\times10$  & $4.39$  & $5.17\times10^{2}$  & $5.17\times10^{2}$  & $3.13\times10^{2}$  \\
\cmidrule{2-9}                                                                                                                                                                            
&V7P \& V8P                & $M_{DM}/\Lambda_D^3$       & $2.50\times10$  & $2.38\times10$  & $1.38$  & $2.72\times10^{2}$  & $2.66\times10^{2}$  & $6.73\times10$  \\
\cmidrule{2-9}                                                                                                                                                                                
&V9M \& V10M               & $M_{DM}/\Lambda_D^3$       & $2.50\times10$  & $2.50\times10$  & $4.50$  & $2.72\times10^{2}$  & $2.72\times10^{2}$  & $1.41\times10^{2}$  \\
\cmidrule{2-9}                                                                                                                                                                                
&V9P \& V10P               & $M_{DM}/\Lambda_D^3$       & $2.50\times10$  & $2.30\times10$  & $6.71\times10^{-1}$  & $2.71\times10^{2}$  & $2.66\times10^{2}$  & $4.37\times10$  \\
\cmidrule{2-9}                                                                                                                                                                            
&V11 \& V11A               & $M_{DM}^2/\Lambda_D^4$     & $2.82\times10^{2}$  & $2.71\times10^{2}$  & $3.35\times10$  & $6.96\times10^{3}$  & $6.83\times10^{3}$  & $2.56\times10^{3}$  \\
\bottomrule 
\end{tabular}
\caption{\label{tab:sigmas} Mono-jet cross-sections in fb for the EFT operators at 8 TeV and 13 TeV. The UV cut-off $\Lambda$ has been set to $1~\rm{TeV}$ for all operators. Operators with same cross-section have been grouped together.}
\end{table}
 
It must be noted that the cross-sections for operators proportional to $m_q$ are always extremely small, at both 8 TeV and 13 TeV.  For this reason, these scenarios are not plotted in Fig.~\ref{fig:sigmas}.

\subsection{Limits for LHC@8TeV}
We will now estimate the significance of the signal by taking into consideration the backgrounds and comparing with the observed data from experimental searches in the mono-jet channel at 8 TeV. For this purpose we will consider 2 mono-jet searches implemented in {\sc CheckMATE}, one from ATLAS~\cite{Aad:2015zva} and one from CMS~\cite{Khachatryan:2014rra}.

The ATLAS analysis selects events through the following main criteria: the leading jet must have $p_T>120$~GeV and $|\eta|<2$, the \MET{} must be larger than 150 GeV, and relations between the \MET{} and jets properties must be satisfied, i.e. $p_{\rm{Tj}}/\MET>0.5$ and $\Delta\phi(\rm{jet},$\textbf{\textit{p}}$_{\rm{T}}^{\rm{miss}})>1$; further cuts on sub-leading jets and vetoes on leptons are imposed; 9 signal regions (SRs) (not statistically independent) are then defined with increasing \MET{} cuts from 150~GeV to 700~GeV.

The CMS analysis has the following criteria: leading jet with $p_T>110$~GeV and $|\eta|<2.4$, \MET{} larger than 250 GeV and further cuts on sub-leading jets kinematic properties; 7 non statistically independent SRs are then defined with increasing \MET{} cuts from 250~GeV to 550~GeV.

\begin{figure}[htbp]
    \includegraphics[width=\textwidth]{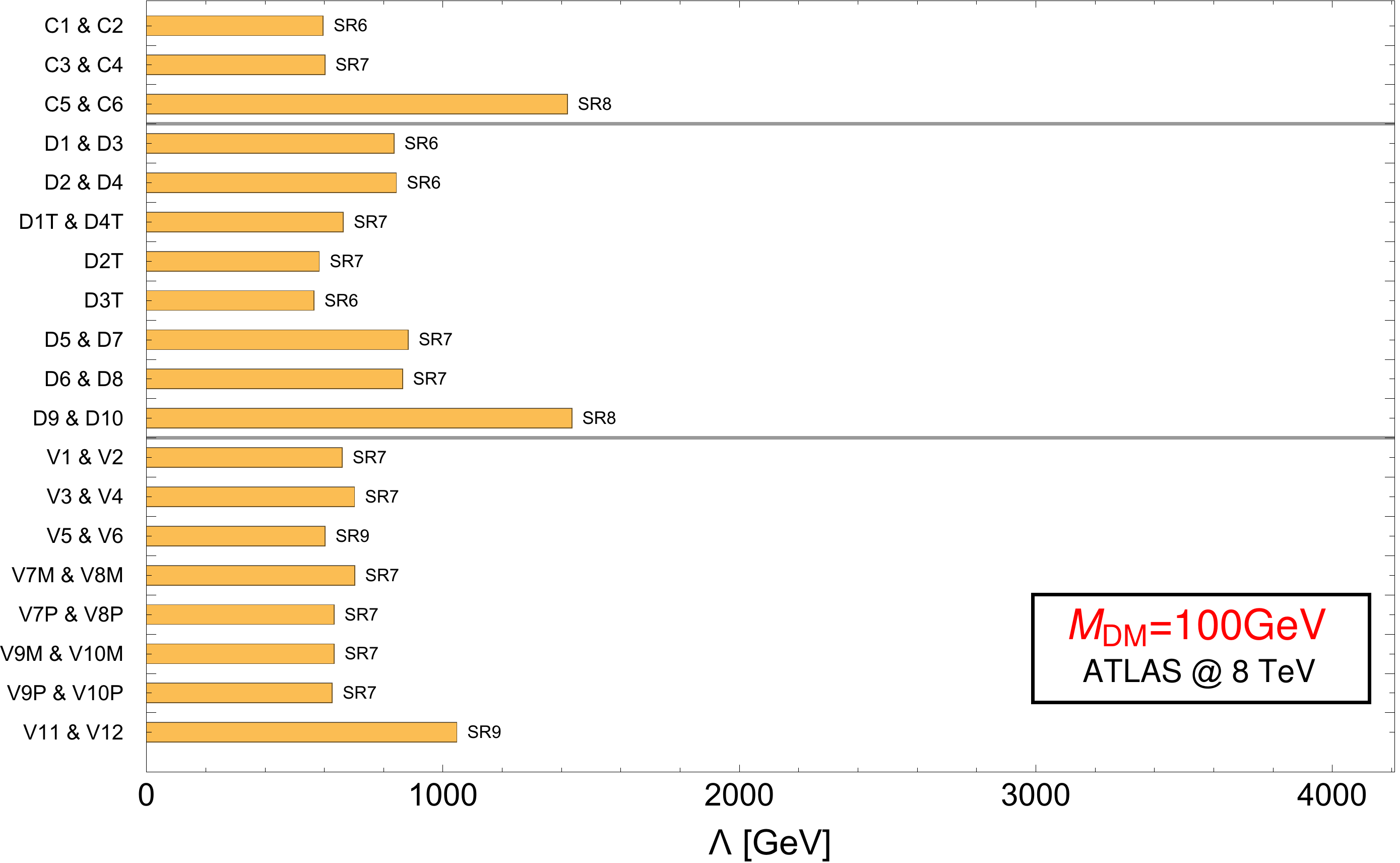}\\
    \includegraphics[width=\textwidth]{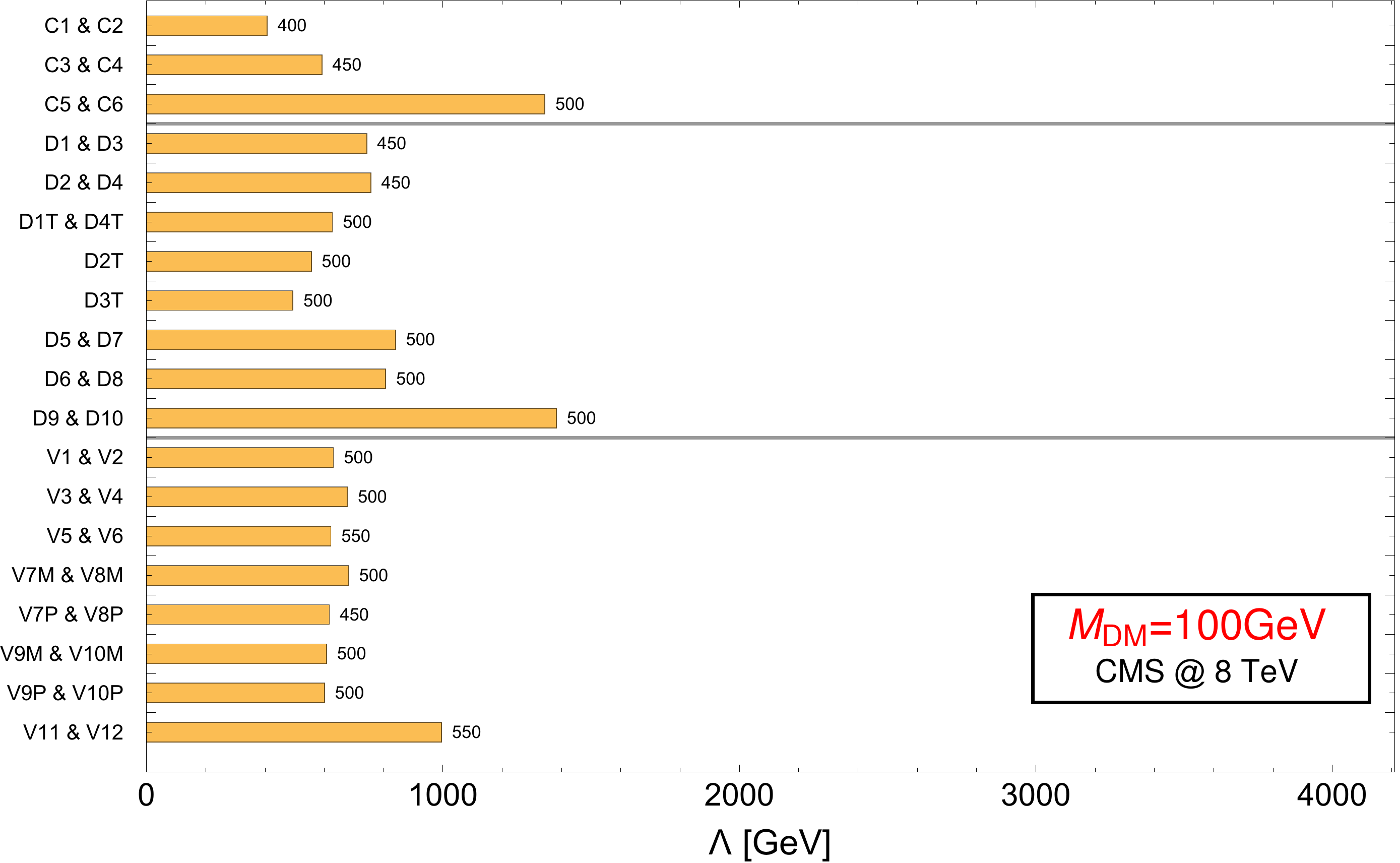}
  \caption{\label{fig:LambdaLimits8TeV}Observed 95\% CL limits on the UV cut-off $\Lambda$ from LHC 8 TeV data: from the ATLAS search(top) of Ref.~\cite{Aad:2015zva} and from the CMS search (bottom) of Ref.~\cite{Khachatryan:2014rra}. The signal regions which determine the strongest constraint are displayed for each operator. Black horizontal lines distinguish groups of operators for same-spin DM. 
  }
\end{figure}

Our results, in terms of 95\%CL limits on $\Lambda$, are provided in Fig.~\ref{fig:LambdaLimits8TeV} for all EFT operators under study. Here and in the following we have assumed a 10\% theoretical error on the signal for the CheckMATE statistical analysis.

From Fig.~\ref{fig:LambdaLimits8TeV} one can see that for $M_{DM}=100$ GeV, the LHC@8TeV limits for $\Lambda$ are between about 0.5 and 1. TeV for all operators.
For the $\Lambda_d$ parameterisation the limits on \VDM operators 
are enhanced as
\begin{equation}
\Lambda_d=\left(\Lambda^{D-4}M_{DM}{d-D}\right)^\frac{1}{d-4}
\end{equation}
as follows from  Eq.\ref{eq:LD}.
From this formula one can see that for $\Lambda_d$ the limits are enhanced in  \VDM case and scales with $M_{DM}$.
For example, for $M_{DM}=100$~GeV for the (V1,V2) and (V5,V6) the limit on $\Lambda_d$ is around 29 TeV and 22 TeV respectively exceeding limits on scalar and fermion DM case
(for which $\Lambda_d=\Lambda_D \equiv \Lambda$)  by more than one order of magnitude.
While confirming these results of Ref.~\cite{Kumar:2015wya}
we believe that this parameterisation which, from our point of view, leads to an artificial enhancement of the limits for \VDM case is not quite physical
and suggest the parameterisation we propose in this paper given in Eq.(\ref{eq:newparam})
for which the LHC limits on $\Lambda$ are of the same order of magnitude.

We would like to note that   $\Lambda$ is related to the mass of the heavy mediator but does not have  necessarily the same value, as we can see this depends on the parameterisation 
and the mechanism of how the new physics is realised at high scale.
Therefore it is hard to judge for how low values of  $\Lambda$ the EFT breaks down,
while of course we expect that $\Lambda$ should be about TeV scale or above.
The only robust criterion one can use within this framework is the unitarity condition.
Using results of Ref.\cite{Kumar:2015wya} we have checked  unitarity
limits for the most `dangerous' \VDM operators with $D=8$. 
The most stringent constraint comes from $(V3,V4)$ operators,
the mean values of invariant DM mass distribution for which is the highest.
The energy at which unitarity is violated, $E_{lim}$  for these operators is about
twice as $\Lambda_{lim}$, the LHC  limit on $\Lambda$, which means that unitarity is violated for $M_{inv}(DM,DM) >  2E_{lim} \simeq 4 \Lambda_{lim}$.
We have checked that $M_{inv}(DM,DM) >  2E_{lim}$ cut lead to about 15\% decrease of the cross section and less than 4\% decrease in the limit on $\Lambda$  which is a quite small correction even for  potentially the most problematic operator.
Here we do not perform detailed study on the unitarity which is out of the scope of this paper.

Results for  $M_{DM}=$10 and 1000~GeV masses are presented in Fig.~\ref{fig:LambdaLimits8TeV-detailed} of Appendix~\ref{app:2}. One can see that for  $M_{DM}=10$~GeV the limits are  very similar to the 100 GeV case, while 
for $M_{DM}=1000$~GeV the limits are visibly weaker especially for the $D=5$ (C1,C2) operators
(for which limits are about factor of 10 weaker) which $\MET$ shape is the most close to the SM BG one
as we demonstrate below.

\subsection{Limits for LHC@13TeV for current and projected luminosities}

\begin{figure}[htb]
     \includegraphics[width=\textwidth]{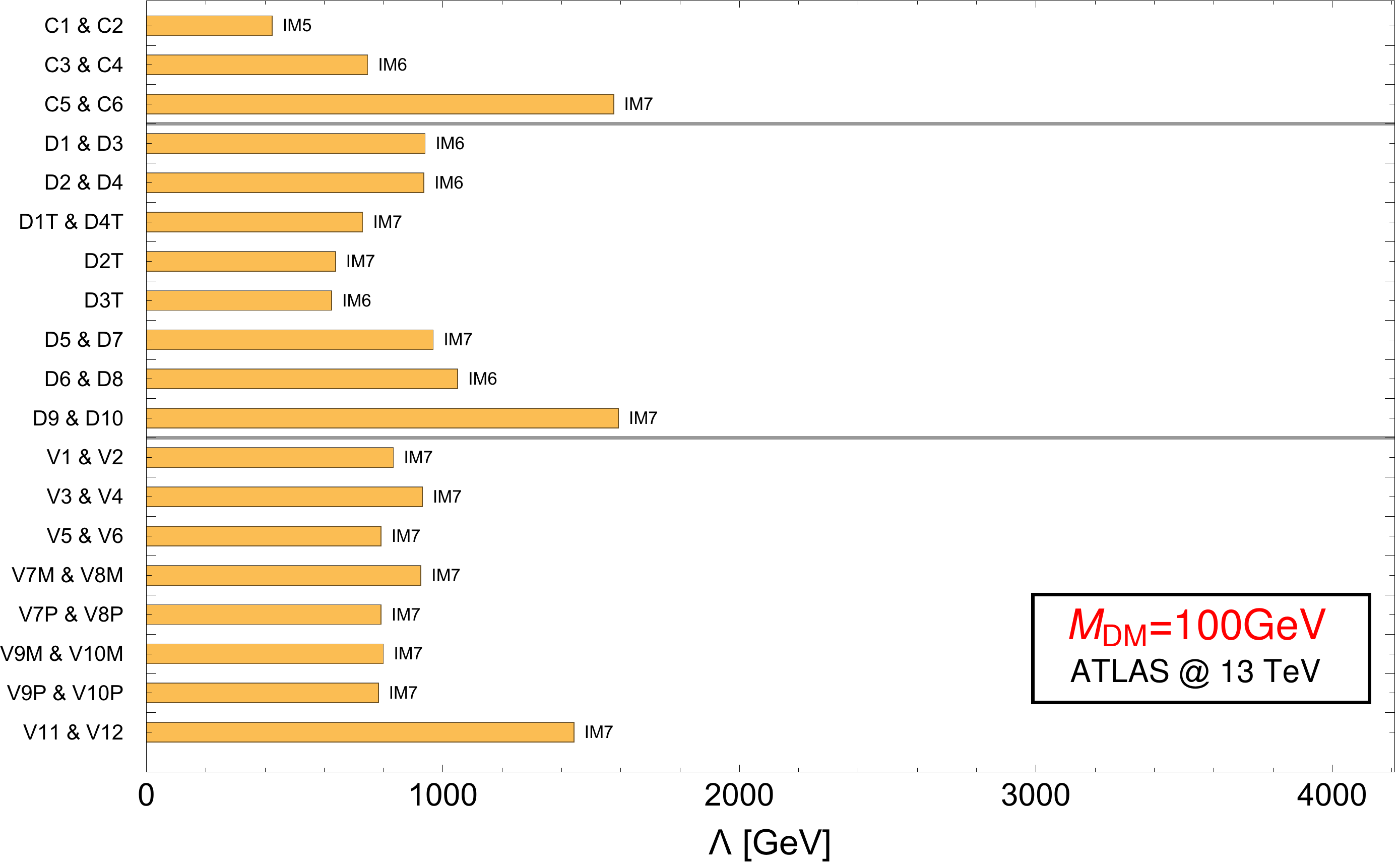}\\
 \caption{\label{fig:LambdaLimits13TeV}
 Observed limits on the UV cut-off $\Lambda$ from the ATLAS search of Ref.~\cite{Aaboud:2016tnv} using 13 TeV data with a luminosity of 3.2~fb$^{-1}$. See the caption of Fig.~\ref{fig:LambdaLimits8TeV} for more details about the interpretation of the plot.}
 \vskip 1cm
\end{figure}

In this section we find the limits for LHC@13TeV considering the ATLAS mono-jet analysis of Ref.~\cite{Aaboud:2016tnv}. This search considers a data sample obtained with a luminosity of 3.2~fb$^{-1}$ and, analogously to the 8 TeV searches, it uses  inclusive and exclusive signal regions, characterised by  cuts on the $\MET$, from 250 GeV to 750 GeV.
For our analysis we have used the recent implementation of this search into {\sc CheckMATE}~v2.0.1.
The limits on $\Lambda$ for the operators under study are presented in Fig.~\ref{fig:LambdaLimits13TeV} for $M_{DM}=100$~GeV (analogous limits for $M_{DM}=10$~GeV and 1000 GeV are presented in Fig.~\ref{fig:LambdaLimits13TeV-detailed} of Appendix~\ref{app:2}).

One can see that the bounds on $\Lambda$ with the 13 TeV data corresponding to a luminosity of  3.2 fb$^{-1}$ are very similar to those for  8 TeV data, corresponding to a luminosity of about a factor of 10 higher. This is expected, as the increase of the signal cross-section from 8 TeV to 13 TeV (about one order of magnitude in most cases) is compensated by the corresponding lack of luminosity.

The enhancement of the cross-section from 8 to 13 TeV combined with the significant increase of luminosity in the near future will indeed open a new potential for the LHC to test different DM theories and hopefully to understand their nature. Therefore, in the rest of this Section we analyse the LHC sensitivity to EFT DM operators for higher luminosities (up to 300 fb$^{-1}$) assuming the same selection and kinematics cuts of the ATLAS analysis of Ref.~\cite{Aaboud:2016tnv}. This will allow us to estimate the potential of current searches for distinguishing these operators and hence characterise DM properties  with the future LHC data. 

In order to be able to distinguish the EFT operators  with higher luminosities at 13 TeV we need 
the following conditions should be satisfied: 
1) given that now the signal is not observed yet, at higher luminosity
the significance of the signal must be large enough to actually claim it in the presence of the SM background (BG);
2) the number of signal events must be large enough to distinguish differences in the \textit{shape} of distributions. 

\begin{figure}[htb]
    \includegraphics[width=\textwidth]{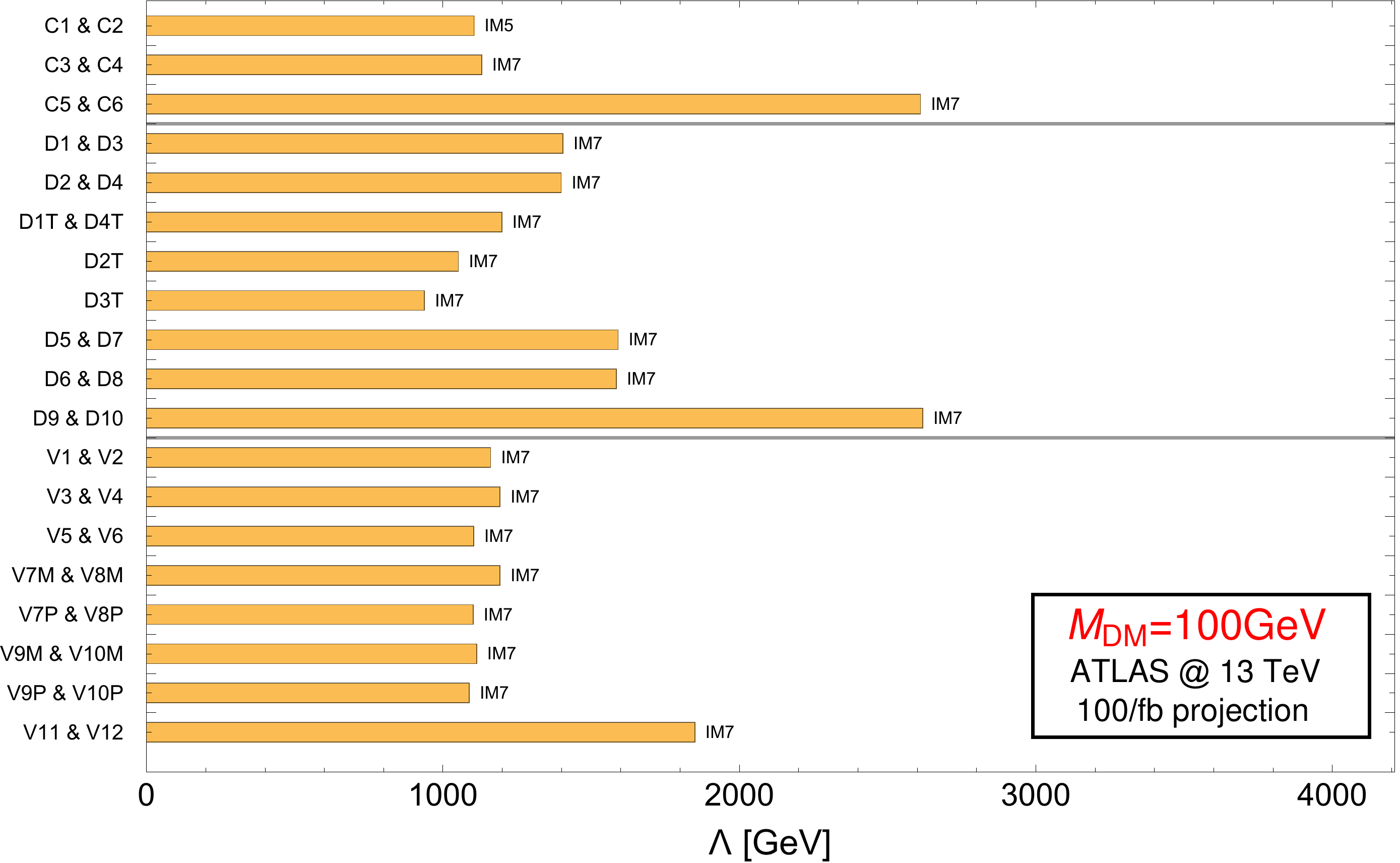}\\
    \includegraphics[width=\textwidth]{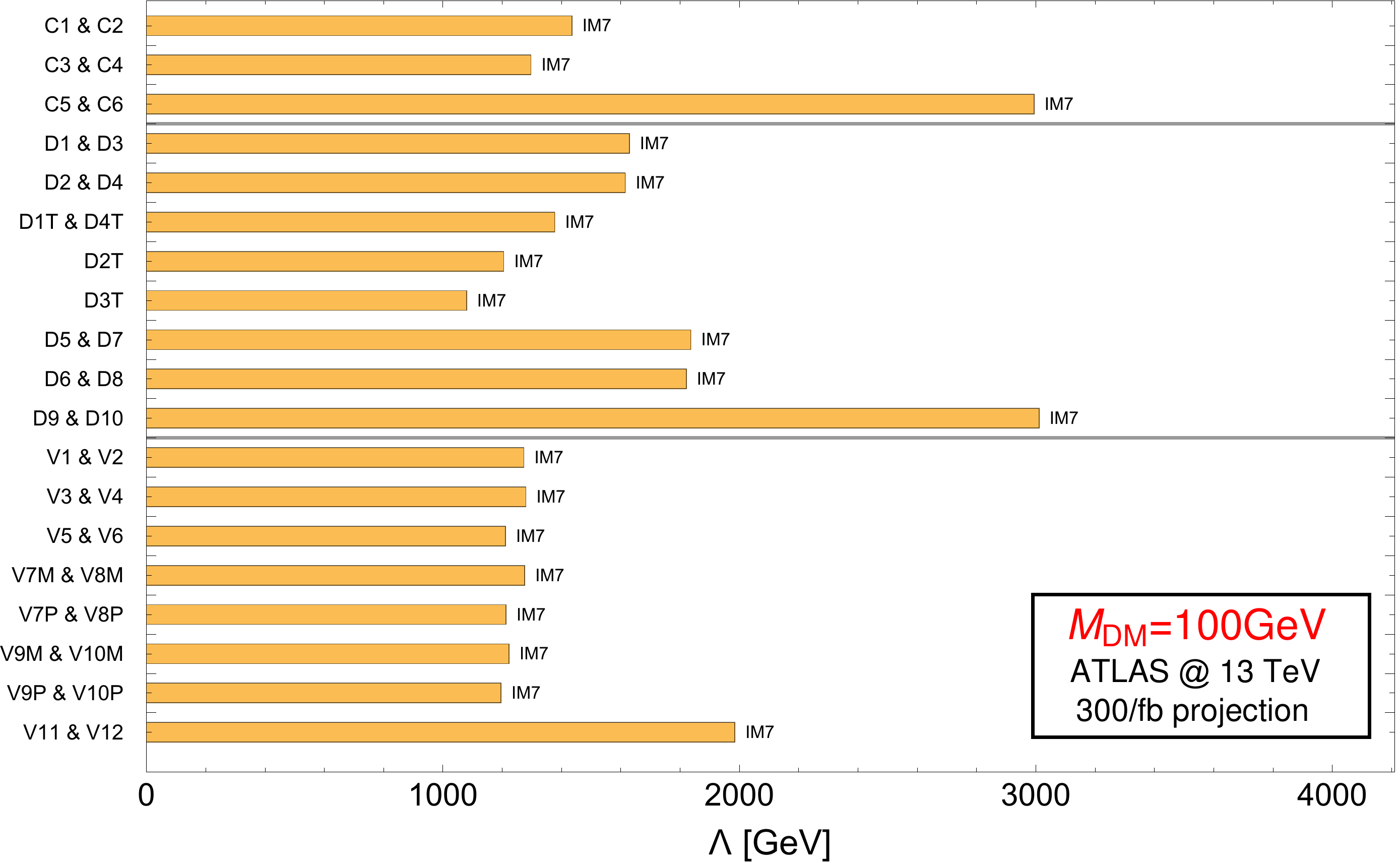}
  \caption{\label{fig:LambdaLimits13TeVRescaled}
Expected limits on the UV cut-off $\Lambda$ considering the selection of the ATLAS search of Ref.~\cite{Aaboud:2016tnv} and re-scaling to luminosities of
100~fb$^{-1}$ (top panel) and 300~fb$^{-1}$ (bottom panel). The exclusion values of $\Lambda$ correspond to the assumptions that the background scales linearly with the luminosity and that the number of observed events matches the background (expected limit).}
\end{figure}
Our estimation is based on the assumptions that the number of BG events scales with the luminosity and that the uncertainty on the BG scales as the square root of the luminosity. However, we set the lower limit for the BG uncertainty to be 1\% of the BG. This choice of 1\% for the limit on BG uncertainty is based on the post-fit numbers with respective BG error provided by ATLAS and CMS 
for \MET{} bins with high statistics, see e.g. 
\cite{Aaboud:2016tnv,CMS:2016tns} together with additional materials provided by CMS collaboration~\cite{EXO-16-013}.
We stress that the 1\% systematic uncertainty floor limit for the BG  plays a very important role. This statistically driven limit for the leading BG is based on the well measured $Zj \to \ell^+\ell^-j$ and $Wj \to \ell^+\nu j$ SM signatures, because the $p_T$ distributions for the observed lepton pair have the same shape as the corresponding \MET{} distributions in case the $Z$ boson decays to neutrinos or the charged lepton from the $W$-boson decay is lost.
Since the statistically driven BG error has a lower limit of 1\%, even for the most stringent cut in the ATLAS signal region IM7, with \MET{} threshold of 700 GeV, the BG uncertainty reaches the 1\% floor already with a luminosity of about 300 fb$^{-1}$. Therefore with this cut there will be no further improvement on the LHC reach with larger luminosities, and for this reason we do not present results for luminosity greater than 300 fb$^{-1}$.

\begin{table}
\small
\centering
\begin{tabular}{ccc|ccc|ccc}
\toprule
&\multirow{4}{*}{Operators}& \multirow{4}{*}{Coefficient} & \multicolumn{3}{c|}{Excluded $\Lambda$ (GeV) at 3.2 fb$^{-1}$} & \multicolumn{3}{c}{Excluded $\Lambda$ (GeV) at 100 fb$^{-1}$} \\
\cmidrule{4-9}
&              &                        & \multicolumn{3}{c|}{DM Mass}  & \multicolumn{3}{c}{DM Mass} \\
&              &                        & 10 GeV & 100 GeV & 1000 GeV & 10 GeV & 100 GeV & 1000 GeV   \\
\midrule                                                                                                                                                                                   
\midrule                                                                                                                                                                           
\multirow{3}{*}{\begin{sideways} \begin{tabular}{c} Complex \\ Scalar DM\end{tabular} \end{sideways}}                                                                                      
&C1 \& C2      & $1/\Lambda$            & 456    & 424     & 98       & 1168   & 1115    & 267        \\
\cmidrule{2-9}                                                                                        
& C3 \& C4     & $1/\Lambda^2$          & 750    & 746     & 400      & 1134   & 1131    & 662        \\
\cmidrule{2-9}                                                                                        
& C5 \& C6     & $1/\Lambda^2$          & 1621   & 1576    & 850      & 2656   & 2611    & 1398       \\
\midrule                                                                                                                                                                                  
\midrule                                                                                                                                                                                  
\multirow{13}{*}{\begin{sideways} Dirac Fermion DM \end{sideways}}                                                                                                                                         
&D1 \& D3      & $1/\Lambda^2$          & 931    & 940     & 522      & 1386   & 1405    & 861        \\
\cmidrule{2-9}                                                                                        
&D2 \& D4      & $1/\Lambda^2$          & 952    & 936     & 620      & 1426   & 1399    & 1022       \\
\cmidrule{2-9}                                                                                        
&D1T \& D4T    & $1/\Lambda^2$          & 735    & 729     & 476      & 1217   & 1199    & 780        \\
\cmidrule{2-9}                                                                                        
&D2T           & $1/\Lambda^2$          & 637    & 638     & 407      & 1053   & 1052    & 670        \\
\cmidrule{2-9}                                                                                        
&D3T           & $1/\Lambda^2$          & 586    & 625     & 391      & 969    & 938     & 644        \\
\cmidrule{2-9}                                                                                        
&D5 \& D7      & $1/\Lambda^2$          & 1058   & 967     & 721      & 1580   & 1591    & 1190       \\
\cmidrule{2-9}                                                                                        
&D6 \& D8      & $1/\Lambda^2$          & 978    & 1050    & 579      & 1608   & 1585    & 955        \\
\cmidrule{2-9}                                                                                        
&D9 \& D10     & $1/\Lambda^2$          & 1587   & 1592    & 958      & 2613   & 2619    & 1580       \\
\midrule                                                                                                                                                                  
\midrule                                                                                                                                                                  
\multirow{15}{*}{\begin{sideways} Complex Vector DM \end{sideways}}                                                                                                                        
&V1 \& V2      & $M_{DM}^2/\Lambda_D^3$ & 831    & 833     & 714      & 1162   & 1161     & 997       \\
\cmidrule{2-9}                                                                                        
&V3 \& V4      & $M_{DM}^2/\Lambda_D^4$ & 930    & 931     & 833      & 1196   & 1193     & 1070      \\
\cmidrule{2-9}                                                                                        
&V5 \& V6      & $M_{DM}^2/\Lambda_D^3$ & 784    & 791     & 711      & 1095   & 1104     & 993       \\
\cmidrule{2-9}                                                                                        
&V7M \& V8M    & $M_{DM}^2/\Lambda_D^4$ & 930    & 926     & 882      & 1195   & 1193     & 1130      \\
\cmidrule{2-9}                                                                                        
&V7P \& V8P    & $M_{DM}/\Lambda_D^3$   & 796    & 791     & 652      & 1112   & 1102     & 911       \\
\cmidrule{2-9}                                                                                        
&V9M \& V10M   & $M_{DM}/\Lambda_D^3$   & 796    & 799     & 737      & 1109   & 1114     & 1027      \\
\cmidrule{2-9}                                                                                        
&V9P \& V10P   & $M_{DM}/\Lambda_D^3$   & 794    & 782     & 609      & 1110   & 1089     & 850       \\
\cmidrule{2-9}                                                                                        
&V11 \& V11A   & $M_{DM}^2/\Lambda_D^4$ & 1435   & 1442    & 1309     & 1844   & 1850     & 1683      \\
\bottomrule                                 
\end{tabular}         
\caption{\label{tab:lambda100} Projections for the exclusion limits for $\Lambda$ with a luminosity of 100 fb$^{-1}$ with the cuts of the ATLAS search of Ref.~\cite{Aaboud:2016tnv}.}
\end{table}

If the shape of the signal \MET{} distribution is flatter than that of the BG, then eventually the LHC reach for the signal can be improved if cuts on \MET{} beyond the present searches are applied (as it was done, for example in~\cite{Barducci:2015ffa}) or if a shape analysis is performed. As stated above, for this study we consider the analysis cuts from the current ATLAS monojet search, but further improvements are the subject of the follow up paper. Finally, for our projections we also assume that the detector parameters in the {\sc Delphes} framework do not significantly change at high luminosities. 
\begin{figure}[h]
      \includegraphics[width=\textwidth]{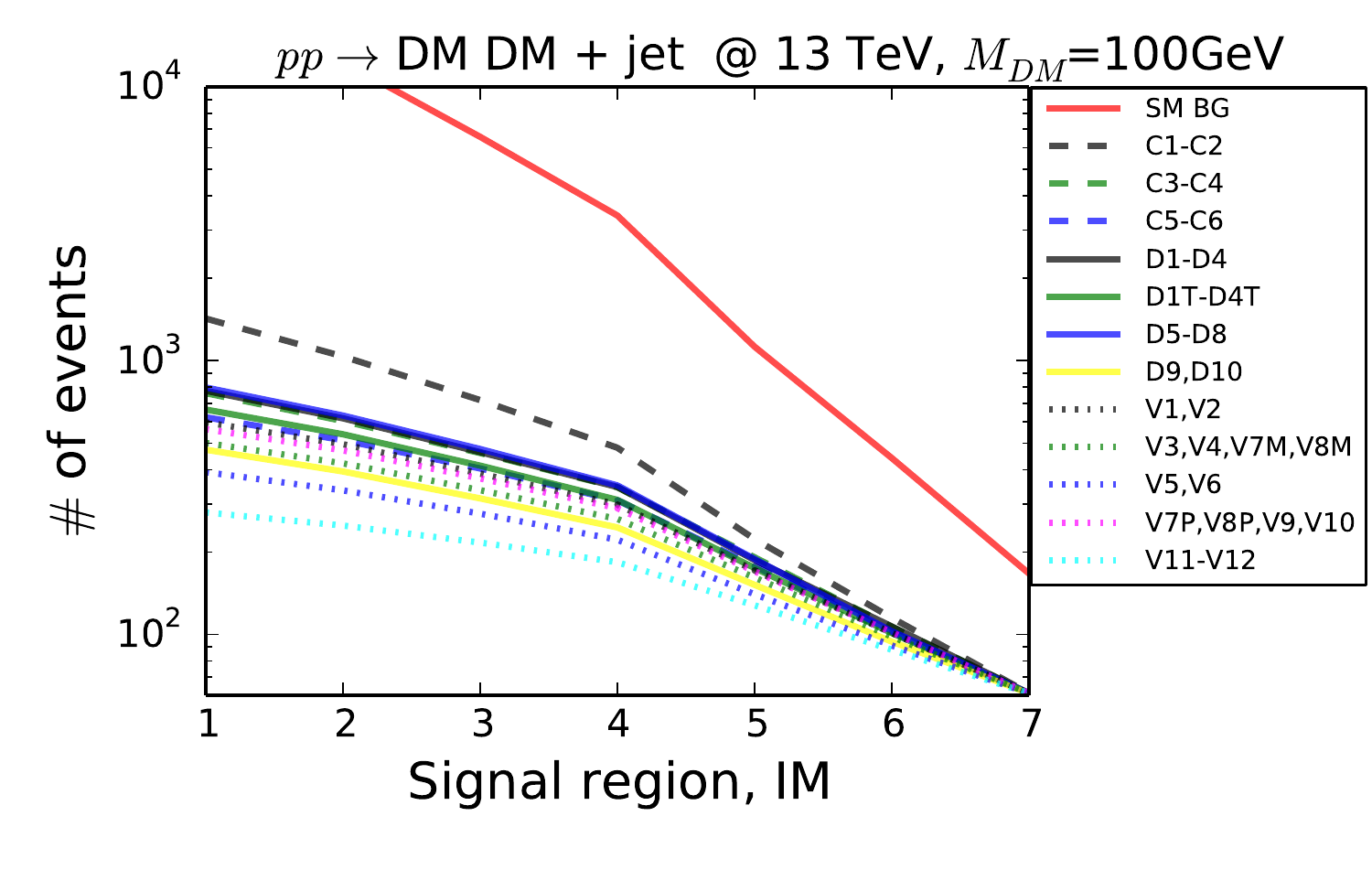}\\
      \includegraphics[width=\textwidth]{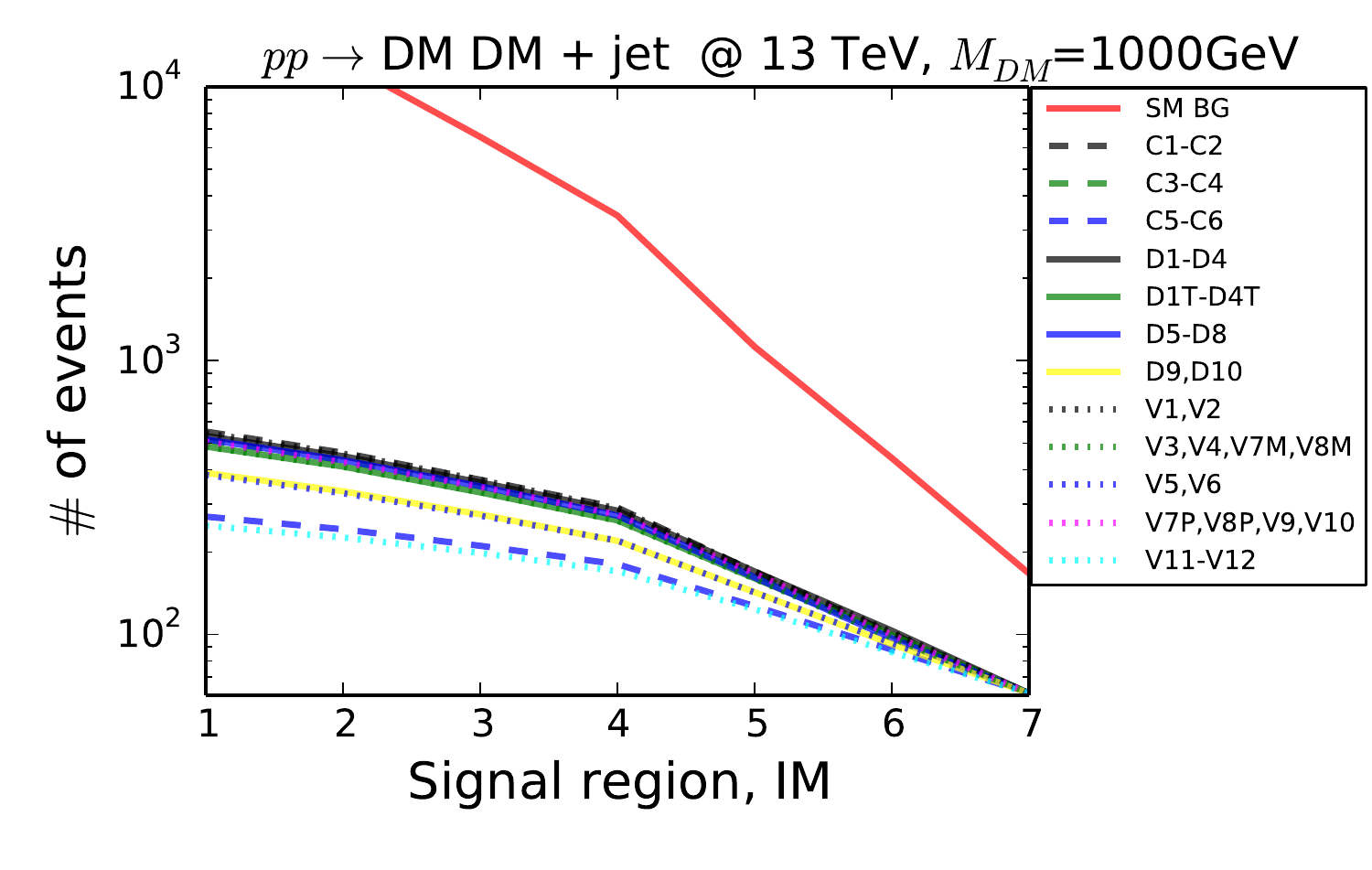}
  \caption{The number of signal 
events for the EFT operators under study for $IM_i$ bins ($i=1-7$) with the signal normalised to 61 events
(maximal allowed deviation from SM BG at 95\% CL) in the  $IM_7$ bin.
The SM BG is normalised to the same amount of events is also presented 
for the sake of the comparison of its shape to the signal. The top and bottom panels of the figure present results for $M_{DM}=100$ and 1000 GeV respectively.\label{fig:ratio} }
\end{figure}

The limits on $\Lambda$ for the projected luminosities of 100~fb$^{-1}$ and 300~fb$^{-1}$ are presented in Fig.~\ref{fig:LambdaLimits13TeVRescaled} for the ATLAS analysis of Ref.~\cite{Aaboud:2016tnv}. The numerical values of the excluded $\Lambda$ for 3.2 fb$^{-1}$ and
projected luminosity of 100 fb$^{-1}$ are presented in Tab.~\ref{tab:lambda100}.
One can see that increase of luminosity would allow to test $\Lambda$ by factor 1.5-3
higher (depending on the operator and DM mass).
One can notice  that the differences between the exclusion limits for 100~fb$^{-1}$ and 300~fb$^{-1}$ are small: this is related to the fact that the systematic error drops very slowly with the increase of the luminosity.

Let us take a look at  the shape difference of  \MET{}  of  BG and  signals, expressed in terms of  signal regions $IM_i$ of the ATLAS analysis~\cite{Aaboud:2016tnv}, where $IM_i=(250,300,350,400,500,600,700)$ define inclusive \MET{} cuts in GeV. 
In Fig.~\ref{fig:ratio} we present the number of signal 
events for the EFT operators under study for 7 $IM_i$ bins with the signal normalised to 61 events
(maximal allowed deviation from SM BG at 95\% CL) in the  $IM_7$ bin.
We also present the number of the expected SM BG events for the sake of comparison of its shape to the signal. The top and bottom panels of the figure present results for $M_{DM}=100$ and 1000 GeV respectively.
One can clearly see that all signal  \MET{} shapes are flatter than the BG
and different between each other. 
This shape difference will cause the respective  difference of
efficiencies for operators and the BG, with the largest difference of about two orders of magnitude occurring between  (V11,V12) and
SM BG.
Moreover, as the BG exhibits a steeper slope, this difference grows with increasing \MET{}: this creates the opportunity to improve the LHC sensitivity to DM models using higher \MET{} cuts in dedicated analysis.

\subsection{LHC@13TeV potential to distinguish EFT DM operators}

In this final  section we give a quantitative answer on the possibility to distinguish EFT DM operators at the LHC. 
We assume the presence of a signal from the EFT operators in the  current data close to  the exclusion limits found in the above sections and verify if these operators can be observed and distinguished at high LHC luminosities.

Let us recall that  in   Fig.~\ref{fig:ratio} we present the numbers of signal 
events $N^k_i$ with $N^k_7$ normalised to 61 events corresponding to the exclusion at 95\% CL 
for $3.2 fb^{-1}$ luminosity as well as number of expected background events $BG_i$, 
 where index $k$ denote the $k^{th}$  EFT operator, while index $i=1-7$
denotes the  $i^{th}$ signal region for the  [{\it EM1,EM2,EM3, EM4, EM5, EM6, IM7}] set.
We assume now  that there are $\frac{1}{2}N^k_7$ number of DM signal  events  present in data,
i.e. just half of those for  the exclusion at 95\% CL in the most sensitive signal region.
Such signal can not be detected at $3.2fb^{-1}$ experiment but for sure will be detected at large 
luminosities.  
At high luminosity both signal and background will be increased by the same factor. 
As we discussed in the previous section, the BG uncertainty  for $L\gtrsim 300fb^{-1}$  is about 1\% of background. 
Because the signal and the  BG uncertainty  are multiplied to the same luminosity factor, 
for the  $\chi^2$   evaluation at high luminosity one can use estimations for signal and background at
 $3.2 fb^{-1}$.
 Taking this into account, the  $\chi^2$ value for differentiating the signals from operator $k$,
and operator $l$ takes the form:
\begin{equation}
\chi^2_{k,l}=  \min_{\kappa} \sum \limits_{i=3}^7 [(\frac{1}{2}N^k_i-\kappa\cdot N^l_i)/(10^{-2} BG_i)]^2 
\label{eq:chi2}
\end{equation}
For all pairs of EFT operators we compare  the obtained value of $\chi^2_{min}$  with the reference value 9.49, corresponding to a 95\%CL for four degrees of freedom: if $\chi^2_{min}>9.49$ the operators can be distinguished
for  DM masses we have considered in this analysis.

The result  is shown in Table~\ref{tab:shapes} where we present the matrix of the $\chi^2$ values for all pairs of C1,C5,D1,D9,V1,V3,V5 and V11 operators (noting the equivalence of C1 and C2, C5 and C6, D1 and D2, D9 and D10, V1 and V2, V3 and V4, V11 and V12 pairs with identical \MET{} distributions) for 100 GeV and 1000 GeV  DM masses. This set of  operators represent all operators under study since it contains
all combinations of $D=5-8$, all structures -- scalar,vector, tensor -- of EFT operators and all partons -- quark and gluons --
which define the shape of $\MET$ distributions. We omit results for 10 GeV DM mass for the sake of simplicity  since those are very similar numerically and identical qualitatively to the 100 GeV case.

This choice results in a $16\times16$ matrix: each row corresponds to the operators for which the normalisation was fixed to provide $61/2$ events in the IM7 ATLAS~\cite{Aaboud:2016tnv} signal region at 3.2 fb$^{-1}$, while the normalisation of the corresponding operator in each column was chosen to minimise the value of the $\chi^2$ according to the Eq(\ref{eq:chi2}).
For values of the $\chi^2$ (for 4 degrees of freedom) above 9.49, the operators which are distinguishable at 95\%CL and highlighted  in red boldface font in Table~\ref{tab:shapes}.

\begin{sidewaystable}
\small
\begin{center}
\setlength{\tabcolsep}{3pt}
\begin{tabular}{ccc||cc|cc||cc|cc||cccc|cccc}
\toprule
&&& \multicolumn{4}{c||}{Complex Scalar DM} & \multicolumn{4}{c||}{Dirac Fermion DM} & \multicolumn{8}{c}{Complex Vector DM} \\
&&& \multicolumn{2}{c|}{100 GeV} & \multicolumn{2}{c||}{1000 GeV} & \multicolumn{2}{c|}{100 GeV} & \multicolumn{2}{c||}{1000 GeV} & \multicolumn{4}{c|}{100 GeV} & \multicolumn{4}{c}{1000 GeV} \\
    &&& C1 & C5 & C1 & C5 & D1 & D9 & D1 & D9 & V1 & V3 & V5 & V11 & V1 & V3 & V5 & V11  \\
\midrule
\midrule
\multirow{4}{*}{\begin{tabular}{c} Complex\\ Scalar\\ DM \end{tabular}} &
\multirow{1}{*}{\begin{tabular}{c} 100 \\ GeV \end{tabular}} &
   	C1        & 0.0  & \tcr{\bf19.7}  & \tcr{\bf25.54}  & \tcr{\bf74.63}  & \tcr{\bf11.73} &  \tcr{\bf41.79} &   \tcr{\bf25.78}  & \tcr{\bf52.58}  & \tcr{\bf22.97} &  \tcr{\bf32.89}  & \tcr{\bf54.35}  & \tcr{\bf73.34} &  \tcr{\bf25.18} &  \tcr{\bf34.61}  &  \tcr{\bf52.34} &  \tcr{\bf80.85} \\
      && 	C5 & \tcr{\bf15.74}  &          0.0  &         0.37  & \tcr{\bf16.25}  &         1.11 &          3.93 &           0.74  &         7.35  &         0.18 &          1.53  &          8.2  & \tcr{\bf15.73} &          0.44 &           1.9  &          7.24 &  \tcr{\bf19.13} \\
\cmidrule{2-19}
& \multirow{1}{*}{\begin{tabular}{c} 1000 \\ GeV \end{tabular} } &
	C1 & \tcr{\bf19.89}  &         0.36  &          0.0  & \tcr{\bf11.82}  &         2.33 &          2.09 &           0.27  &         4.58  &         0.06 &          0.45  &         5.29  & \tcr{\bf11.41} &          0.06 &          0.68  &          4.42 &  \tcr{\bf14.36} \\
&& 	C5& \tcr{\bf50.86}  & \tcr{\bf13.86}  & \tcr{\bf10.34}  &          0.0  & \tcr{\bf21.03} &           3.7 &   \tcr{\bf11.18}  &         1.53  & \tcr{\bf11.57} &          6.82  &         1.26  &         0.01 &  \tcr{\bf10.84} &           6.1  &          1.61 &          0.14 \\
\midrule
\midrule
\multirow{4}{*}{\begin{tabular}{c} Dirac \\ Fermion \\ DM \end{tabular} } &
\multirow{1}{*}{\begin{tabular}{c} 100 \\ GeV \end{tabular}} &
	D1  & \tcr{\bf9.88}  &         1.17  &         2.52  & \tcr{\bf25.99}  &          0.0 &          9.23 &            2.4  & \tcr{\bf14.17}  &         1.85 &          5.09  & \tcr{\bf15.34}  & \tcr{\bf25.37} &          2.29 &          5.85  &  \tcr{\bf13.85} &  \tcr{\bf29.81} \\
 && 	D9   & \tcr{\bf30.49}  &         3.59  &         1.96  &         3.96  &         7.99 &           0.0 &           2.71  &         0.52  &         2.49 &          0.62  &         0.73  &         3.69 &          2.31 &          0.39  &          0.56 &          5.36 \\
\cmidrule{2-19}
& \multirow{1}{*}{\begin{tabular}{c} 1000 \\ GeV \end{tabular} } &
        D1  & \tcr{\bf20.31}  &         0.73  &         0.27  & \tcr{\bf12.92}  &         2.25 &          2.93 &            0.0  &         5.42  &         0.32 &          0.82  &         6.33  & \tcr{\bf12.58} &          0.08 &          1.18  &          5.08 &  \tcr{\bf15.7} \\
&& 	D9 & \tcr{\bf37.38}  &         6.54  &         4.18  &          1.6  & \tcr{\bf11.96} &           0.5 &           4.89  &          0.0  &         4.98 &          2.02  &         0.06  &         1.44 &          4.56 &          1.61  &          0.04 &          2.55 \\
\midrule
\midrule
\multirow{11}{*}{\begin{tabular}{c} Complex \\ Vector \\ DM \end{tabular}} &
\multirow{4}{*}{\begin{tabular}{c} 100 \\ GeV \end{tabular}} &
	V1 & \tcr{\bf18.06}  &         0.17  &         0.06  & \tcr{\bf13.34}  &         1.72 &          2.68 &           0.32  &          5.5  &          0.0 &          0.77  &         6.25  & \tcr{\bf12.9} &           0.1 &          1.06  &          5.34 &  \tcr{\bf16.03} \\
&&      V3 & \tcr{\bf24.86}  &         1.45  &         0.44  &         7.57  &         4.57 &          0.65 &           0.79  &         2.14  &         0.74 &           0.0  &         2.68  &         7.25 &          0.57 &          0.03  &          2.04 &  \tcr{\bf9.59} \\
&& 	V5 & \tcr{\bf38.36}  &         7.24  &         4.79  &          1.3  & \tcr{\bf12.86} &           0.7 &           5.67  &         0.06  &         5.61 &           2.5  &          0.0  &         1.14 &          5.24 &          2.04  &          0.13 &          2.13 \\
&& 	V11 & \tcr{\bf50.03}  & \tcr{\bf13.43}  & \tcr{\bf10.0}  &         0.01  & \tcr{\bf20.55} &          3.45 &   \tcr{\bf10.89}  &         1.39  & \tcr{\bf11.2} &          6.54  &         1.11  &          0.0 &  \tcr{\bf10.52} &          5.83  &          1.49 &          0.16 \\
\cmidrule{2-19}
& \multirow{4}{*}{\begin{tabular}{c} 1000 \\ GeV \end{tabular}} &
	V1  & \tcr{\bf19.73}  &         0.43  &         0.06  & \tcr{\bf12.46}  &         2.13 &          2.48 &           0.08  &         5.02  &          0.1 &          0.59  &         5.83  & \tcr{\bf12.09} &           0.0 &          0.89  &          4.78 &  \tcr{\bf15.14} \\
&&       V3 & \tcr{\bf25.96}  &         1.78  &         0.65  &         6.72  &         5.21 &           0.4 &           1.12  &          1.7  &         1.01 &          0.03  &         2.17  &         6.41 &          0.85 &           0.0  &          1.65 &           8.6 \\
&& 	V5  & \tcr{\bf37.33}  &         6.47  &         4.04  &         1.68  & \tcr{\bf11.72} &          0.55 &           4.59  &         0.04  &         4.84 &          1.93  &         0.14  &         1.55 &          4.34 &          1.57  &           0.0 &          2.72 \\
&& 	V11 & \tcr{\bf54.48}  & \tcr{\bf16.14}  & \tcr{\bf12.42}  &         0.13  & \tcr{\bf23.85} &          4.95 &   \tcr{\bf13.43}  &         2.41  & \tcr{\bf13.74} &          8.55  &         2.03  &         0.16 &  \tcr{\bf13.01} &          7.73  &          2.57 &           0.0 \\
\bottomrule
\end{tabular}
\end{center}
\caption{\label{tab:shapes}The matrix of $\chi^2$ values for all pairs of  C1,C5,D1,D9,V1,V3,V5 and V11 operators and 100 GeV and 1000 GeV masses. For the operator in each row the normalisation was fixed to provide $61/2$ events in the IM7 ATLAS~\cite{Aaboud:2016tnv} signal region at 3.2~fb$^{-1}$, while the normalisation of the corresponding operator in each column was chosen to minimize the value of the $\chi^2$ according to Eq(\ref{eq:chi2}).
If the $\chi^2$ values (for 4 degrees of freedom) are above 9.49, corresponding to operators which are distinguishable at 95\%CL, the entries in have been highlighted in red boldface font.}
\end{sidewaystable}

In particular, from Table~\ref{tab:shapes} one can see that:
\begin{itemize}
\item The C1  operator can be  distinguished from all other operators, with same or different DM masses.
\item The C1 operator with  100 GeV DM and 1000 GeV masses can be distinguished,
the same true also for C5 operator,
contrary to all other operators, which means that the shape of $\MET$ only for C1 and C5 operators
significantly changes with the increase of DM mass.
\item For $M_{DM}=100$~GeV the C5  operators can be distinguished (in addition to C1) from   D1, D9 and V1
\item  For $M_{DM}=100$~GeV  the D1 operator  can be distinguished (in addition to C1 and C5)  from D9, V5 and V11
\item all vector DM operators can be distinguished from C1, but not from each other, with the only exception of V1 and V11, which are clearly distinguishable
from each other (as well as V3 and V11 for the cases of some masses)
\end{itemize}

Therefore, the certain sets of DM EFT operators can be distinguished.
This is especially true for the C1 operator which can be distinguished from all others, which gives the possibility to link a C1-like signal with the spin of the DM.
One can expect that further exploration of the LHC potential beyond the cuts defined in the IM7 signal region should lead to a substantial improvement of the LHC sensitivity for the distinction of operators in EFT scenarios and in the characterisation of DM properties.
 
\clearpage

\section{Conclusions}\label{sec:conclusion}

We have analysed mono-jet signals from Dark Matter (DM)
in the Effective Field Theory (EFT) approach
and the LHC potential to
distinguish  EFT operators and DM properties.

We studied the complete set of dimension-5 and dimension-6 effective operators
involving interactions of scalar, fermion and vector DM with SM quarks and gluons,
implemented the models into CalcHEP and Madgraph,
fully validated them, and made them publicly available at the HEPMDB model database.

We have found that the main observable  \MET{} which allows to distinguish DM EFT operators
is fully defined by  the effective dimension of the operator, $D$, the structure of the operator
(scalar, vector or tensor) and the parton densities of the SM partons (quarks or gluons) of the operator.
The effective dimension of \VDM operators is different from the naive one because of the $E/M_{DM}$ enhancement 
factor for each longitudinal  \VDM polarisation, such that $D=7$ or 8 for \VDM operators.
Because of this fact we have suggested a new parameterisation for \VDM operators given by the Eq.(\ref{eq:newparam}).

We have found that if the invariant  mass of the DM pair, $M_{\rm inv}(DM,DM)$, is fixed then \MET{}  is defined by the SM part of the EFT operator (as presented in Fig.~\ref{fig:DMDM-MET}) and that the larger the invariant DM pair mass, the less steep is the resulting \MET{} distribution. 
$M_{\rm inv}(DM,DM)$ distributions are not observable but they are correlated with the \MET{} distributions  (Fig.~ \ref{fig:AllOperators-10-100-parton}) and, 
since the effective dimension $D$, the structure of the operator and the parton densities uniquely define $M_{\rm inv}(DM,DM)$ distributions (Fig.~ \ref{fig:inv-mass}), operators for which one or more of these factors are different are potentially distinguishable at the LHC.
Since DM spin is partly correlated with these factors, LHC can potentially shed light also on DM spin. 
For large $M_{DM}\gtrsim 1$~TeV, the  DM pair is produced close to threshold, so the sensitivity to $D$ of the operators is 
suppressed and the \MET{} distribution is completely defined by the SM component of the operator.   
Using Fierz transformations we show how this analysis can also be applied to operators which are not a product of SM and DM bilinears.

We have shown that the pattern of \MET{} distributions initially observed at the parton level
is not changed at the detector level (Fig.~\ref{fig:ComparisonDetector}) and have analysed the LHC sensitivity
to EFT DM operators and assessed the LHC potential to distinguish them at high luminosities.

We have found that at the LHC with a luminosity of 300 fb$^{-1}$ or higher and using the kinematic cuts of the current ATLAS monojet analysis~\cite{Aaboud:2016tnv} it is  possible to distinguish certain classes of EFT operators among each other, such as (C1,C2), (C5,C6) , (D1,D2), (D9,D10), (V1,V2), (V3,V4), (V5,V6) and (V11,V12) (Table~\ref{tab:shapes}).
A further exploration of the LHC potential beyond the $\MET{}>700$ GeV cut of the current analysis 
and beyond 300 fb$^{-1}$ should lead to a substantial improvement of the LHC sensitivity to the DM models
and to the characterisation of DM properties including its spin.
We would like to stress the LHC has a sensitivity to the main three factors which uniquely define the $\MET{}$
shape -- effective dimension $D$, the structure of the operator and the involved parton densities --
and not directly to DM spin. However, for some operators, the spin of DM is correlated with these factors, so scenarios with one or even two DM spins can be excluded in some cases as one can see from Table~\ref{tab:shapes}.

We have also found a drastic difference in the efficiencies (up to two orders of magnitude) for large \MET{} cuts and for
scenarios with different operators. This makes a further step forward beyond the results obtained at the LHC DM forum~\cite{Abercrombie:2015wmb}.
Finally, our analysis could be generically applicable to different scenarios, not necessarily in the EFT approach,
where the mediator is not produced on-the-mass-shell, such as the case of t-channel mediator, or mediators with mass below
$2M_{DM}$, where the $M_{\rm inv}(DM,DM)$ is not fixed.

\section*{Acknowledgements}
The authors would like to thank John Ellis, Thiago Tomei and Chang-Seong Moon for useful discussions.
Authors also acknowledge summer students: J. Pillow, J. Blandford and T. V. B. Claringbold
for useful interaction and participation at the very early stage of the project.
AB acknowledges partial support from the STFC grant ST/L000296/1, the NExT Institute , Royal Society Leverhulme Trust Senior Research Fellowship LT140094 and Soton-FAPESP grant.
MCT acknowledges partial support from Soton-FAPESP grant.
AP acknowledges partial support from SOTON Jubilee visiting grant as well from Royal Society International Exchanges grant IE150682.
AB and MT would like to thank FAPESP grant 2011/11973-4 for funding their visit to ICTP-SAIFR
where part of this work was done.
Authors would like to thank  Referees of the paper, critique and comments of whom helped to improve the quality of the paper and further highlight its main message.
Also, authors are greateful to Giovanni Grilli di Cortona for the cross-check of  Eq.(2.1)
and finding typo in it.
\clearpage
\appendix
\appendix
\section{Fierz Identities}\label{app:Fierz}
The Fierz identities can be used to rewrite the product of two Dirac bilinears as a linear combination of other bilinears with the Dirac spinors in a different order.
These are well known and discussed (e.g. \cite{Nishi:2004st}). Here we provide a brief derivation of the identity required to transform between different dimension-6 operators, mainly using the notation of \cite{Itzykson:1980rh}.

$4 \times 4$ complex matrices are spanned by 16 basis matrices. Different bases can be chosen to simplify the calculations depending mainly on whether your initial dimension-6 operators contain chiral projection operators or not. For the purpose of our calculation, we use the basis $\Gamma^\alpha = \{I,\gamma^\mu, \sigma^{\mu \nu}, \gamma^5 \gamma^\mu, i \gamma^5 \}$, where $\sigma^{\mu \nu} = \frac{i}{2}\left[\gamma^\mu, \gamma^\nu \right]$, and defining $\Gamma_{\alpha} = (\Gamma^\alpha)^{-1}$.

The basis is chosen to satisfy,
\begin{align}
\mathrm{Tr} \left( \Gamma^\alpha \Gamma_\beta \right) = 4 \delta^\alpha_\beta
\end{align}
and this orthogonality allows us to expand a general $4 \times 4$ matrix, $X$, in terms of this basis,
\begin{align}
X = x_\alpha \Gamma^\alpha = x_\beta \delta^\beta_\alpha \Gamma^\alpha = \frac{1}{4} \mathrm{Tr}(X \Gamma_\alpha) \Gamma^\alpha.
\end{align}
Writing this explicitly in terms of matrix elements, 
\begin{align}
X_{ij} = \frac{1}{4} X_{kl} [\Gamma_\alpha]_{lk} [\Gamma^\alpha]_{ij},
\end{align}
we see that,
\begin{align}
\frac{1}{4} [\Gamma_\alpha]_{lk} [\Gamma^\alpha]_{ij} = \delta_{lj} \delta_{ki}. \label{eqn:FierzIdent}
\end{align}
Inserting Kronecker deltas into our dimension-6 operator allows Equation~\ref{eqn:FierzIdent} to be used to derive a useful identity for Fierz transformations. If $X$ and $Y$ are any $4 \times 4$ matrix we have,
\begin{align}
(\bar{\chi} X q) (\bar{q} Y \chi) &= (\bar{\chi}_i X_{ij} \delta_{jk} q_k) (\bar{q}_l \delta_{lm} Y_{mn} \chi_n)\\
&= \frac{1}{4} (\bar{\chi}_i X_{ij} [\Gamma^\alpha]_{jm} Y_{mn} \chi_n) (\bar{q}_l [\Gamma_\alpha]_{lk} q_k)\\
&= \sum_{\alpha} \frac{1}{4} (\bar{\chi} X \Gamma^\alpha Y \chi) (\bar{q} \Gamma_\alpha q)
\end{align}
where the sum over indices has been made explicit in the final line.

To Fierz transform D1T, we set $X$ and $Y$ to the identity matrix yielding Equation~\ref{eqn:D1T}. Similar results apply for operators D2T to D4T.

\section{Plots for the LHC reach for DM masses of 10 GeV and 1000 GeV.}
\label{app:2}
\begin{figure}[ht!]
  \begin{subfigure}[t]{1.\textwidth}
    \includegraphics[width=0.5\textwidth]{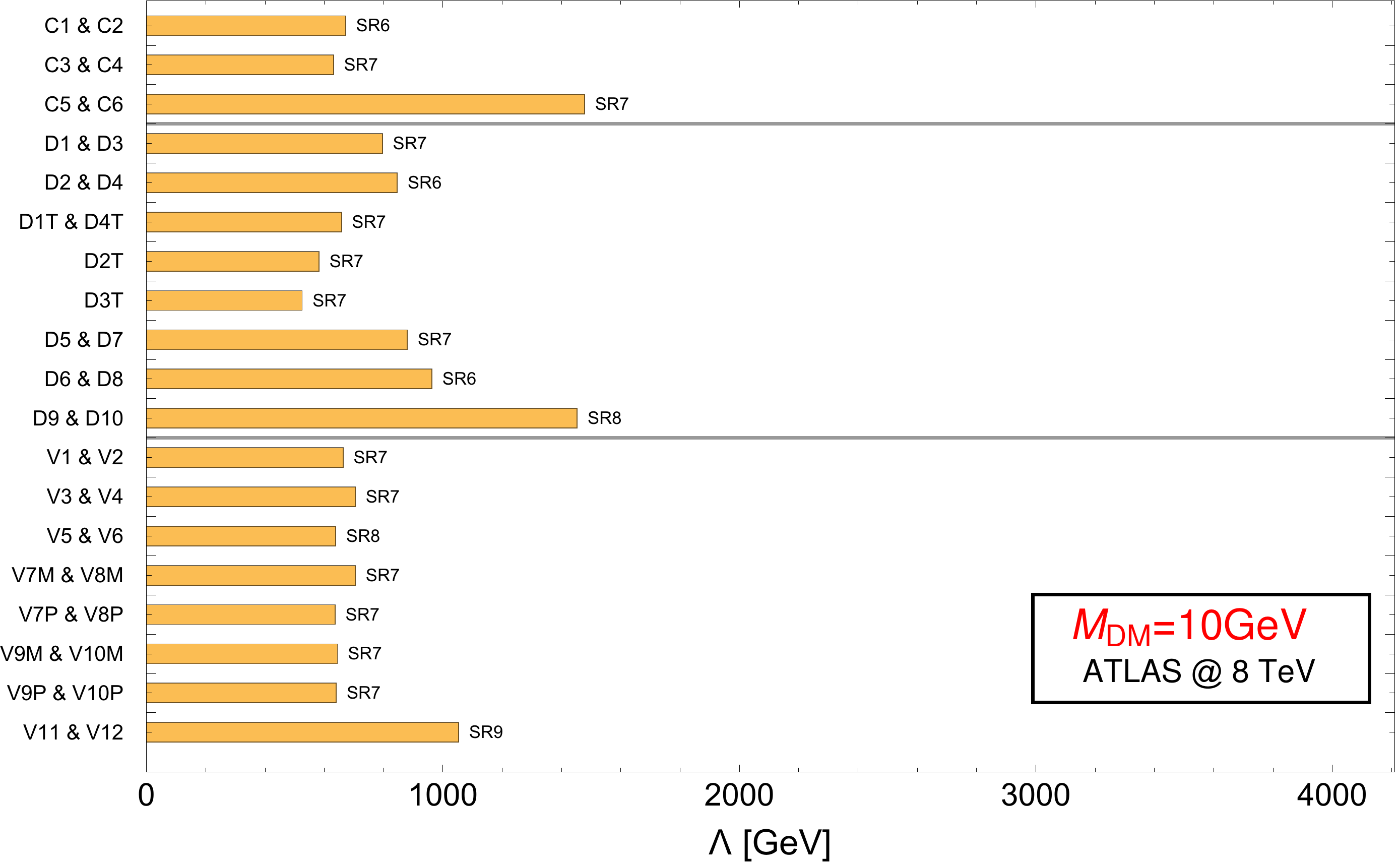}\quad
    \includegraphics[width=0.5\textwidth]{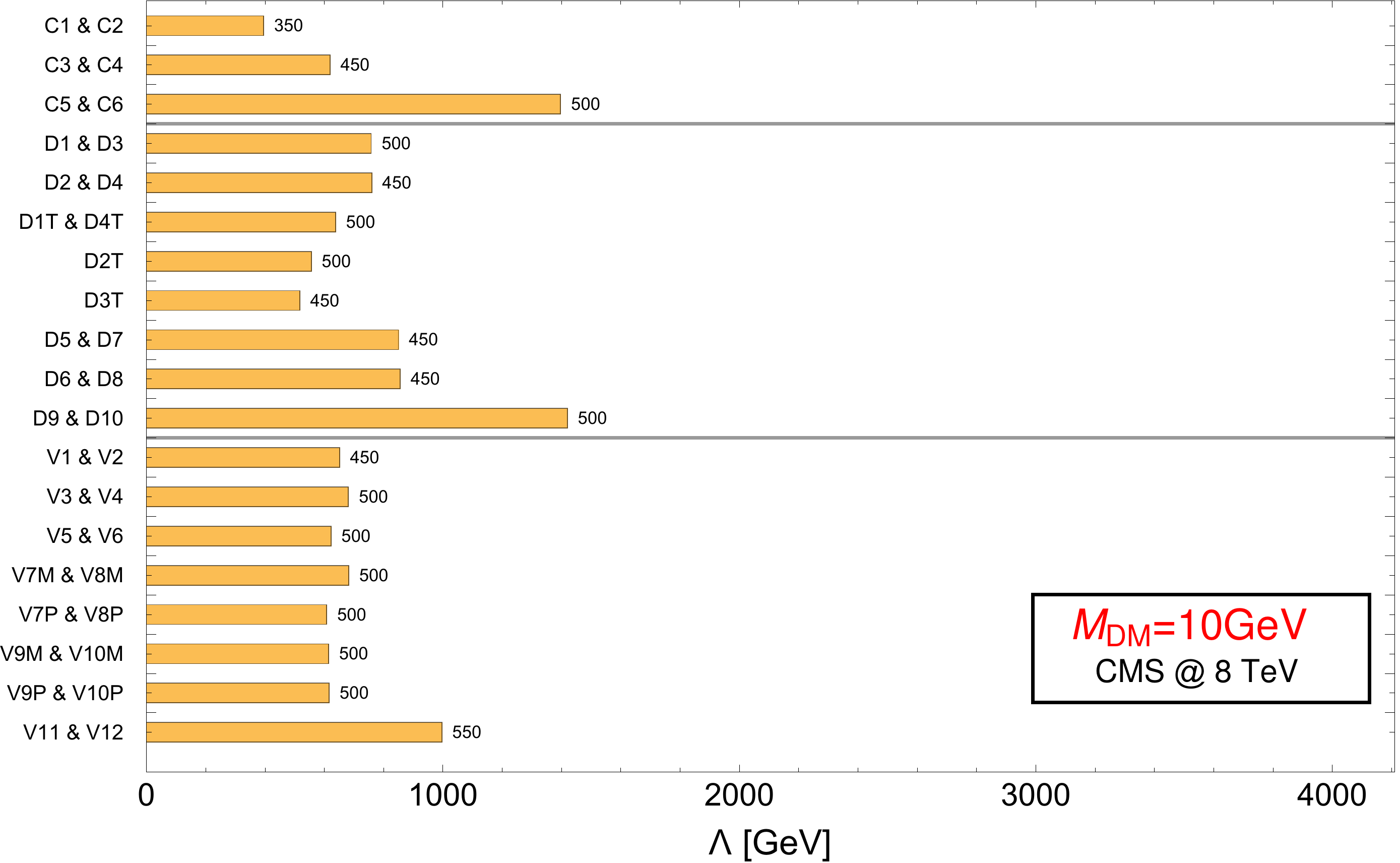}    
    \subcaption{$M_{DM} = 10$ GeV}
    \vspace{4mm}
  \end{subfigure}
  \begin{subfigure}[t]{1.\textwidth}
    \includegraphics[width=0.5\textwidth]{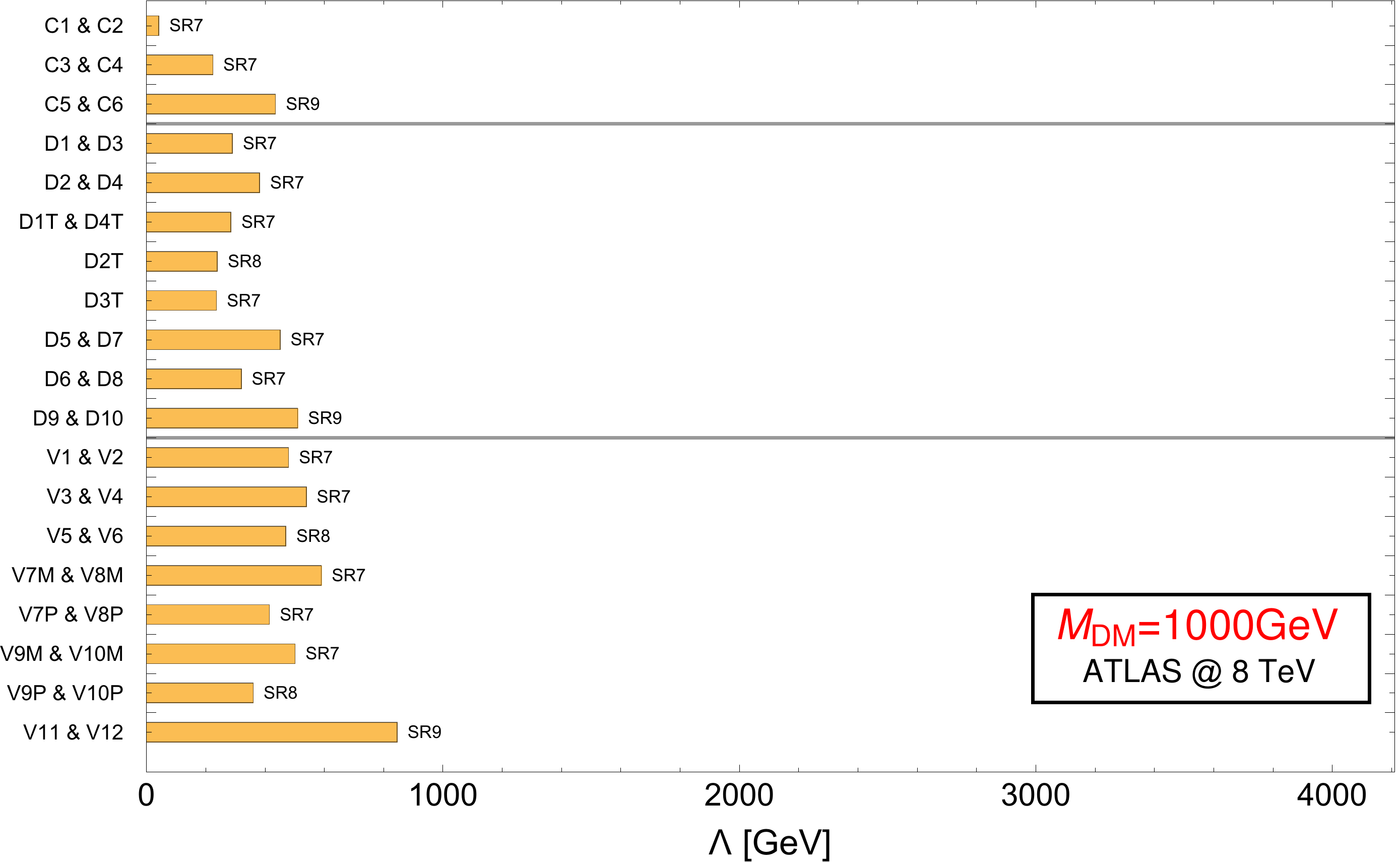}\quad
    \includegraphics[width=0.5\textwidth]{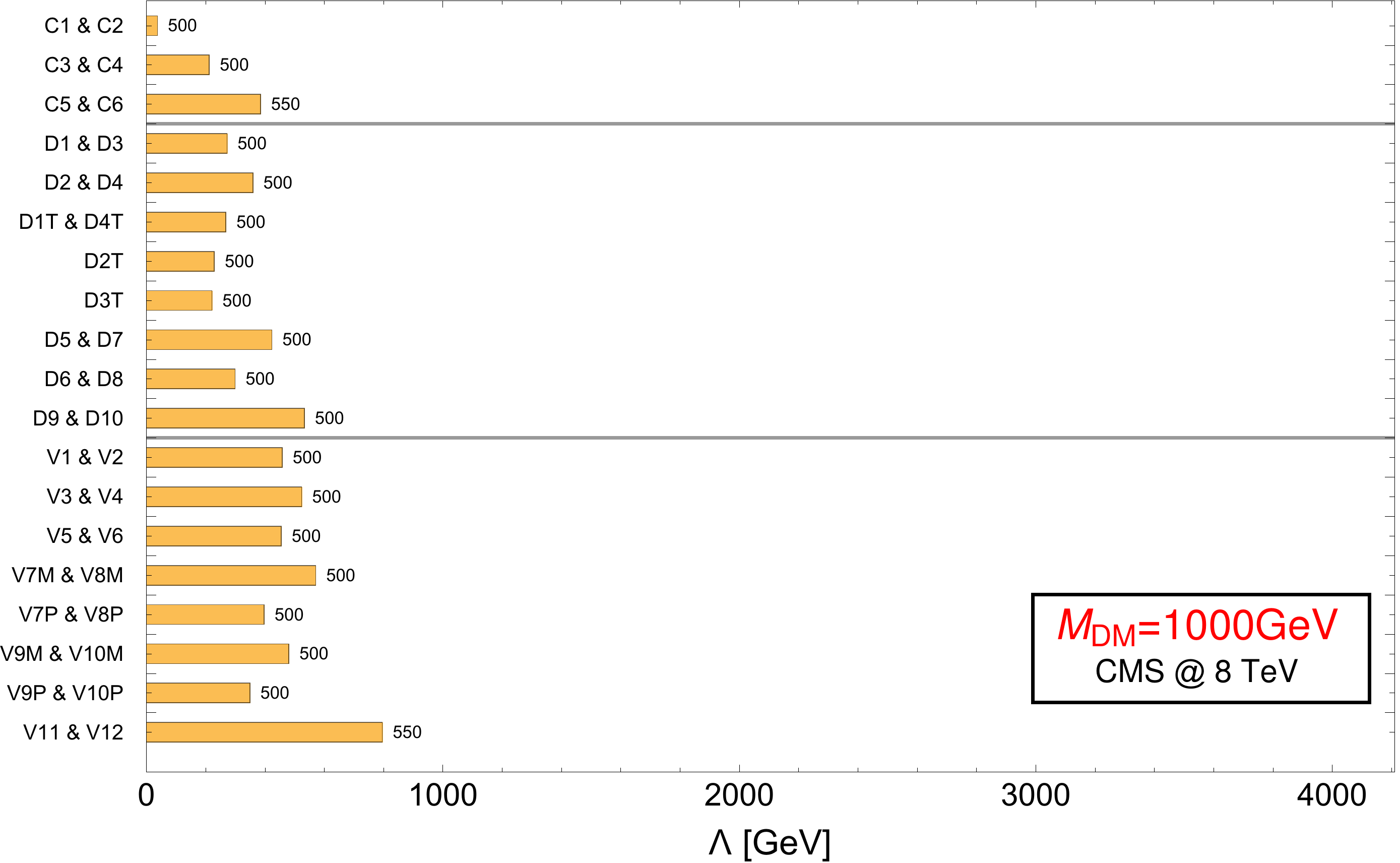}
    \subcaption{$M_{DM} = 1000$ GeV}
    \vspace{4mm}
  \end{subfigure}
  \caption{\label{fig:LambdaLimits8TeV-detailed}Observed 95\% CL limits on the UV cut-off $\Lambda$ from LHC 8 TeV data: from the ATLAS search(left) of Ref.~\cite{Aad:2015zva} and from the CMS search (right) of Ref.~\cite{Khachatryan:2014rra}. See the caption of Fig.~\ref{fig:LambdaLimits8TeV} for more details about the interpretation of the plot. }
\end{figure}

\begin{figure}[ht!]
  \centering
  \begin{subfigure}[t]{1.1\textwidth}
    \centering
    \includegraphics[width=0.51\textwidth]{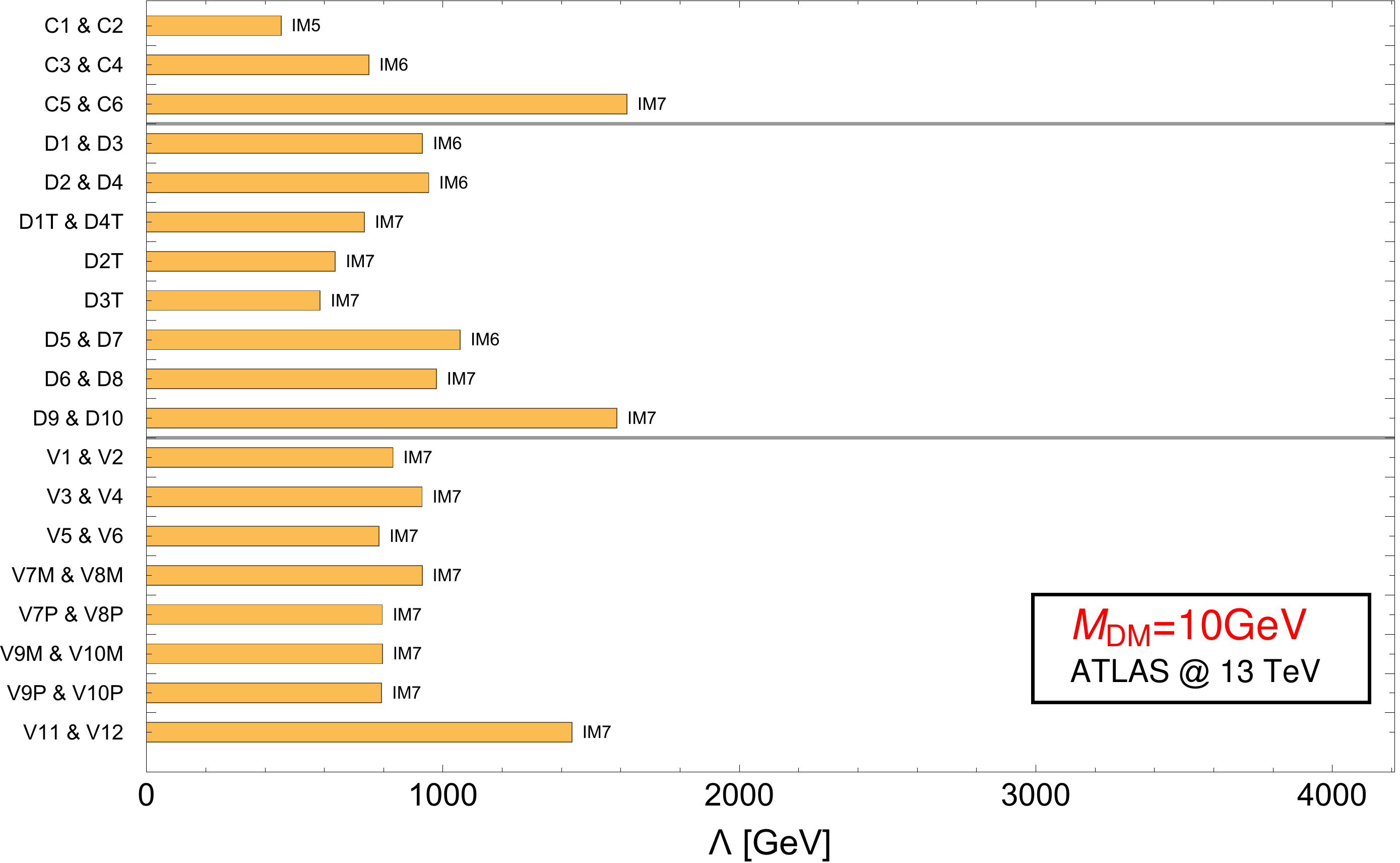}
    \vspace{2mm}
    \subcaption{$M_{\rm{DM}} = 10$ GeV}
    \vspace{4mm}
  \end{subfigure}
  \begin{subfigure}[t]{1.1\textwidth}
    \centering
    \includegraphics[width=0.51\textwidth]{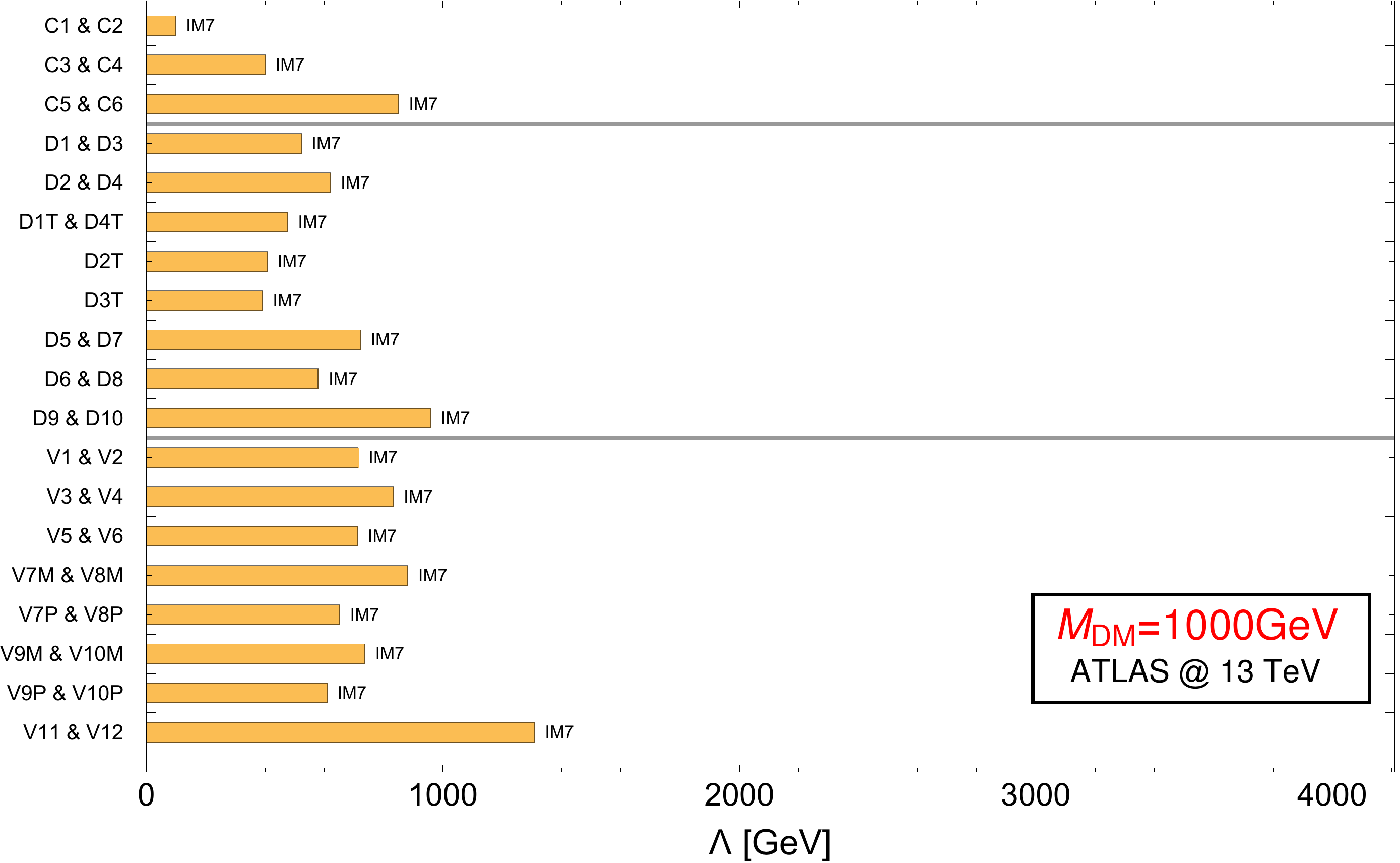}
    \vspace{2mm}
    \subcaption{$M_{\rm{DM}} = 1000$ GeV}
    \vspace{4mm}
  \end{subfigure}
  \caption{\label{fig:LambdaLimits13TeV-detailed}Observed limits on the UV cut-off $\Lambda$ from the ATLAS search of Ref.~\cite{Aaboud:2016tnv} using 13 TeV data with a luminosity of 3.2~fb$^{-1}$. See the caption of Fig.~\ref{fig:LambdaLimits8TeV} for more details about the interpretation of the plot.}
\end{figure}

\begin{figure}[ht!]
  \captionsetup[subfigure]{skip=-1mm}
  \begin{subfigure}[t]{1.1\textwidth}
   \includegraphics[width=0.5\textwidth]{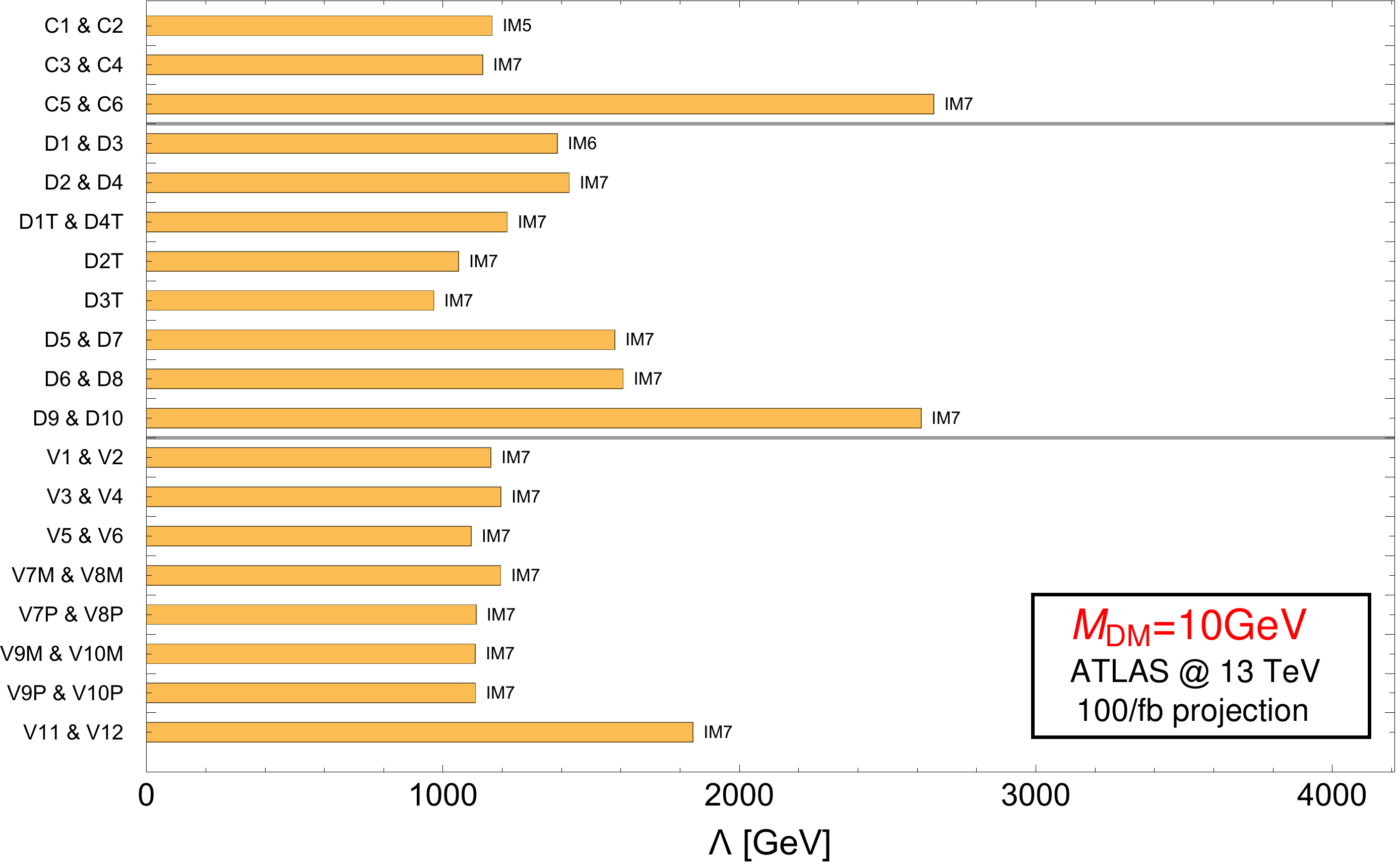}\quad
    \includegraphics[width=0.5\textwidth]{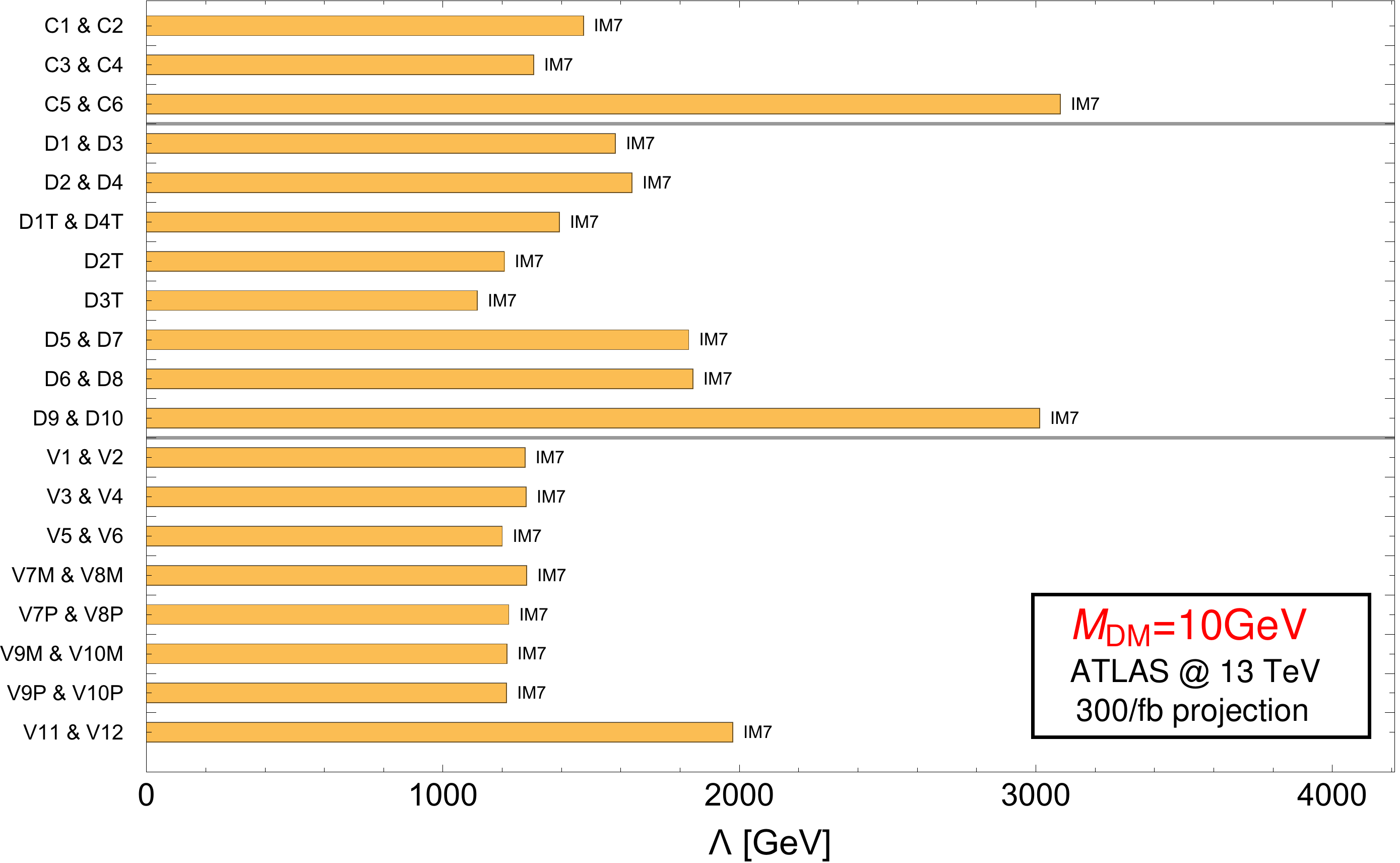}
    \subcaption{$M_{\rm{DM}} = 10$ GeV}
    \vspace{4mm}
  \end{subfigure}
  \begin{subfigure}[t]{1.1\textwidth}
    \includegraphics[width=0.5\textwidth]{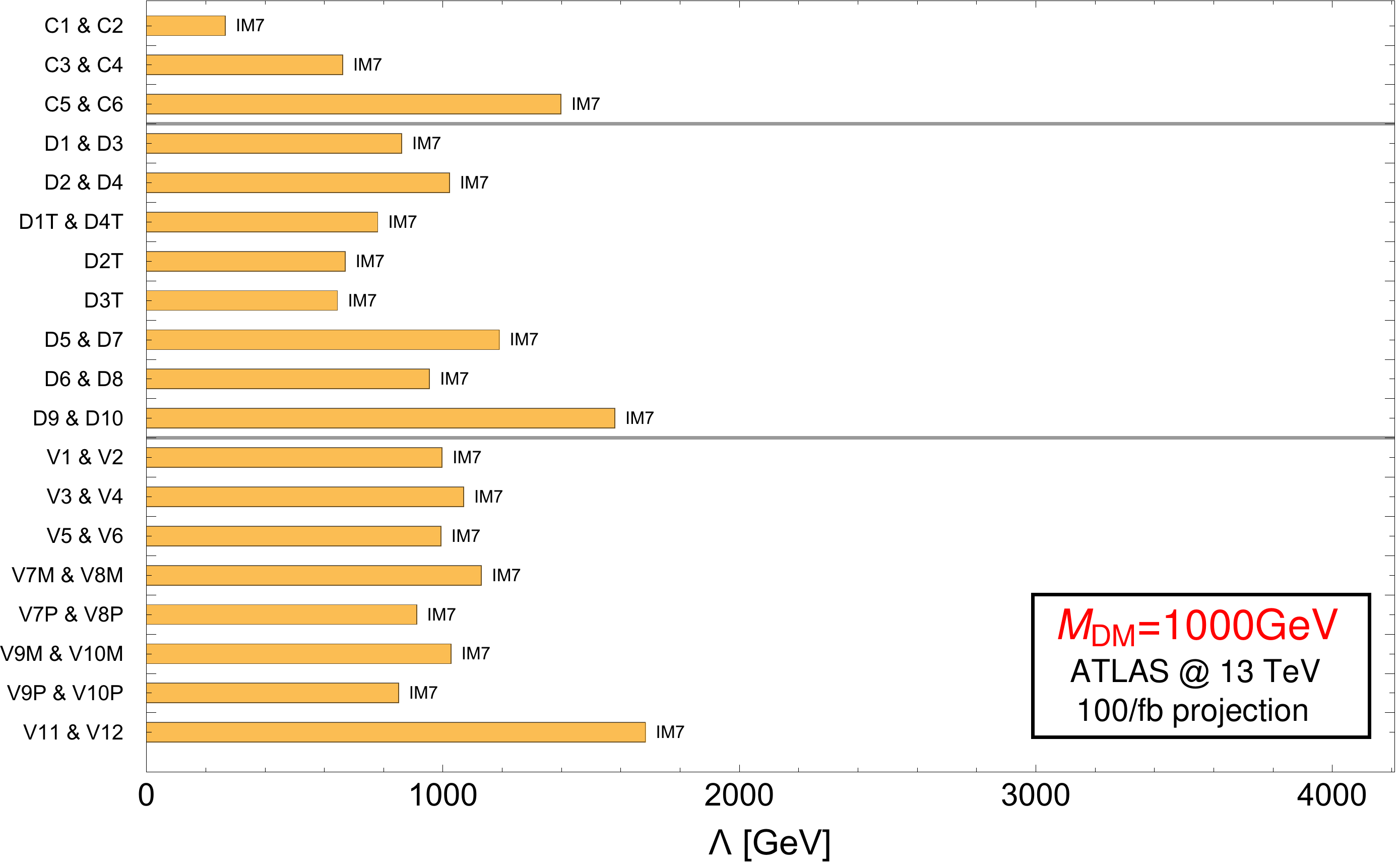}\quad
    \includegraphics[width=0.5\textwidth]{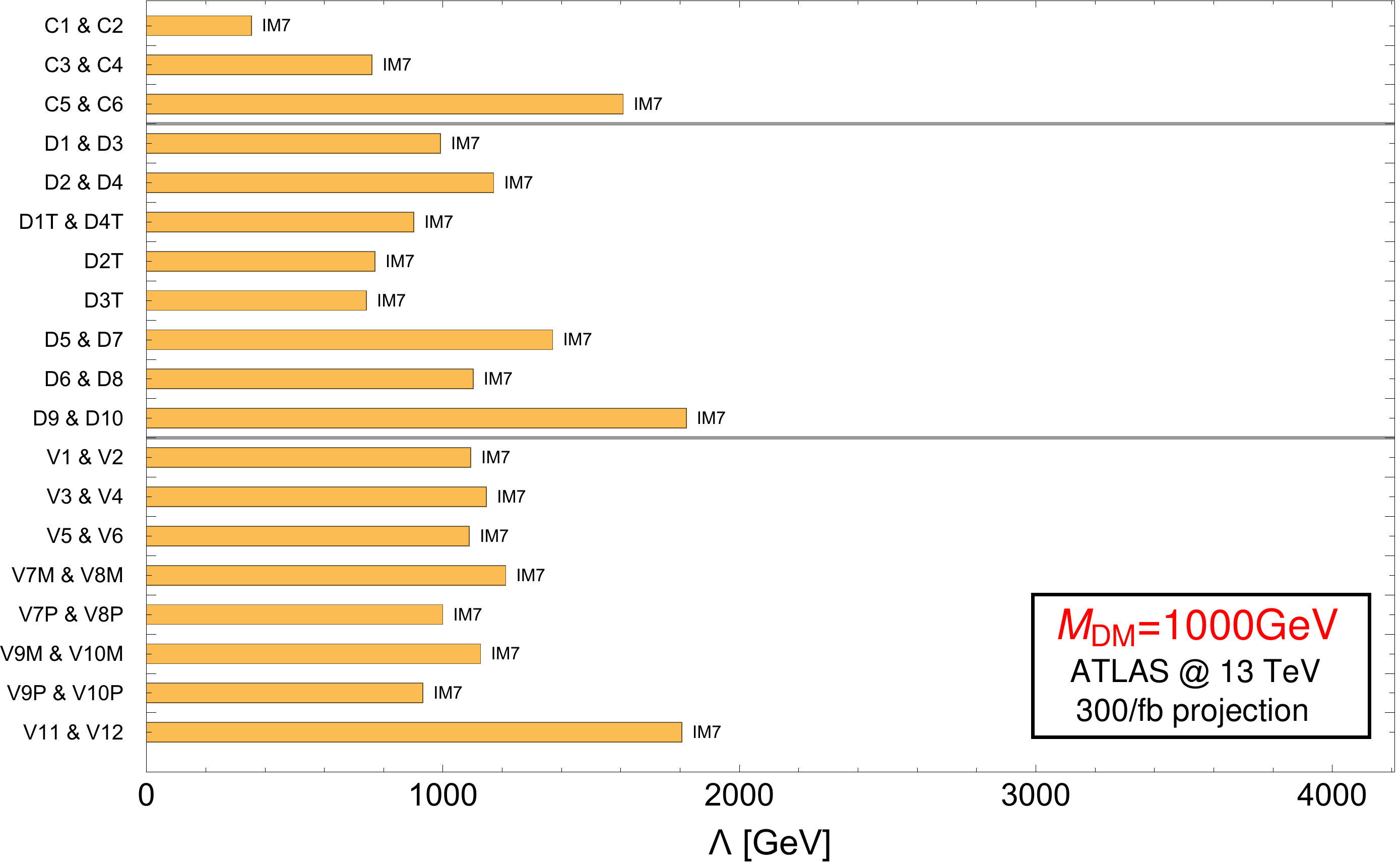}
    \subcaption{$M_{\rm{DM}} = 1000$ GeV}
    \vspace{4mm}
  \end{subfigure}
  \caption{\label{fig:LambdaLimits13TeVRescaled-detailed}Expected limits on the UV cut-off $\Lambda$ considering the selection of the ATLAS search of Ref.~\cite{Aaboud:2016tnv} and rescaling to luminosities of 100~fb$^{-1}$ (left panels) and 300~fb$^{-1}$ (right panels).}
\end{figure}

In this section we present plots for additional $M_{DM}=10$ and 1000 GeV
for the LHC reach complementary to those presented in Section~\ref{sec:comparison}.
In Fig~\ref{fig:LambdaLimits8TeV-detailed} a),b),c) we present results, in terms of 95\%CL on $\Lambda$
for LHC@8TeV from ATLAS(left) and CMS(right).
In Fig~\ref{fig:LambdaLimits13TeV-detailed}  we present analogous  results
for ATLAS analysis for  LHC@13TeV 3.2fb$^{-1}$ data.
In Fig~\ref{fig:LambdaLimits13TeVRescaled-detailed}  we present results
for high luminocity  projections for 100 fb$^{-1}$(left) and 300 fb$^{-1}$(right)
for  LHC@13TeV.

\clearpage
\newpage
\bibliography{bib}

\providecommand{\href}[2]{#2}\begingroup\raggedright\begin{thebibliography}{10}

\bibitem{Aad:2015zva}
{\scshape ATLAS} collaboration, G.~Aad et~al., \emph{{Search for new phenomena
  in final states with an energetic jet and large missing transverse momentum
  in pp collisions at $\sqrt{s}=$8 TeV with the ATLAS detector}},
  \href{http://dx.doi.org/10.1140/epjc/s10052-015-3517-3,
  10.1140/epjc/s10052-015-3639-7}{\emph{Eur. Phys. J.} {\bf C75} (2015) 299},
  [\href{http://arxiv.org/abs/1502.01518}{{\tt 1502.01518}}].

\bibitem{Khachatryan:2014rra}
{\scshape CMS} collaboration, V.~Khachatryan et~al., \emph{{Search for dark
  matter, extra dimensions, and unparticles in monojet events in
  proton–proton collisions at $\sqrt{s} = 8$ TeV}},
  \href{http://dx.doi.org/10.1140/epjc/s10052-015-3451-4}{\emph{Eur. Phys. J.}
  {\bf C75} (2015) 235}, [\href{http://arxiv.org/abs/1408.3583}{{\tt
  1408.3583}}].

\bibitem{Aaboud:2016tnv}
{\scshape ATLAS} collaboration, M.~Aaboud et~al., \emph{{Search for new
  phenomena in final states with an energetic jet and large missing transverse
  momentum in $pp$ collisions at $\sqrt{s}=13$ TeV using the ATLAS detector}},
  \href{http://arxiv.org/abs/1604.07773}{{\tt 1604.07773}}.

\bibitem{CMS:2016flr}
{\scshape CMS} collaboration, C.~Collaboration, \emph{{Search for new physics
  in a boosted hadronic monotop final state using $12.9~\mathrm{fb}^{-1}$ of
  $\sqrt{s}=13~\mathrm{TeV}$ data}}, {\emph{CMS-PAS-EXO-16-040} (2016) }.

\bibitem{Goldberg:1983nd}
H.~Goldberg, \emph{{Constraint on the Photino Mass from Cosmology}},
  \href{http://dx.doi.org/10.1103/PhysRevLett.50.1419}{\emph{Phys. Rev. Lett.}
  {\bf 50} (1983) 1419}.

\bibitem{Ellis:1983ew}
J.~R. Ellis, J.~S. Hagelin, D.~V. Nanopoulos, K.~A. Olive and M.~Srednicki,
  \emph{{Supersymmetric Relics from the Big Bang}},
  \href{http://dx.doi.org/10.1016/0550-3213(84)90461-9}{\emph{Nucl. Phys.} {\bf
  B238} (1984) 453--476}.

\bibitem{Antoniadis:1990ew}
I.~Antoniadis, \emph{{A Possible new dimension at a few TeV}},
  \href{http://dx.doi.org/10.1016/0370-2693(90)90617-F}{\emph{Phys.Lett.} {\bf
  B246} (1990) 377--384}.

\bibitem{Appelquist:2000nn}
T.~Appelquist, H.-C. Cheng and B.~A. Dobrescu, \emph{{Bounds on universal extra
  dimensions}},
  \href{http://dx.doi.org/10.1103/PhysRevD.64.035002}{\emph{Phys.Rev.} {\bf
  D64} (2001) 035002}, [\href{http://arxiv.org/abs/hep-ph/0012100}{{\tt
  hep-ph/0012100}}].

\bibitem{Servant:2002aq}
G.~Servant and T.~M.~P. Tait, \emph{{Is the lightest Kaluza-Klein particle a
  viable dark matter candidate?}},
  \href{http://dx.doi.org/10.1016/S0550-3213(02)01012-X}{\emph{Nucl. Phys.}
  {\bf B650} (2003) 391--419}, [\href{http://arxiv.org/abs/hep-ph/0206071}{{\tt
  hep-ph/0206071}}].

\bibitem{Csaki:2003sh}
C.~Csaki, C.~Grojean, J.~Hubisz, Y.~Shirman and J.~Terning, \emph{{Fermions on
  an interval: Quark and lepton masses without a Higgs}},
  \href{http://dx.doi.org/10.1103/PhysRevD.70.015012}{\emph{Phys.Rev.} {\bf
  D70} (2004) 015012}, [\href{http://arxiv.org/abs/hep-ph/0310355}{{\tt
  hep-ph/0310355}}].

\bibitem{ArkaniHamed:2002qx}
N.~Arkani-Hamed, A.~Cohen, E.~Katz, A.~Nelson, T.~Gregoire et~al., \emph{{The
  Minimal moose for a little Higgs}},
  \href{http://dx.doi.org/10.1088/1126-6708/2002/08/021}{\emph{JHEP} {\bf 0208}
  (2002) 021}, [\href{http://arxiv.org/abs/hep-ph/0206020}{{\tt
  hep-ph/0206020}}].

\bibitem{Cheng:2003ju}
H.-C. Cheng and I.~Low, \emph{{TeV symmetry and the little hierarchy problem}},
  \href{http://dx.doi.org/10.1088/1126-6708/2003/09/051}{\emph{JHEP} {\bf 09}
  (2003) 051}, [\href{http://arxiv.org/abs/hep-ph/0308199}{{\tt
  hep-ph/0308199}}].

\bibitem{Cheng:2004yc}
H.-C. Cheng and I.~Low, \emph{{Little hierarchy, little Higgses, and a little
  symmetry}},
  \href{http://dx.doi.org/10.1088/1126-6708/2004/08/061}{\emph{JHEP} {\bf 08}
  (2004) 061}, [\href{http://arxiv.org/abs/hep-ph/0405243}{{\tt
  hep-ph/0405243}}].

\bibitem{Low:2004xc}
I.~Low, \emph{{T parity and the littlest Higgs}},
  \href{http://dx.doi.org/10.1088/1126-6708/2004/10/067}{\emph{JHEP} {\bf 10}
  (2004) 067}, [\href{http://arxiv.org/abs/hep-ph/0409025}{{\tt
  hep-ph/0409025}}].

\bibitem{Hubisz:2004ft}
J.~Hubisz and P.~Meade, \emph{{Phenomenology of the littlest Higgs with
  T-parity}}, \href{http://dx.doi.org/10.1103/PhysRevD.71.035016}{\emph{Phys.
  Rev.} {\bf D71} (2005) 035016},
  [\href{http://arxiv.org/abs/hep-ph/0411264}{{\tt hep-ph/0411264}}].

\bibitem{Cheng:2005as}
H.-C. Cheng, I.~Low and L.-T. Wang, \emph{{Top partners in little Higgs
  theories with T-parity}},
  \href{http://dx.doi.org/10.1103/PhysRevD.74.055001}{\emph{Phys. Rev.} {\bf
  D74} (2006) 055001}, [\href{http://arxiv.org/abs/hep-ph/0510225}{{\tt
  hep-ph/0510225}}].

\bibitem{Hubisz:2005tx}
J.~Hubisz, P.~Meade, A.~Noble and M.~Perelstein, \emph{{Electroweak precision
  constraints on the littlest Higgs model with T parity}},
  \href{http://dx.doi.org/10.1088/1126-6708/2006/01/135}{\emph{JHEP} {\bf 01}
  (2006) 135}, [\href{http://arxiv.org/abs/hep-ph/0506042}{{\tt
  hep-ph/0506042}}].

\bibitem{Nussinov:1985xr}
S.~Nussinov, \emph{{Technocosmology: could a technibaryon excess provide a
  'natural' missing mass candidate?}},
  \href{http://dx.doi.org/10.1016/0370-2693(85)90689-6}{\emph{Phys. Lett.} {\bf
  B165} (1985) 55--58}.

\bibitem{Barr:1990ca}
S.~M. Barr, R.~S. Chivukula and E.~Farhi, \emph{{Electroweak Fermion Number
  Violation and the Production of Stable Particles in the Early Universe}},
  \href{http://dx.doi.org/10.1016/0370-2693(90)91661-T}{\emph{Phys. Lett.} {\bf
  B241} (1990) 387--391}.

\bibitem{Gudnason:2006ug}
S.~B. Gudnason, C.~Kouvaris and F.~Sannino, \emph{{Towards working technicolor:
  Effective theories and dark matter}},
  \href{http://dx.doi.org/10.1103/PhysRevD.73.115003}{\emph{Phys. Rev.} {\bf
  D73} (2006) 115003}, [\href{http://arxiv.org/abs/hep-ph/0603014}{{\tt
  hep-ph/0603014}}].

\bibitem{Fox:2011pm}
P.~J. Fox, R.~Harnik, J.~Kopp and Y.~Tsai, \emph{{Missing Energy Signatures of
  Dark Matter at the LHC}},
  \href{http://dx.doi.org/10.1103/PhysRevD.85.056011}{\emph{Phys. Rev.} {\bf
  D85} (2012) 056011}, [\href{http://arxiv.org/abs/1109.4398}{{\tt
  1109.4398}}].

\bibitem{Rajaraman:2011wf}
A.~Rajaraman, W.~Shepherd, T.~M.~P. Tait and A.~M. Wijangco, \emph{{LHC Bounds
  on Interactions of Dark Matter}},
  \href{http://dx.doi.org/10.1103/PhysRevD.84.095013}{\emph{Phys. Rev.} {\bf
  D84} (2011) 095013}, [\href{http://arxiv.org/abs/1108.1196}{{\tt
  1108.1196}}].

\bibitem{Goodman:2010ku}
J.~Goodman, M.~Ibe, A.~Rajaraman, W.~Shepherd, T.~M. Tait et~al.,
  \emph{{Constraints on Dark Matter from Colliders}},
  \href{http://dx.doi.org/10.1103/PhysRevD.82.116010}{\emph{Phys.Rev.} {\bf
  D82} (2010) 116010}, [\href{http://arxiv.org/abs/1008.1783}{{\tt
  1008.1783}}].

\bibitem{Bai:2010hh}
Y.~Bai, P.~J. Fox and R.~Harnik, \emph{{The Tevatron at the Frontier of Dark
  Matter Direct Detection}},
  \href{http://dx.doi.org/10.1007/JHEP12(2010)048}{\emph{JHEP} {\bf 12} (2010)
  048}, [\href{http://arxiv.org/abs/1005.3797}{{\tt 1005.3797}}].

\bibitem{Beltran:2010ww}
M.~Beltran, D.~Hooper, E.~W. Kolb, Z.~A.~C. Krusberg and T.~M.~P. Tait,
  \emph{{Maverick dark matter at colliders}},
  \href{http://dx.doi.org/10.1007/JHEP09(2010)037}{\emph{JHEP} {\bf 09} (2010)
  037}, [\href{http://arxiv.org/abs/1002.4137}{{\tt 1002.4137}}].

\bibitem{Goodman:2010yf}
J.~Goodman, M.~Ibe, A.~Rajaraman, W.~Shepherd, T.~M.~P. Tait and H.-B. Yu,
  \emph{{Constraints on Light Majorana dark Matter from Colliders}},
  \href{http://dx.doi.org/10.1016/j.physletb.2010.11.009}{\emph{Phys. Lett.}
  {\bf B695} (2011) 185--188}, [\href{http://arxiv.org/abs/1005.1286}{{\tt
  1005.1286}}].

\bibitem{Buchmueller:2013dya}
O.~Buchmueller, M.~J. Dolan and C.~McCabe, \emph{{Beyond Effective Field Theory
  for Dark Matter Searches at the LHC}},
  \href{http://dx.doi.org/10.1007/JHEP01(2014)025}{\emph{JHEP} {\bf 1401}
  (2014) 025}, [\href{http://arxiv.org/abs/1308.6799}{{\tt 1308.6799}}].

\bibitem{Fox:2011fx}
P.~J. Fox, R.~Harnik, J.~Kopp and Y.~Tsai, \emph{{LEP Shines Light on Dark
  Matter}}, \href{http://dx.doi.org/10.1103/PhysRevD.84.014028}{\emph{Phys.
  Rev.} {\bf D84} (2011) 014028}, [\href{http://arxiv.org/abs/1103.0240}{{\tt
  1103.0240}}].

\bibitem{Shoemaker:2011vi}
I.~M. Shoemaker and L.~Vecchi, \emph{{Unitarity and Monojet Bounds on Models
  for DAMA, CoGeNT, and CRESST-II}},
  \href{http://dx.doi.org/10.1103/PhysRevD.86.015023}{\emph{Phys. Rev.} {\bf
  D86} (2012) 015023}, [\href{http://arxiv.org/abs/1112.5457}{{\tt
  1112.5457}}].

\bibitem{Fox:2012ru}
P.~J. Fox and C.~Williams, \emph{{Next-to-Leading Order Predictions for Dark
  Matter Production at Hadron Colliders}},
  \href{http://dx.doi.org/10.1103/PhysRevD.87.054030}{\emph{Phys. Rev.} {\bf
  D87} (2013) 054030}, [\href{http://arxiv.org/abs/1211.6390}{{\tt
  1211.6390}}].

\bibitem{Haisch:2012kf}
U.~Haisch, F.~Kahlhoefer and J.~Unwin, \emph{{The impact of heavy-quark loops
  on LHC dark matter searches}},
  \href{http://dx.doi.org/10.1007/JHEP07(2013)125}{\emph{JHEP} {\bf 07} (2013)
  125}, [\href{http://arxiv.org/abs/1208.4605}{{\tt 1208.4605}}].

\bibitem{Busoni:2013lha}
G.~Busoni, A.~De~Simone, E.~Morgante and A.~Riotto, \emph{{On the Validity of
  the Effective Field Theory for Dark Matter Searches at the LHC}},
  \href{http://dx.doi.org/10.1016/j.physletb.2013.11.069}{\emph{Phys. Lett.}
  {\bf B728} (2014) 412--421}, [\href{http://arxiv.org/abs/1307.2253}{{\tt
  1307.2253}}].

\bibitem{Busoni:2014haa}
G.~Busoni, A.~De~Simone, T.~Jacques, E.~Morgante and A.~Riotto, \emph{{On the
  Validity of the Effective Field Theory for Dark Matter Searches at the LHC
  Part III: Analysis for the $t$-channel}},
  \href{http://dx.doi.org/10.1088/1475-7516/2014/09/022}{\emph{JCAP} {\bf 1409}
  (2014) 022}, [\href{http://arxiv.org/abs/1405.3101}{{\tt 1405.3101}}].

\bibitem{Busoni2014a}
G.~Busoni, A.~De~Simone, J.~Gramling, E.~Morgante and A.~Riotto, \emph{{On the
  Validity of the Effective Field Theory for Dark Matter Searches at the LHC,
  Part II: Complete Analysis for the $s$-channel}},
  \href{http://dx.doi.org/10.1088/1475-7516/2014/06/060}{\emph{JCAP} {\bf 1406}
  (2014) 060}, [\href{http://arxiv.org/abs/1402.1275}{{\tt 1402.1275}}].

\bibitem{Abercrombie:2015wmb}
D.~Abercrombie et~al., \emph{{Dark Matter Benchmark Models for Early LHC Run-2
  Searches: Report of the ATLAS/CMS Dark Matter Forum}},
  \href{http://arxiv.org/abs/1507.00966}{{\tt 1507.00966}}.

\bibitem{Endo:2014mja}
M.~Endo and Y.~Yamamoto, \emph{{Unitarity Bounds on Dark Matter Effective
  Interactions at LHC}},
  \href{http://dx.doi.org/10.1007/JHEP06(2014)126}{\emph{JHEP} {\bf 06} (2014)
  126}, [\href{http://arxiv.org/abs/1403.6610}{{\tt 1403.6610}}].

\bibitem{Busoni:2014sya}
G.~Busoni, A.~De~Simone, J.~Gramling, E.~Morgante and A.~Riotto, \emph{{On the
  Validity of the Effective Field Theory for Dark Matter Searches at the LHC,
  Part II: Complete Analysis for the $s$-channel}},
  \href{http://dx.doi.org/10.1088/1475-7516/2014/06/060}{\emph{JCAP} {\bf 1406}
  (2014) 060}, [\href{http://arxiv.org/abs/1402.1275}{{\tt 1402.1275}}].

\bibitem{Buchmueller:2014yoa}
O.~Buchmueller, M.~J. Dolan, S.~A. Malik and C.~McCabe, \emph{{Characterising
  dark matter searches at colliders and direct detection experiments: Vector
  mediators}}, \href{http://dx.doi.org/10.1007/JHEP01(2015)037}{\emph{JHEP}
  {\bf 1501} (2015) 037}, [\href{http://arxiv.org/abs/1407.8257}{{\tt
  1407.8257}}].

\bibitem{Buckley:2014fba}
M.~R. Buckley, D.~Feld and D.~Goncalves, \emph{{Scalar Simplified Models for
  Dark Matter}},
  \href{http://dx.doi.org/10.1103/PhysRevD.91.015017}{\emph{Phys. Rev.} {\bf
  D91} (2015) 015017}, [\href{http://arxiv.org/abs/1410.6497}{{\tt
  1410.6497}}].

\bibitem{Abdallah:2015ter}
J.~Abdallah et~al., \emph{{Simplified Models for Dark Matter Searches at the
  LHC}}, \href{http://dx.doi.org/10.1016/j.dark.2015.08.001}{\emph{Phys. Dark
  Univ.} {\bf 9-10} (2015) 8--23}, [\href{http://arxiv.org/abs/1506.03116}{{\tt
  1506.03116}}].

\bibitem{Abdallah:2014hon}
J.~Abdallah, A.~Ashkenazi, A.~Boveia, G.~Busoni, A.~De~Simone et~al.,
  \emph{{Simplified Models for Dark Matter and Missing Energy Searches at the
  LHC}},  \href{http://arxiv.org/abs/1409.2893}{{\tt 1409.2893}}.

\bibitem{Englert:2016joy}
C.~Englert, M.~McCullough and M.~Spannowsky, \emph{{S-Channel Dark Matter
  Simplified Models and Unitarity}},
  \href{http://dx.doi.org/10.1016/j.dark.2016.09.002}{\emph{Phys. Dark Univ.}
  {\bf 14} (2016) 48--56}, [\href{http://arxiv.org/abs/1604.07975}{{\tt
  1604.07975}}].

\bibitem{Kahlhoefer:2015bea}
F.~Kahlhoefer, K.~Schmidt-Hoberg, T.~Schwetz and S.~Vogl, \emph{{Implications
  of unitarity and gauge invariance for simplified dark matter models}},
  \href{http://dx.doi.org/10.1007/JHEP02(2016)016}{\emph{JHEP} {\bf 02} (2016)
  016}, [\href{http://arxiv.org/abs/1510.02110}{{\tt 1510.02110}}].

\bibitem{Kumar:2015wya}
J.~Kumar, D.~Marfatia and D.~Yaylali, \emph{{Vector dark matter at the LHC}},
  \href{http://dx.doi.org/10.1103/PhysRevD.92.095027}{\emph{Phys. Rev.} {\bf
  D92} (2015) 095027}, [\href{http://arxiv.org/abs/1508.04466}{{\tt
  1508.04466}}].

\bibitem{Buchmuller:1985jz}
W.~Buchmuller and D.~Wyler, \emph{{Effective Lagrangian Analysis of New
  Interactions and Flavor Conservation}},
  \href{http://dx.doi.org/10.1016/0550-3213(86)90262-2}{\emph{Nucl. Phys.} {\bf
  B268} (1986) 621--653}.

\bibitem{Arzt:1993gz}
C.~Arzt, \emph{{Reduced effective Lagrangians}},
  \href{http://dx.doi.org/10.1016/0370-2693(94)01419-D}{\emph{Phys. Lett.} {\bf
  B342} (1995) 189--195}, [\href{http://arxiv.org/abs/hep-ph/9304230}{{\tt
  hep-ph/9304230}}].

\bibitem{Alwall:2011uj}
J.~Alwall, M.~Herquet, F.~Maltoni, O.~Mattelaer and T.~Stelzer, \emph{{MadGraph
  5 : Going Beyond}},
  \href{http://dx.doi.org/10.1007/JHEP06(2011)128}{\emph{JHEP} {\bf 1106}
  (2011) 128}, [\href{http://arxiv.org/abs/1106.0522}{{\tt 1106.0522}}].

\bibitem{Alwall:2014hca}
J.~Alwall, R.~Frederix, S.~Frixione, V.~Hirschi, F.~Maltoni, O.~Mattelaer
  et~al., \emph{{The automated computation of tree-level and next-to-leading
  order differential cross sections, and their matching to parton shower
  simulations}}, \href{http://dx.doi.org/10.1007/JHEP07(2014)079}{\emph{JHEP}
  {\bf 07} (2014) 079}, [\href{http://arxiv.org/abs/1405.0301}{{\tt
  1405.0301}}].

\bibitem{Belyaev:2012qa}
A.~Belyaev, N.~D. Christensen and A.~Pukhov, \emph{{CalcHEP 3.4 for collider
  physics within and beyond the Standard Model}},
  \href{http://dx.doi.org/10.1016/j.cpc.2013.01.014}{\emph{Comput.Phys.Commun.}
  {\bf 184} (2013) 1729--1769}, [\href{http://arxiv.org/abs/1207.6082}{{\tt
  1207.6082}}].

\bibitem{Semenov:2010qt}
A.~Semenov, \emph{{LanHEP - a package for automatic generation of Feynman rules
  from the Lagrangian. Updated version 3.1}},
  \href{http://arxiv.org/abs/1005.1909}{{\tt 1005.1909}}.

\bibitem{Alloul:2013bka}
A.~Alloul, N.~D. Christensen, C.~Degrande, C.~Duhr and B.~Fuks,
  \emph{{FeynRules 2.0 - A complete toolbox for tree-level phenomenology}},
  \href{http://dx.doi.org/10.1016/j.cpc.2014.04.012}{\emph{Comput. Phys.
  Commun.} {\bf 185} (2014) 2250--2300},
  [\href{http://arxiv.org/abs/1310.1921}{{\tt 1310.1921}}].

\bibitem{hepmdb}
M.~Bondarenko, A.~Belyaev, L.~Basso, E.~Boos, V.~Bunichev et~al., \emph{{High
  Energy Physics Model Database : Towards decoding of the underlying theory
  (within Les Houches 2011: Physics at TeV Colliders New Physics Working Group
  Report)}},  \href{http://arxiv.org/abs/1203.1488}{{\tt 1203.1488}}.

\bibitem{hepmdbmodel}
\url{https://hepmdb.soton.ac.uk/index.php?mod=user&act=showmodel&id=166}.

\bibitem{Pumplin:2002vw}
J.~Pumplin, D.~R. Stump, J.~Huston, H.~L. Lai, P.~M. Nadolsky and W.~K. Tung,
  \emph{{New generation of parton distributions with uncertainties from global
  QCD analysis}},
  \href{http://dx.doi.org/10.1088/1126-6708/2002/07/012}{\emph{JHEP} {\bf 07}
  (2002) 012}, [\href{http://arxiv.org/abs/hep-ph/0201195}{{\tt
  hep-ph/0201195}}].

\bibitem{Backovic:2015soa}
M.~Backovi\u0107, M.~Kramer, F.~Maltoni, A.~Martini, K.~Mawatari and M.~Pellen,
  \emph{{Higher-order QCD predictions for dark matter production at the LHC in
  simplified models with s-channel mediators}},
  \href{http://dx.doi.org/10.1140/epjc/s10052-015-3700-6}{\emph{Eur. Phys. J.}
  {\bf C75} (2015) 482}, [\href{http://arxiv.org/abs/1508.05327}{{\tt
  1508.05327}}].

\bibitem{Sjostrand:2006za}
T.~Sjostrand, S.~Mrenna and P.~Z. Skands, \emph{{PYTHIA 6.4 Physics and
  Manual}}, \href{http://dx.doi.org/10.1088/1126-6708/2006/05/026}{\emph{JHEP}
  {\bf 0605} (2006) 026}, [\href{http://arxiv.org/abs/hep-ph/0603175}{{\tt
  hep-ph/0603175}}].

\bibitem{deFavereau:2013fsa}
{\scshape DELPHES 3} collaboration, J.~de~Favereau, C.~Delaere, P.~Demin,
  A.~Giammanco, V.~Lemaître, A.~Mertens et~al., \emph{{DELPHES 3, A modular
  framework for fast simulation of a generic collider experiment}},
  \href{http://dx.doi.org/10.1007/JHEP02(2014)057}{\emph{JHEP} {\bf 02} (2014)
  057}, [\href{http://arxiv.org/abs/1307.6346}{{\tt 1307.6346}}].

\bibitem{Cacciari:2005hq}
M.~Cacciari and G.~P. Salam, \emph{{Dispelling the $N^{3}$ myth for the $k_t$
  jet-finder}},
  \href{http://dx.doi.org/10.1016/j.physletb.2006.08.037}{\emph{Phys. Lett.}
  {\bf B641} (2006) 57--61}, [\href{http://arxiv.org/abs/hep-ph/0512210}{{\tt
  hep-ph/0512210}}].

\bibitem{Cacciari:2011ma}
M.~Cacciari, G.~P. Salam and G.~Soyez, \emph{{FastJet User Manual}},
  \href{http://dx.doi.org/10.1140/epjc/s10052-012-1896-2}{\emph{Eur. Phys. J.}
  {\bf C72} (2012) 1896}, [\href{http://arxiv.org/abs/1111.6097}{{\tt
  1111.6097}}].

\bibitem{Drees:2013wra}
M.~Drees, H.~Dreiner, D.~Schmeier, J.~Tattersall and J.~S. Kim,
  \emph{{CheckMATE: Confronting your Favourite New Physics Model with LHC
  Data}}, \href{http://dx.doi.org/10.1016/j.cpc.2014.10.018}{\emph{Comput.
  Phys. Commun.} {\bf 187} (2014) 227--265},
  [\href{http://arxiv.org/abs/1312.2591}{{\tt 1312.2591}}].

\bibitem{CMS:2016tns}
{\scshape CMS} collaboration, C.~Collaboration, \emph{{Search for dark matter
  production in association with jets, or hadronically decaying W or Z boson at
  $\sqrt{s} = 13$ TeV}}, {\emph{CMS-PAS-EXO-16-013} (2016) }.

\bibitem{EXO-16-013}
CMS:EXO-16-013.
  \url{http://cms-results.web.cern.ch/cms-results/public-results/preliminary-results/EXO-16-013/#AddFig},
  2016.

\bibitem{Barducci:2015ffa}
D.~Barducci, A.~Belyaev, A.~K.~M. Bharucha, W.~Porod and V.~Sanz,
  \emph{{Uncovering Natural Supersymmetry via the interplay between the LHC and
  Direct Dark Matter Detection}},
  \href{http://dx.doi.org/10.1007/JHEP07(2015)066}{\emph{JHEP} {\bf 07} (2015)
  066}, [\href{http://arxiv.org/abs/1504.02472}{{\tt 1504.02472}}].

\bibitem{Nishi:2004st}
C.~C. Nishi, \emph{{Simple derivation of general Fierz-like identities}},
  \href{http://dx.doi.org/10.1119/1.2074087}{\emph{Am. J. Phys.} {\bf 73}
  (2005) 1160--1163}, [\href{http://arxiv.org/abs/hep-ph/0412245}{{\tt
  hep-ph/0412245}}].

\bibitem{Itzykson:1980rh}
C.~Itzykson and J.~B. Zuber, \emph{{Quantum Field Theory}}.
\newblock International Series In Pure and Applied Physics. McGraw-Hill, New
  York, 1980.

\end{thebibliography}\endgroup
\bibliographystyle{JHEP}
\end{document}